\documentclass[12pt,eadjoint tfnpsf]{article}
\usepackage{footnote}

\catcode`\@=11
\@addtoreset{equation}{section}

\global\arraycolsep=1pt

\usepackage{amsbsy,amssymb,latexsym,amsfonts, amsmath}
\usepackage{mathrsfs}
\usepackage{graphicx}
\usepackage{youngtab}
\usepackage{marvosym}
\usepackage{ytableau}

\usepackage{hyperref}
\hypersetup{
     colorlinks   = true,
     linkcolor	  = black, 
     citecolor    = gray,
     urlcolor	  = blue	
}
\usepackage{accents}
\newcommand\thickbar[1]{\accentset{\rule{.7em}{1.4pt}}{#1}}
\newcommand{\ubar}[1]{\underaccent{\bar}{#1}}

\newcommand{\beq}{\begin{equation}}
\newcommand{\eeq}{\end{equation}}
\newcommand{\bea}{\begin{eqnarray}}
\newcommand{\eea}{\end{eqnarray}}

\newcommand\C{{\mathbb{C}}}

\def\E{{\epsilon_{1}}}
\def\EE{{\epsilon_{2}}}

\renewcommand{\thefootnote}{\fnsymbol{footnote}}

\def\XXint#1#2#3{{\setbox0=\hbox{$#1{#2#3}{\int}$}
     \vcenter{\hbox{$#2#3$}}\kern-.5\wd0}}

\begin{document}

\begin{titlepage}

\begin{flushright}
\normalsize
~~~~
SISSA  23/2015/FISI-MATE
\end{flushright}

\vspace{80pt}

\begin{center}
{\LARGE Quantum Cohomology and Quantum Hydrodynamics \\ \vspace{0.5 cm}} 
{\LARGE from Supersymmetric Quiver Gauge Theories}
\end{center}

\vspace{25pt}

\begin{center}
{Giulio Bonelli, Antonio Sciarappa, Alessandro Tanzini and Petr Vasko }\\

\vspace{15pt}

International School of Advanced Studies (SISSA) \\via Bonomea 265, 34136 Trieste, Italy 
and INFN, Sezione di Trieste \footnote{email: bonelli,asciara,tanzini,vaskop@sissa.it}\\


\end{center}

\vspace{20pt}

Abstract:
We study the connection between $\mathcal{N}=2$ supersymmetric gauge theories,
quantum cohomology and quantum integrable systems of hydrodynamic type.
We consider gauge theories on ALE spaces of $A$ and $D$-type and discuss how they 
describe the quantum cohomology of the corresponding Nakajima's quiver varieties.
We also discuss how the exact evaluation of local BPS observables in the gauge theory
can be used to calculate the spectrum of quantum Hamiltonians of
spin Calogero integrable systems and spin Intermediate Long Wave hydrodynamics.      
This is explicitly obtained by a Bethe Ansatz Equation provided by the quiver gauge theory
in terms of its adjacency matrix.
\vfill

\setcounter{footnote}{0}
\renewcommand{\thefootnote}{\arabic{footnote}}

\end{titlepage}


\tableofcontents


\section{Introduction and Discussion} \label{chap1}

The study of BPS correlators in $\mathcal{N}=2$ supersymmetric gauge theories reveals to be a rich source of results in various branches of modern mathematical physics, ranging from classical \cite{Donagi:1995cf,Gorsky:1995zq,Martinec:1995by}
and quantum \cite{Nekrasov:2009rc} integrable systems to topological invariants \cite{1991hep.th...12056W,Witten:1988ze,COM:501896}. In this paper we use exact results in supersymmetric gauge theories 
to highlight new connections between quantum cohomology of algebraic varieties \cite{2012arXiv1211.1287M}
and quantum integrable systems, focusing on Nakajima's quiver varieties \cite{Nakajima}. 
These have a manifold interest, since they host representations of infinite dimensional Lie algebrae of Kac-Moody type; 
moreover, they naturally describe moduli spaces of Yang-Mills instantons on ALE spaces \cite{Kron-Naka}
and are linked to free \cite{nakajima1999lectures,Vafa:1994tf,2003hep.th....6238N} and interacting \cite{Alday:2009aq,Belavin:2011pp,Nishioka:2011jk,Bonelli:2011jx} two-dimensional conformal field theories.   
This reflects in the special nature of the corresponding quantum integrable systems, which reveal to be of {\it hydrodynamical} type, namely admitting an infinite complete set of quantum Hamiltonians in involution. 
The vev of BPS local operators in the gauge theory on the $\Omega$-background captures the spectrum of these Hamiltonians \cite{Bonelli:2009zp,Ntalk,2014JHEP...07..141B}. The prototypical example for six-dimensional gauge theories is the quantum $gl_N$ Intermediate Long Wave system \cite{LR2,2013JHEP...11..155L} associated to the equivariant quantum cohomology of the ADHM quiver variety \cite{Ntalk,Otalk,NOinp,2014JHEP...07..141B}.
In the four dimensional limit this reduces to the correspondence between instanton counting and Benjamin-Ono
quantum system discussed in \cite{Alba:2010qc,Estienne:2011qk}. Our considerations are based on the
intriguing interplay taking place between the description of supersymmetric vacua of D-brane systems and quiver representation theory \cite{Douglas:1996sw}. The superstring background hosting the D-brane system fixes the quiver type, while the D-branes dimensions select the abelian category where the quiver representation is realized.  
We analyse in detail the D1-D5 system on ALE spaces and show that it provides a description of the equivariant 
quantum cohomology of the associated Nakajima's quiver varietes and links it to quantum spin Calogero and spin 
Intermediate Long Wave integrable systems.
More precisely, we study the supersymmetric gauged linear sigma model (GLSM) on $S^2$ which describes the low-energy 
D1-branes dynamics by calculating its exact partition function. 
In the {\em Higgs phase}, this model flows in the infrared to a non-linear sigma model with target space the Nakajima's quiver variety, naturally describing its equivariant quantum cohomology. An equivalent description of the gauged linear sigma model can be obtained in the {\em Coulomb phase}, giving rise to a Landau-Ginzburg model whose twisted superpotential is the Yang-Yang function \cite{Yang:1968rm} of a quantum integrable system which we identify with spin $gl_N$ Intermediate Long Wave system \cite{2014arXiv1411.3313A}.
The Bethe Ansatz Equation are provided by the general quiver gauge theory in terms of its adjacency matrix which reduces to the Cartan matrix if $\epsilon_1=\epsilon_2$. Once the BAE are formulated in these terms, they naturally extend to the affine ADE general case. 

Thus the equivariant quantum cohomology/quantum integrable system correspondence is realized as an incarnation of mirror symmetry for {\em non-abelian} GLSMs.


The gauge theories we consider are obtained as low energy theories of D-branes configurations. The generic D-brane set up we consider is realized in the ten dimensional target space $\widetilde{\left(\C^2/\Gamma\right)}\times {\cal O}_{\mathbb{P}^1}(-2) \times \C$ where $\Gamma$ is the ADE discrete group defining the ALE space as the quotient $\C^2/\Gamma$,
$\widetilde{\left(\C^2/\Gamma\right)}$ is its minimal resolution and ${\cal O}_{\mathbb{P}^1}(-2)$ is the total space of the canonical bundle over $\mathbb{P}^1$. 
We place $N$ D5-branes on $\widetilde{\left(\C^2/\Gamma\right)}\times \mathbb{P}^1$ and $k$ D1-branes on $\mathbb{P}^1$ and consider the resulting quiver GLSM. It corresponds to the affine quiver $\hat\Gamma$, where the nodes are representing the 
GLSM input data corresponding to the dynamics on the common $\mathbb{P}^1$. Notice that, if $\Gamma$ is trivial, then the resulting quiver gauge theory is the ADHM quiver.
The resulting affine quiver has indeed a natural interpretation from the string theory point of view in that it keeps into account all the different low energy open string sectors.


In section 2 we review the $S^2$ partition function and explain the quantum cohomology/quantum hydrodynamics
mirror symmetry in a general set-up. We then focus on the ADHM quiver whose Higgs phase is described in section 3 
along with the description of its associated quantum cohomology problem in the non linear sigma model, 
while its Coulomb phase is described in section 4 along with its quantum hydrodynamical associated model. 
The generalization to ALE quiver gauge theories is explained in section 5, where we consider $A_{p-1}$ and $D_p$ quivers and the respective mirror phases. In particular in subsection 5.3 we discuss the Bethe Ansatz equations for quantum spin $gl_N$ ILW arising from the LG model and the connection with quantum spin Calogero model.
Let us finally present some open issues. 
\begin{itemize}
\item D1-D5 systems are naturally related to Donaldson-Thomas (DT) invariants.
In particular, for the ADHM quiver, our results directly link to the ones obtained by Diaconescu \cite{2012JGP....62..763D} for non abelian local DT invariants on $\C^2\times \mathbb{P}^1$. 
Therefore it would be useful to explore the relevance of our results for the computation of non abelian DT invariants on $\widetilde{\left(\C^2/\Gamma\right)}\times \mathbb{P}^1$.
\item We did not consider $E_n$ quivers, although it would be interesting to analyse this class of models too.  
\item The extension of our computations to the presence of gauge theory defects -- as surface operators -- along the way of \cite{2014arXiv1412.6081B} could pave the way to further applications of supersymmetric gauge theories to the quantization of integrable systems. For example, this could provide explicit expressions for the quantum ILW eigenfunctions and the corresponding quantum Hamiltonians.
\item In the context of the AGT correspondence it was shown \cite{Alba:2010qc} that the basis of Virasoro descendants reproducing instanton counting on $\C^2$ has the special property of diagonalizing the Benjamin-Ono quantum Hamiltonians.
Our results point towards an analogous connection between spin $gl_N$ Benjamin-Ono quantum Hamiltonians and parafermionic ${\cal W}_N$-algebras that would be interesting to further explore. 
\item Finally, it would be interesting to understand how far the above correspondence goes in using
quiver gauge theories to describe the quantization of integrable models of hydrodynamical type.
\end{itemize}


\section{Gauged Linear Sigma Models on $S^2$: generalities} \label{chap2}

Since in this paper we will be working with supersymmetric $\mathcal{N}=(2,2)$ gauge theories on $S^2$, in this section we review the main points concerning localization on an euclidean two-sphere of radius $r$ along the lines of \cite{BC,DFGL}, to which we refer for further details. After briefly reviewing the computation of the partition function $Z_{S^2}$ for these theories, we will discuss how $Z_{S^2}$ is related to Givental's approach to genus zero Gromov-Witten theory for K\"{a}hler manifolds when we consider the Higgs branch of the theory. On the other hand, the Coulomb branch has strong connections to quantum integrable systems via the so-called Bethe/Gauge correspondence; in the last subsection we will discuss the meaning of the partition function in that context.


\subsection{$\mathcal{N}=(2,2)$ gauge theories on $S^2$} \label{sub2.1}

The two-sphere $S^2$ is a conformally flat space; it does not admit Killing spinors, but it admits four complex conformal Killing spinors, which realize the $\mathfrak{osp}(2 \vert 2, \mathbb{C})$ superconformal algebra on $S^2$. We take as $\mathcal{N}=(2,2)$ supersymmetry algebra on $S^2$ the subalgebra $\mathfrak{su}(2 \vert 1) \subset \mathfrak{osp}(2 \vert 2, \mathbb{C})$, realized by two out of the four conformal Killing spinors, which does not contain conformal nor superconformal transformations; its bosonic subalgebra $\mathfrak{su}(2) \oplus \mathfrak{u}(1)_R \subset \mathfrak{su}(2 \vert 1)$ generates the isometries of $S^2$ and an abelian vector R-symmetry, which is now part of the algebra and not an outer automorphism of it. \\

The basic multiplets (i.e. representations of the supersymmetry algebra) of two-dimensional $\mathcal{N}=(2,2)$ supersymmetry are vector and chiral multiplets, which arise by dimensional reduction of four dimensional $\mathcal{N} = 1$ vector and chiral multiplets. In detail
\begin{equation}
\begin{split}
\text{vector multiplet}:& \; (A_{\mu}, \sigma, \eta, \lambda, \bar{\lambda}, D) \\
\text{chiral multiplet}:& \; (\phi, \bar{\phi}, \psi, \bar{\psi}, F, \bar{F})
\end{split}
\end{equation}
with ($\lambda, \bar{\lambda},\psi, \bar{\psi}$) two component complex Dirac spinors, ($\sigma, \eta, D$) real scalar fields and ($\phi, \bar{\phi}, F, \bar{F}$) complex scalar fields. In two dimensions there also exist twisted chiral multiplets, whose matter content is the same as a chiral multiplet, but they satisfy different constraints:
\begin{equation}
\text{twisted chiral multiplet}: \; (Y, \bar{Y}, \chi, \bar{\chi}, G, \bar{G})
\end{equation}
The superfield strength
\begin{equation}
\Sigma = (\sigma - i \eta, \lambda, \bar{\lambda}, D - i F_{12})
\end{equation}
i.e. the supermultiplet containing the field strength is a particular case of twisted chiral multiplet.

The theories we are interested in are $\mathcal{N}=(2,2)$ gauged linear sigma models (GLSM) on $S^2$; these are made out of chiral and vector multiplets only, with canonical kinetic term for the fields. A GLSM is specified by the choice of the gauge group $G$, the representation $R$ of $G$ for the matter fields, and the matter interactions contained in the superpotential $W (\Phi)$, which is an R-charge 2 gauge-invariant holomorphic function of the chiral multiplets $\Phi = (\phi, \psi, F)$. If the gauge group admits an abelian term, we can also add a Fayet-Iliopoulos term $\xi$ and theta-angle $\theta$.\\

The most general renormalizable $\mathcal{N} = (2, 2)$ Lagrangian density of a GLSM on $S^2$ can be written down as
\begin{equation}
\mathcal{L} = \mathcal{L}_{\text{vec}} + \mathcal{L}_{\text{chiral}} + \mathcal{L}_W + \mathcal{L}_{FI}
\end{equation}
where
\begin{equation}
\begin{split}
\mathcal{L}_{\text{vec}} = \frac{1}{g^{2}}\text{Tr} \bigg\{ & \dfrac{1}{2} \left( F_{12} - \dfrac{\eta}{r} \right)^2 + \dfrac{1}{2} \left( D + \dfrac{\sigma}{r} \right)^2 + \dfrac{1}{2} D_{\mu}\sigma D^{\mu} \sigma + \dfrac{1}{2} D_{\mu}\eta D^{\mu} \eta \\
& - \dfrac{1}{2} [\sigma,\eta]^2 + \dfrac{i}{2} \bar{\lambda} \gamma^{\mu} D_{\mu} \lambda + \dfrac{i}{2} \bar{\lambda} [\sigma,\lambda] + \dfrac{1}{2} \bar{\lambda} \gamma_3 [\eta,\lambda] \bigg\}
\end{split}
\end{equation}
\begin{equation}
\begin{split}
\mathcal{L}_{\text{chiral}} = &D_\mu \bar\phi D^\mu \phi + \bar\phi \sigma^2 \phi + \bar\phi \eta^2 \phi + i \bar\phi D \phi + \bar F F + \frac{iq}r \bar\phi \sigma \phi + \frac{q(2-q)}{4r^2} \bar\phi \phi \\
&- i \bar\psi \gamma^\mu D_\mu \psi + i \bar\psi \sigma \psi - \bar\psi \gamma_3 \eta \psi + i \bar\psi \lambda \phi - i \bar\phi \bar\lambda \psi - \frac q{2r} \bar\psi \psi \label{chi}
\end{split}
\end{equation}
\begin{equation}
\mathcal{L}_W = \sum_j \frac{\partial W}{\partial\phi_j} F_j - \sum_{j,k} \frac12 \frac{\partial^2 W}{\partial\phi_j \partial \phi_k} \psi_j \psi_k
\end{equation}
\begin{equation}
\mathcal{L}_{FI} = \text{Tr} \left[ - i \xi D + i \frac\theta{2\pi} F_{12} \right]
\end{equation}
where $q$ is the R-charge of the chiral multiplet. In addition, if there is a global (flavour) symmetry group $G_F$ it is possible to turn on in a supersymmetric way \textit{twisted masses} for the chiral multiplets. These are obtained by first weakly gauging $G_F$, then coupling the matter fields to a vector multiplet for $G_F$, and finally giving a supersymmetric background VEV $\sigma^{\text{ext}}$, $\eta^{\text{ext}}$ for the scalar fields in that vector multiplet. Supersymmetry on $S^2$ requires $\sigma^{\text{ext}}$, $\eta^{\text{ext}}$ being constants and in the Cartan of $G_F$; in particular $\eta^{\text{ext}}$ should be quantized, and in the following we will only consider $\eta^{\text{ext}} = 0$. The twisted mass terms can be obtained simply by substituting $\sigma \rightarrow \sigma + \sigma^{\text{ext}}$ in \eqref{chi}.


\subsection{Localization on $S^2$} \label{sub2.2}

\noindent \textit{Coulomb branch localization} \\

The computation of the partition function of a GLSM on the two-sphere can be performed via equivariant localization \cite{Witten:1988ze,1991hep.th...12056W}. Following \cite{BC,DFGL}, in order to localize the path integral we consider an $\mathfrak{su}(1\vert 1) \subset \mathfrak{su}(2 \vert 1)$ subalgebra generated by two fermionic charges $\delta_{\epsilon}$ and $\delta_{\bar{\epsilon}}$. In terms of
\begin{equation}
\delta_{\mathcal{Q}} = \delta_{\epsilon} + \delta_{\bar{\epsilon}}
\end{equation} 
this subalgebra is given by\footnote{$\delta^2_{\mathcal{Q}}$ also generates gauge and flavour transformations.}
\begin{equation}
\delta^2_{\mathcal{Q}} = J_3 + \dfrac{R_V}{2} \;\;\;\;,\;\;\;\; \left[ J_3 + \dfrac{R_V}{2}, \delta_{\mathcal{Q}} \right] = 0
\end{equation}
In particular, we notice that the choice of $\delta_{\mathcal{Q}}$ breaks the $SU(2)$ isometry group of $S^2$ to a $U(1)$ subgroup, thus determining a North and South pole on the two-sphere. \\
\noindent It turns out that $\mathcal{L}_{\text{vec}}$ and $\mathcal{L}_{\text{chiral}}$ are $\delta_{\mathcal{Q}}$-exact terms:
\begin{equation}
\begin{split}
\bar{\epsilon} \epsilon \mathcal{L}_{\text{vec}} &= \delta_{\mathcal{Q}} \delta_{\bar{\epsilon}} \text{Tr} \left( \dfrac{1}{2} \bar{\lambda} \lambda - 2 D \sigma - \dfrac{1}{r} \sigma^2 \right) \\
\bar{\epsilon} \epsilon \mathcal{L}_{\text{chiral}} &= \delta_{\mathcal{Q}} \delta_{\bar{\epsilon}} \text{Tr} \left( \bar{\psi} \psi - 2 i \bar{\phi} \sigma \phi + \dfrac{q-1}{r} \bar{\phi} \phi \right) 
\end{split}
\end{equation}
This means that the partition function will not depend on the gauge coupling constant, since it is independent of $\delta_{\mathcal{Q}}$-exact terms; for the same reason it will not depend on the superpotential parameters, $\mathcal{L}_W$ being also $\delta_{\mathcal{Q}}$-exact (although the presence of a superpotential constrains the value of the R-charges). This choice of localizing action is referred to as the \textit{Coulomb branch} localization scheme, since the localization locus coincides in this case with the Coulomb branch of the theory; in particular, in this case the localization locus is given by
\begin{equation}
0 = \phi = \bar{\phi} = F = \bar{F}
\end{equation}
(for generic R-charges) and
\begin{equation}
0 = F_{12} - \dfrac{\eta}{r} = D + \dfrac{\sigma}{r} = D_{\mu} \sigma  = D_{\mu} \eta = [\sigma, \eta] 
\end{equation}
These equations imply that $\sigma$ and $\eta$ are constant and in the Cartan of the gauge group; moreover, since the gauge flux is quantized on $S^2$
\begin{equation}
\dfrac{1}{2 \pi} \int F = 2 r^2 F_{12} = \mathfrak{m}
\end{equation}
we remain with 
\begin{equation}
F_{12} = \dfrac{\mathfrak{m}}{2 r^2} \;\;\;,\;\;\; \eta = \dfrac{\mathfrak{m}}{2 r}
\end{equation}
The localization argument \cite{Witten:1988ze,1991hep.th...12056W} implies that the partition function is a 1-loop exact quantity. One can therefore compute the one-loop determinants for vector and chiral multiplets around the localization locus; the final result is
\begin{equation}
Z_{\text{vec}}^{\text{1l}} = \prod_{\alpha > 0} \left( \dfrac{\alpha(\mathfrak{m})^2}{4} + r^2 \alpha (\sigma)^2 \right)
\end{equation}
\begin{equation}
Z_{\Phi}^{\text{1l}} = \prod_{\rho \in R} \dfrac{\Gamma \left( \frac{q}{2} - i r \rho (\sigma) - \frac{\rho (\mathfrak{m})}{2} \right)}{\Gamma \left( 1 - \frac{q}{2} + i r \rho (\sigma) - \frac{\rho (\mathfrak{m})}{2} \right)} \label{matter1l}
\end{equation}
with $\alpha > 0$ positive roots of the gauge group $G$ and $\rho$ weights of the representation $R$ of the chiral multiplet. Twisted masses for the chiral multiplet can be added by shifting $\rho (\sigma) \rightarrow \rho (\sigma) + \tilde{\rho}(\sigma^{\text{ext}})$ and multiplying over the roots of the representation $\tilde{\rho}$ of the flavour group $G_F$. 
The classical part of the action is simply given by the Fayet-Iliopoulos term:
\begin{equation}
S_{FI} = 4 \pi i r \xi_{\text{ren}} \text{Tr} (\sigma) + i \theta_{\text{ren}} \text{Tr} (\mathfrak{m})
\end{equation}
where we are taking into account that in general the Fayet-Iliopoulos parameter runs \cite{DFGL} and the $\theta$-angle gets a shift from integrating out the $W$-bosons \cite{2013arXiv1308.2438H}, according to
\begin{equation}
\xi_{\text{ren}} = \xi - \frac{1}{2 \pi} \sum_l Q_l \log (r M) \;\;\;,\;\;\; \theta_{\text{ren}} = \theta + (s-1)\pi \label{ren}
\end{equation}
Here $M$ is a SUSY-invariant ultraviolet cut-off, $s$ is the rank of the gauge group and $Q_l$ are the charges of the chiral fields with respect to the abelian part of the gauge group. In the Calabi-Yau case the sum of the charges is zero, therefore $\xi_{\text{ren}} = \xi$. \\
All in all, the partition function for an $\mathcal{N} = (2,2)$ GLSM on $S^2$ reads
\begin{equation}
Z_{S^2} = \dfrac{1}{\vert \mathcal{W} \vert} \sum_{\mathfrak{m} \in \mathbb{Z}} \int \left( \prod_{s = 1}^{\text{rk} G} \dfrac{d \sigma_s}{2 \pi} \right) e^{- 4 \pi i r \xi_{\text{ren}} \text{Tr} (\sigma) - i \theta_{\text{ren}} \text{Tr} (\mathfrak{m})}
Z_{\text{vec}}^{\text{1l}}(\sigma, \mathfrak{m}) \prod_{\Phi} Z_{\Phi}^{\text{1l}}(\sigma, \mathfrak{m}, \sigma^{\text{ext}})
\label{partitionfunction}
\end{equation}
where $\vert \mathcal{W} \vert$ is the order of the Weyl group of $G$. If $G$ has many abelian components, we will have more Fayet-Iliopoulos terms and $\theta$-angles. \\

\noindent \textit{Higgs branch localization} \\

As we saw, equation \eqref{partitionfunction} gives a representation of the partition function as an integral over Coulomb branch vacua. For the theories we will consider another representation of $Z_{S^2}$ is possible, in which the BPS configurations dominating the path integral are a finite number of points on the Higgs branch, supporting point-like vortices at the North pole and anti-vortices at the South pole; we will call this \textit{Higgs branch} representation. 

Starting from the localization technique, the Higgs branch representation can be obtained by adding another $\delta_{\mathcal{Q}}$-exact term to the action which introduces a parameter $\chi$ acting as a Fayet-Iliopoulos. At $q = 0$ the localization locus admits a Higgs branch, given by
\begin{equation}
0 = F = D_{\mu}\phi = \eta \phi = (\sigma + \sigma{^{\text{ext}}}) \phi = \phi \phi^{\dagger} - \chi \mathbf{1} 
\end{equation}
\begin{equation}
0 = F_{12} - \dfrac{\eta}{r} = D + \dfrac{\sigma}{r} = D_{\mu} \sigma  = D_{\mu} \eta = [\sigma, \eta] 
\end{equation}
According to the matter content of the theory, this set of equations can have a solution with $\eta = F_{12} = 0$ and $\sigma = - \sigma{^{\text{ext}}}$, so that for generic twisted masses the Higgs branch consists of a finite number of isolated vacua, which could be different for $\chi \gtrless 0$. 

On top of each classical Higgs vacuum there are vortex solutions at the North pole $\theta = 0$ satisfying
\begin{equation}
D + \dfrac{\sigma}{r} = - i (\phi \phi^{\dagger} - \chi \mathbf{1}) = i F_{12} \;\;\;,\;\;\; D_-\phi = 0
\end{equation}
and anti-vortex solutions at the South pole $\theta = \pi$
\begin{equation}
D + \dfrac{\sigma}{r} = - i (\phi \phi^{\dagger} - \chi \mathbf{1}) = - i F_{12} \;\;\;,\;\;\; D_+\phi = 0
\end{equation}
The size of vortices depends on $\chi$ and tends to zero for $\vert \chi \vert \rightarrow \infty$.

All in all, the partition function $Z_{S^2}$ in the Higgs branch can be schematically written in the form
\begin{equation}
Z_{S^2} = \sum_{\sigma = - \sigma{^{\text{ext}}}} Z_{\text{cl}} Z_{\text{1l}} Z_{\text{v}} Z_{\text{av}} \label{hbr} 
\end{equation}
Apart from the usual classical and 1-loop terms, we have the vortex / anti-vortex partition functions $Z_{\text{v}}$, $Z_{\text{av}}$; they coincide with the ones computed on $\mathbb{R}^2$ with $\Omega$-background, where the $\Omega$-background parameter $\hbar$ depends on the $S^2$ radius as $\hbar = \frac{1}{r}$.

As a final remark, let us stress that although the explicit expressions for $Z_{S^2}$ in the Higgs and Coulomb branch might look very different, they are actually the same because of the localization argument, and in fact the Higgs branch representation \eqref{hbr} can be recovered from the Coulomb branch one \eqref{partitionfunction} by residue evaluation of the integral. 


\subsection{Quantum cohomology from $Z_{S^2}$} \label{sub2.3}

At the classical level, the space $X$ of supersymmetric vacua in the Higgs branch of the theory is given by the set of constant VEVs for the chiral fields minimizing the scalar potential, i.e. solving the $F$- and $D$-equations, modulo the action of the gauge group:
\begin{equation}
X = \{ \text{constant } \langle \phi \rangle / F = 0, D = 0 \} / G
\end{equation}
This space is always a K\"{a}hler manifold with first Chern class $c_1 \geqslant 0$; a very important subcase is when $c_1 = 0$, in which $X$ is a Calabi-Yau manifold. In the following we will refer to $X$ as the \textit{target manifold} of the GLSM.

From the physics point of view, the most interesting GLSMs are those whose target is a Calabi-Yau three-fold, since they provide (in the infra-red) a very rich set of four-dimensional vacua of string theory. The study of Calabi-Yau sigma models led to great discoveries both in mathematics and in physics such as mirror symmetry \cite{Dixon:1987bg,Lerche:1989uy,Candelas1990383,Aspinwall:1990xe,Greene:1990ud}, 
an extremely important tool to understand world-sheet quantum corrections to the moduli space of Calabi-Yau three-folds.
These non-perturbative quantum corrections form a power series whose coefficients, known as Gromov-Witten invariants \cite{gromov,Dine:1986zy,witten1988}, are related to the counting of holomorphic maps of fixed degree from the world-sheet to the Calabi-Yau.
The physical interpretation is that these terms capture
Yukawa couplings in the four-dimensional effective theory obtained from string theory after compactification on the Calabi-Yau.  
Unfortunately, mirror symmetry can only be applied when the Calabi-Yau three-fold under consideration has a known mirror
construction; this is the case for complete intersections in a toric variety and few other exceptions. \\

The exact expression for $Z_{S^2}$ in subsection \ref{sub2.2} can be used to compute these non-perturbative corrections without having to resort to mirror symmetry. As conjectured in \cite{2012arXiv1208.6244J} and further discussed in 
\cite{2013JHEP...04..019G} building on \cite{1991NuPhB.367..359C,Hori:2000kt}, the partition function $Z_{S^2}$ for an $\mathcal{N}=(2,2)$ GLSM computes the vacuum amplitude of the associated infrared non-linear sigma model:
\begin{equation}
Z_{S^2} (t_a, \bar{t}_a) = \langle \bar{0} \vert 0 \rangle = e^{- \mathcal{K}_K(t_a, \bar{t}_a)} \label{pot}
\end{equation} 
where $\mathcal{K}_K$ is the exact K\"ahler potential on the quantum K\"ahler moduli space $\mathcal{M}_K$ of the corresponding Calabi-Yau target $X$. The $t_a$ are coordinates in $\mathcal{M}_K$ parametrizing the K\"ahler moduli of $X$, and correspond to the complexified Fayet-Iliopoulos parameters of the GLSM. Since $\mathcal{K}_K(t_a, \bar{t}_a)$ contains all the necessary information about the Gromov-Witten invariants of the target, this allows us to compute them for targets more generic than those whose mirror is known, and in particular for non-abelian quotients. More in detail, the exact expression reads
\begin{equation}
\begin{split}
e^{-K(t,\bar t)} & = -\frac{i}{6} \sum_{l,m,n}\kappa_{l m n} (t^l -\bar t^l)(t^m -\bar t^m)(t^n -\bar t^n) + \frac{\zeta(3)}{4\pi^3} \chi(X) \\
& + \frac{2i}{(2\pi i)^3} \sum_{\eta} N_{\eta} \Big( \text{Li}_3(q^{\eta}) + \text{Li}_3(\bar q^{\eta}) \Big) 
- \frac{i}{(2\pi i)^2}\sum_{\eta,l} N_{\eta} \Big( \text{Li}_2(q^{\eta}) + \text{Li}_2(\bar q^{\eta}) \Big)\eta_l  
(t^l -\bar t^l) \label{standard}
\end{split}
\end{equation}
Here $\chi(X)$ is the Euler characteristic of $X$, and
\begin{equation}
\text{Li}_k(q) = \sum_{n=1}^{\infty} \dfrac{q^n}{n^k} \;\;\;,\;\;\; q^{\eta} = e^{2 \pi i \sum_l \eta_l t^l} \;\;\;,\;\;\; 
\end{equation}
with $\eta_l$ an element of the second homology group of the target Calabi-Yau three-fold, while $N_{\eta}$ are the genus zero Gromov-Witten invariants. \\

In \cite{2013arXiv1307.5997B} we took a different approach to the same problem, by re-interpreting $Z_{S^2}$ in terms of Givental's formalism \cite{1996alg.geom..3021G} and its extension to non-abelian quotients in terms of quasi-maps \cite{2011arXiv1106.3724C}. More in general we considered both Calabi-Yau and Fano manifolds, as well as both compact and non-compact targets; in the latter case we have to turn on twisted masses to regularize the infinite volume, while Gromov-Witten invariants and quantum cohomology become equivariant. A good review of Givental's formalism can be found in \cite{2001math.....10142C}, here we will only discuss basic facts which will be needed in the following. \\

In order to introduce Givental's formalism we consider the flat sections $V_a$ of the Gauss-Manin connection spanning the vacuum bundle of the theory and satisfying \cite{1993NuPhB.405..279B,1994hep.th....7018D}
\begin{equation}
\left(\hbar D_a \delta_b^c + C_{ab}^c\right)V_c=0.
\label{1}
\end{equation}
Here $D_a$ is the covariant derivative on the vacuum line bundle and $C_{ab}^c$ are the coefficients of the OPE in the chiral ring of observables $\phi_a\phi_b =C_{ab}^c \phi_c$; the observables $\{\phi_a\}$ provide a basis for $H^0(X)\oplus H^2(X)$ with $a=0,1,\ldots,b^2(X)$, $\phi_0$ being the identity operator.\footnote{For non-compact targets we work in the context of equivariant cohomology $H^\bullet_T(X)$, where $T$ is the torus acting on $X$. The values of the twisted masses assign the weights of the torus action.} 
The parameter $\hbar$ is the spectral parameter of the Gauss-Manin connection. When $b=0$ in (\ref{1}) we find that $V_a=-\hbar D_aV_0$, which means that the flat sections are all generated by the fundamental solution ${\cal J}:=V_0$ of the equation
\begin{equation}
\left(\hbar D_a D_b + C_{ab}^c D_c\right){\cal J}=0
\label{2}
\end{equation}
The metric on the vacuum bundle is given by a symplectic pairing of the flat sections $g_{\bar a b}= \langle \bar a|b\rangle = V_{\bar a}^t E V_b$; in particular, the vacuum-vacuum amplitude can be written as  
\begin{equation}
Z_{S^2} = \langle\bar 0 | 0 \rangle = {\cal J}^t E {\cal J}
\label{vac}
\end{equation}
for a suitable symplectic form $E$ \cite{1993NuPhB.405..279B} that will be specified later.

Givental's small $\mathcal{J}$-function is the $H_T^0(X)\oplus H^2_T(X)$-valued generating function of holomorphic maps of degree $d \in \mathbb{N}_{>0}$ from the sphere with one marked point to the target space $X$. The world-sheet instanton corrections are labelled by the parameter $Q^d = \prod_{i=1}^{b_2(X)} Q_i^{d_i}$ with $Q_i = e^{- t^i}$, $t^i$ being the complexified K\"ahler parameters. 
This function can be recovered from a set of oscillatory integrals, called ``$\mathcal{I}$-functions'', which are generating functions of hypergeometric type in the variables $\hbar$ and $z_i = e^{-\tau_i}$; for abelian quotients the $\mathcal{I}$-function is the generating function of solutions of the Picard-Fuchs equations for the mirror manifold $\check X$ of $X$ and can be expressed in terms of periods on $\check X$, with $\tau_i$ complex structure moduli of $\check X$.

In order to calculate the equivariant Gromov-Witten invariants from the above functions, one has to consider their asymptotic
expansion in $\hbar$. The ${\cal J}$ function expands as 
\begin{equation}
1 + \frac{J^{(2)}}{\hbar^2} +  \frac{J^{(3)}}{\hbar^3} + \ldots
\label{Jexp}
\end{equation}
Each coefficient is a cohomology-valued formal power series in the $Q$-variables. In particular $J^{(2)a} = \eta^{ab}\partial_b{\cal F}$, where $\eta^{ab}$ is the inverse topological metric and ${\cal F}$ the Gromov-Witten prepotential. Higher order terms in \eqref{Jexp} are related to gravitational descendant insertions. \\ 
The expansion for ${\cal I}_X(q,\hbar)$ reads 
\begin{equation}
I^{(0)} + \frac{I^{(1)}}{\hbar}
+  \frac{I^{(2)}}{\hbar^2} + \ldots
\label{Iexp}
\end{equation} 
The coefficients $I^{(0)}$, $I^{(1)}$ provide the change of variables (\textit{mirror map}) and normalization (\textit{equivariant mirror map}) which transform ${\cal I}$ into ${\cal J}$. To be more specific, 
let us split $I^{(1)} = \sum_s p_s g^s(z) + \sum_i\widetilde{p}_i h^i(z)$, with $p_s$ cohomology generators and $\widetilde{p}_i$ equivariant parameters of $H^2_T(X)$. The functions $\mathcal{I}$ and $\mathcal{J}$ are related by \begin{equation}
\mathcal{J}(\hbar, q) = e^{-f_0(z)/\hbar} e^{- \sum_i \widetilde{p}_i h^i (z)/\hbar} \mathcal{I}(\hbar, z(Q))
\end{equation}
where $f_0(z)/\hbar = \ln I^{(0)}$. In the simple example with just one $p$ and $\widetilde{p}$, the mirror map is given by $Q = \ln z + \frac{g(z)}{I^{(0)}(z)}$; the equivariant mirror map is given instead by the factor $e^{- \sum_i \widetilde{p}_i h^i (z)/\hbar}$. If ${I}^{(1)}=0$ the mirror maps are trivial and the two functions coincide. \\

We are now ready to illustrate the relation between Givental's formalism and the spherical partition function.
First of all, as shown in many examples in \cite{2012arXiv1208.6244J,2013arXiv1307.5997B} we can factorize the expression \eqref{partitionfunction} in a form similar to \eqref{hbr} even before performing the integral; schematically, we will have
\begin{equation}
{Z}^{S^2} = \oint 
d\lambda Z_{\text{1l}} \left(z^{-r|\lambda|} Z_{\rm v}\right) \left(\bar z^{-r|\lambda|} Z_{\rm av}\right)
\label{ciao} 
\end{equation}
Here $d\lambda=\prod_{\alpha=1}^{\rm rank}d\lambda_\alpha$ and $|\lambda|=\sum_\alpha \lambda_\alpha$, while $z=e^{-2\pi\vec\xi+i\vec\theta}$ labels the different vortex sectors. The contribution $(z \bar{z})^{-r \lambda_{r}}$ comes from the classical action, $Z_{\text{v}}$ is the equivariant vortex partition function on the North pole patch, $Z_{\text{av}}$ is the equivariant vortex partition function on the South pole patch and $Z_{\text{1l}}$ is the remnant one-loop measure. 

The claim is that $Z_{\rm v}$ coincides with the ${\cal I}$-function of the target space $X$ once we identify the vortex counting parameter $z$ with $Q$, $\lambda_\alpha$ with the generators of the cohomology, twisted masses with equivariant parameters in the cohomology, and $r=1/\hbar$. 
The choice of FI parameters and integration contours determines the chamber structure of the GIT quotient. In particular in the geometric phase the vortex counting parameters are identified with the exponentiated complex K\"ahler parameters, while in the orbifold phase they label the twisted sectors of the orbifold itself (i.e. the basis of orbifold cohomology).

The 1-loop term $Z_{\text{1l}}$, even if not discussed by Givental, can be interpreted as the symplectic pairing in \eqref{vac}. In order to reproduce the classical intersection cohomology on the target manifold we need to normalize  $Z_{\text{1l}}$ appropriately, as discussed in \cite{2013arXiv1307.5997B,2014JHEP...01..038B} and reviewed in Chapter \ref{chap3}.


\subsection{Quantum integrable systems from $Z_{S^2}$} \label{sub2.4}

Mirror symmetry for two-dimensional $\mathcal{N} = (2,2)$ gauge theories is a statement about the equivalence of two theories, a GLSM and a twisted Landau-Ginzburg (LG) model (known as \textit{mirror theory}). A twisted LG model is a theory made out of twisted chiral fields $Y$ only (possibly including superfield strengths $\Sigma$), specified by a holomorphic function $\mathcal{W}(Y, \Sigma)$ which contains the information about interactions among the fields.

The Coulomb branch of a twisted LG model is related to quantum integrable systems via Bethe/Gauge correspondence \cite{2009NuPhS.192...91N,2009PThPS.177..105N}. It can be recovered by integrating out the matter fields $Y$ and the massive $W$-bosons: from
\begin{equation}
\dfrac{\partial \mathcal{W}}{\partial Y} = 0 
\end{equation} 
we obtain $Y = Y(\Sigma)$, and substituting back in $\mathcal{W}$ we remain with a purely abelian gauge theory in the infrared, described in terms of the \textit{twisted effective superpotential}
\begin{equation}
\mathcal{W}_{\text{eff}} (\Sigma) = \mathcal{W} (\Sigma, Y(\Sigma))
\end{equation}
The effect of integrating out the $W$-bosons results in a shift of the $\theta$-angle. From the Bethe/Gauge correspondence, the twisted effective superpotential of a 2d $\mathcal{N} = (2,2)$ gauge theory coincides with the Yang-Yang function of a quantum integrable system (QIS); this implies that the quantum supersymmetric vacua equations
\begin{equation}
\dfrac{\partial \mathcal{W}_{\text{eff}}}{\partial \Sigma_s} = 2 \pi i n_s \label{sadd}
\end{equation}
can be identified, after exponentiation, with the Bethe Ansatz Equations (BAE) which determine the spectrum and eigenfunctions of the QIS:
\begin{equation}
\exp \left( \dfrac{\partial \mathcal{W}_{\text{eff}}}{\partial \Sigma_s} \right) = 1 \;\;\; \Longleftrightarrow \;\;\; \text{Bethe Ansatz Equations} \label{randomeq}
\end{equation}
In particular, to each solution of the BAE is associated an eigenstate of the QIS, and its eigenvalues with respect to the set of quantum Hamiltonians of the system can be expressed as functions of the gauge theory observables $\text{Tr}\, \Sigma^n$ evaluated at the solution:
\begin{equation}
\text{quantum Hamiltonians QIS} \;\;\; \longleftrightarrow \;\;\; \text{Tr}\, \Sigma^n \big\vert_{\text{solution BAE}} \label{above} 
\end{equation}

The Coulomb branch representation of the partition function \eqref{partitionfunction} for a GLSM contains all the information about the mirror LG model. We can start by defining
\begin{equation}
\Sigma_s = \sigma_s - i\dfrac{m_s}{2 r}
\end{equation}
which is the twisted chiral superfield corresponding to the superfield strength for the $s$-th component of the vector supermultiplet in the Cartan of the gauge group $G$. We can now use the procedure described in \cite{2013JHEP...04..019G}: each ratio of Gamma functions can be rewritten as 
\begin{equation}
\dfrac{\Gamma(-i r \Sigma)}{\Gamma(1 + i r \overline{\Sigma})} = \int \dfrac{d^2Y}{2\pi}  \text{exp} \Big\{ - e^{-Y} + i r \Sigma Y + e^{-\overline{Y}} + i r \overline{\Sigma} \overline{Y} \Big\} \label{go}
\end{equation}
Here $Y$, $\overline{Y}$ are interpreted as the twisted chiral fields for the matter sector of the mirror Landau-Ginzburg model. The partition function \eqref{partitionfunction} then becomes 
\begin{equation}
Z_{S^2} = \Bigg\vert \int d \Sigma \, d Y \, \, e^{-\mathcal{W}(\Sigma, Y)} \Bigg\vert^2 \label{pretutto}
\end{equation}
from which we can read $\mathcal{W}(\Sigma, Y)$ of the mirror LG theory; this is a powerful method to recover the twisted superpotential of the mirror theory, when it is not known previously. Here $d\Sigma = \prod_{s} d\Sigma_s$ and $d Y = \prod_{j} d Y_j$ collect all the integration variables. 

To recover the IR Coulomb branch of this theory we integrate out the $Y$, $\overline{Y}$ fields by performing a semiclassical approximation of \eqref{go}, which gives
\begin{equation}
Y = - \ln (- i r \Sigma) \;\;\;\;,\;\;\;\; \overline{Y} = - \ln (i r \overline{\Sigma})
\end{equation}
so that we are left with
\begin{equation}
\dfrac{\Gamma(-i r \Sigma)}{\Gamma(1 + i r \overline{\Sigma})} \,\sim \, \text{exp} \Big\{ \omega(- i r \Sigma) - \dfrac{1}{2} \ln (-i r \Sigma) -  \omega(i r \overline{\Sigma}) - \dfrac{1}{2} \ln( i r \overline{\Sigma}) \Big\} \label{exp}
\end{equation}
in terms of the function $\omega(x) = x (\ln x - 1)$. The effect of integrating out the $W$-fields results in having to consider $\theta_{\text{ren}}$ instead of $\theta$ as in \eqref{ren}. As discussed in \cite{2014JHEP...07..141B,2015JHEP...01..100N} the functions $\omega(\Sigma)$ enter in $\mathcal{W}_{\text{eff}}$, while the logarithmic terms in \eqref{exp} (which modify the effective twisted superpotential with respect to the one on $\mathbb{R}^2$) enter into the integration measure.
 
Alternatively, the same results for the IR Coulomb branch can be obtained by taking a large $r$ limit, since $\frac{1}{r}$ sets the energy scale of the theory. In fact Stirling's approximation
\begin{equation}
\begin{split}
\Gamma(z) \,\sim\, \sqrt{2 \pi}\, z^{z-\frac{1}{2}}\,e^{-z} \,(1+ o(z^{-1})) \;\;\;,\;\;\; z \to \infty \\
\Gamma(1+z) \,\sim\, \sqrt{2 \pi}\, z^{z+\frac{1}{2}}\,e^{-z} \,(1+ o(z^{-1})) \;\;\;,\;\;\; z \to \infty 
\end{split}
\end{equation}
implies
\begin{equation}
\begin{split}
\ln \Gamma(z) \,\sim\, \omega(z) - \dfrac{1}{2} \ln z + \dfrac{1}{2}\ln 2 \pi + o(z^{-1}) \;\;\;,\;\;\; z \to \infty \\
\ln \Gamma(1+z) \,\sim\, \omega(z) + \dfrac{1}{2} \ln z + \dfrac{1}{2}\ln 2 \pi + o(z^{-1}) \;\;\;,\;\;\; z \to \infty 
\end{split}
\end{equation} 
from which we recover \eqref{exp}.

After this procedure has been implemented, \eqref{partitionfunction} becomes 
\begin{equation}
Z_{S^2} = \Bigg\vert \int d \Sigma \, Z_{\text{meas}} (\Sigma) \, e^{-\mathcal{W}_{\text{eff}}(\Sigma)} \Bigg\vert^2 \label{pretutto}
\end{equation}
with $Z_{\text{meas}}$ integration measure determined by the logarithms in \eqref{exp}. We can now perform a semiclassical analysis around the saddle points of $\mathcal{W}_{\text{eff}}$.
As we know, the saddle points $\Sigma_{\text{cr}}$ are solutions of the equations \eqref{sadd}\footnote{The $2 \pi i n_s$ comes from the symmetry $\theta \rightarrow \theta + 2 \pi n$}, and coincide with the Bethe ansatz equations governing the spectrum of the associated quantum integrable system; moreover, to each solution $\Sigma_{\text{cr}}^{(a)}$ it corresponds an eigenfunction $\psi^{(a)}$. Up to quadratic fluctuations, the semiclassical approximation of \eqref{pretutto} around $\Sigma_{\text{cr}}^{(a)}$ reads
\begin{equation}
Z_{S^2}^{(a)} = \Bigg{\vert} e^{-\mathcal{W}_{\text{eff,cr}}} Z_{\text{meas}} (\Sigma) \left(\text{Det} \dfrac{\partial^2 \mathcal{W}_{\text{eff}}}{\partial \Sigma_s \partial \Sigma_t} \right)^{-\frac{1}{2}} \Bigg{\vert}^2_{\Sigma = \Sigma_{\text{cr}}^{(a)}}  \label{no}
\end{equation}
The total partition function will be obtained by summing the contributions coming from all vacua. As noticed in \cite{2014JHEP...07..141B,2015JHEP...01..100N}, apart from the classical term $\vert e^{-\mathcal{W}_{\text{eff,cr}}} \vert^2 = \vert e^{-\mathcal{W}_{\text{eff}}(\Sigma_{\text{cr}})} \vert^2$ \eqref{no} can be seen as the inverse norm square of the eigenstates $\psi^{(a)}$:
\begin{equation}
Z_{S^2}^{(a)} = \dfrac{\vert e^{-\mathcal{W}_{\text{eff,cr}}} \vert^2}{\langle \psi^{(a)} \vert \psi^{(a)} \rangle}
\end{equation}
In fact by comparison with \eqref{no} we find
\begin{equation}
\dfrac{1}{\langle \psi^{(a)} \vert \psi^{(a)} \rangle} = \Bigg{\vert} Z_{\text{meas}} (\Sigma) \left(\text{Det} \dfrac{\partial^2 \mathcal{W}_{\text{eff}}}{\partial \Sigma_s \partial \Sigma_t} \right)^{-\frac{1}{2}} \Bigg{\vert}^2_{\Sigma = \Sigma_{\text{cr}}^{(a)}} \label{fin}
\end{equation}
which is the expression for the norm of the Bethe states proposed by Gaudin.


\section{ADHM Gauged Linear Sigma Model:\\ Higgs branch and quantum cohomology} \label{chap3}

The main character of this paper is the ADHM moduli space $\mathcal{M}_{k,N}$ of $k$ instantons for a pure $U(N)$ gauge theory. In this chapter we will describe how this moduli space can be obtained from a system of $k$ D$p$ $-$ $N$ D$(p+4)$ branes in type II string theory on $\mathbb{C}^2 \times \mathbb{C}^2/\mathbb{Z}_2 \times \mathbb{C}$. When $p=1$, resolving the singular space $\mathbb{C}^2/\mathbb{Z}_2$ to $T^* S^2$ naturally leads us to consider a GLSM on $S^2$ whose Higgs branch target space coincides with $\mathcal{M}_{k,N}$; we will study this GLSM and its partition function $Z_{k,N}^{S^2}$, which as discussed in Section \ref{chap2} contains all the information about the equivariant quantum cohomology of the instanton moduli space.


\subsection{The ADHM Gauged Linear Sigma Model} \label{sub3.1}
 
The ADHM moduli space of instantons admits a natural brane construction in type II string theory on $\mathbb{C}^2 \times \mathbb{C}^2/\mathbb{Z}_2 \times \mathbb{C}$ \cite{1995hep.th...12077D,1998JGP....28..255D,1995JGP....15..215W}. We consider a stack of $N$ D$(p+4)$-branes ($p \geqslant -1$) at the $\mathbb{C}^2/\mathbb{Z}_2$ singularity and wrapping $\mathbb{C}^2$; at low energy their world-volume dynamics is described by a $(p+5)-$dimensional pure $U(N)$ super Yang-Mills theory with 8 supercharges. A $k$-instanton configuration in this theory can be thought of as introducing a set of $k$ D$p$-branes on top of the D$(p+4)$-branes. In order to derive the ADHM construction from branes, we have to consider the theory living on the D$p$-branes: at low energy this will be a $(p+1)-$dimensional $U(k)$ gauge theory with matter fields in the adjoint, fundamental and antifundamental representations, coming from D$p-$D$p$ and D$p-$D$(p+4)$ open strings. The key point is that its Higgs branch moduli space of classical supersymmetric vacua is described exactly by the same equations defining $\mathcal{M}_{k,N}$.

To be more specific in the following we will restrict to the $p = 1$ case. If we resolve the singular space $\mathbb{C}^2/\mathbb{Z}_2$ to $T^* S^2$, we can wrap our $N$ D5-branes on $\mathbb{C}^2 \times S^2$ and our $k$ D1-branes on $S^2$; this is the set-up considered in \cite{2014JHEP...01..038B}. From the D5 point of view we have a 6d $\mathcal{N} = 1$ pure $U(N)$ Yang-Mills theory on $\mathbb{C}^2 \times S^2$ at low energy, while the D1 system provides a GLSM on $S^2$ with gauge group $U(k)$ and matter content summarized in Table \ref{ADHMtable}.

\begin{table}[h!]
\begin{center}
\begin{tabular}{c|c|c|c|c|c}
{} & $\chi$ & $B_{1}$ & $B_{2}$ & $I$ & $J$ \\ \hline
D-brane sector & D1/D1 & D1/D1 & D1/D1 & D1/D5 & D5/D1 \\ \hline
gauge $U(k)$ & $Adj$ & $Adj$ & $Adj$ & $\mathbf{k}$ & $\mathbf{\bar{k}}$ \\ \hline
flavor $U(N)\times U(1)^{2}$ & $\mathbf{1}_{(-1,-1)}$ & $\mathbf{1}_{(1,0)}$ & $\mathbf{1}_{(0,1)}$ & $\mathbf{\bar{N}}_{(1/2,1/2)}$ & $\mathbf{N}_{(1/2,1/2)}$ \\ \hline
twisted masses & $\epsilon_1 + \epsilon_2$ & $-\epsilon_{1}$ & $-\epsilon_{2}$ & $-a_{j} - \frac{\epsilon}{2}$ & $a_{j}-\frac{\epsilon}{2}$ \\ \hline
$R$-charge & $2-2q$ & $q$ & $q$ & $q+p$ & $q-p$ \\ \hline
\end{tabular} 
\caption{ADHM gauged linear sigma model} \label{ADHMtable}
\end{center}
\end{table} 

\noindent The superpotential of our model is $W=\textrm{Tr}_{k}\left\{\chi\left([B_{1},B_{2}]+IJ\right)\right\}$, which leaves a global symmetry group $U(N)\times U(1)^{2}$; we denote as $(a_{j},-\epsilon_{1},-\epsilon_{2})$ the twisted masses corresponding to the maximal torus $U(1)^{N+2}$ which acts on $\mathcal{M}_{k,N}$. In the $R$-charges assignment we require $1>q>p>0\,,\, q<1$, so that the integration contour in $\sigma$ is along the real line; for negative $R$-charges we can perform an analytic continuation by deforming the contour. 
We have now all the necessary ingredients to compute the partition function $Z_{k,N}^{S^{2}}$ for our ADHM GLSM. This is given by
\begin{eqnarray}
Z_{k,N}^{S^2} &=& \frac{1}{k!}\sum_{\vec{m}\in\mathbb{Z}^{k}} \int_{\mathbb{R}^{k}} \prod_{s=1}^{k} \frac{\mathrm{d} (r\sigma_{s})}{2\pi} e^{-4 \pi i \xi r \sigma_{s}-i\theta_{\text{ren}} m_{s}} 
Z_{\text{gauge}} Z_{IJ}\, Z_{\text{adj}}\label{sipp}
\end{eqnarray}
with gauge one-loop determinant
\begin{equation}
Z_{\text{gauge}} = \prod_{s<t}^{k}\left(\dfrac{m_{st}^{2}}{4} + r^2\sigma_{st}^{2}\right) \label{g1ld}
\end{equation}
and matter fields one-loop determinants 
\begin{eqnarray}
Z_{IJ} &=& \prod_{s=1}^{k}\prod_{j=1}^{N}
\frac{\Gamma\left(-i r \sigma_{s}+i r a_{j} + i r \frac{\epsilon}{2} -\frac{m_{s}}{2}\right)}{\Gamma\left(1+i r \sigma_{s}-i r a_{j}-i r \frac{\epsilon}{2}-\frac{m_{s}}{2}\right)}
\frac{\Gamma\left(i r \sigma_{s}-i r a_{j} + i r \frac{\epsilon}{2} + \frac{m_{s}}{2}\right)}{\Gamma\left(1-i r \sigma_{s}+i r a_{j}-i r \frac{\epsilon}{2} +\frac{m_{s}}{2}\right)} \label{f} \\
Z_{\text{adj}} &=& \prod_{s,t=1}^{k}\frac{\Gamma\left(1-i r \sigma_{st}-i r \epsilon-\frac{m_{st}}{2}\right)}{\Gamma\left(i r \sigma_{st}+i r \epsilon-\frac{m_{st}}{2}\right)}
\frac{\Gamma\left(-i r \sigma_{st}+i r \epsilon_{1}-\frac{m_{st}}{2}\right)}{\Gamma\left(1+i r \sigma_{st}-i r \epsilon_{1}-\frac{m_{st}}{2}\right)}
\frac{\Gamma\left(-i r \sigma_{st}+i r \epsilon_{2}-\frac{m_{st}}{2}\right)}{\Gamma\left(1+i r \sigma_{st}-i r \epsilon_{2}-\frac{m_{st}}{2}\right)}\nonumber
\end{eqnarray}
Here we defined $\epsilon=\epsilon_1 + \epsilon_2$, $\sigma_{st} = \sigma_s - \sigma_t$ and $m_{st} = m_s - m_t$.
The $Z_{IJ}$ term represents the contributions from $I$ and $J$, while $Z_{\text{adj}}$ contains $\chi$, $B_1$, $B_2$. 
The partition function \eqref{sipp} has been referred to as the \textit{stringy instanton partition function} in \cite{2014JHEP...01..038B}. Let us point out that the gauge one-loop determinant \eqref{g1ld}, together with the shift of $\theta$ in $\theta_{\text{ren}} = \theta + (k-1)\pi$, can be thought of as the one-loop determinant for the $W$-bosons if we consider them as fields with R-charge 2, since
\begin{equation}
\prod_{s\neq t}^{k}\frac{\Gamma\left(1-i r \sigma_{st}-\frac{m_{st}}{2}\right)}{\Gamma\left(i r \sigma_{st} -\frac{m_{st}}{2}\right)} = e^{-i \pi (k-1) \sum_{s=1}^k m_s} \prod_{s<t}^{k}\left(\dfrac{m_{st}^{2}}{4} + r^2\sigma_{st}^{2}\right)
\end{equation}\vspace*{0.2 cm}

When the radius $r$ of $S^2$ goes to zero we recover a system of D(-1)-D3 branes; the partition function of the theory living on the D(-1)-branes will be the one considered in \cite{2002hep.th....6161N,Bruzzo:2002xf,2003hep.th....6238N}, see also \cite{2000CMaPh.209...97M}. In particular we expect $Z_{k,N}^{S^2} \rightarrow Z_{k,N}$ for $r \rightarrow 0$, with $Z_{k,N}$ contour integral representation of the instanton part of the Nekrasov partition function $Z_{N} = \sum_{k} \Lambda^{2 N k} Z_{k,N}$ for a 4d $\mathcal{N} = 2$ pure $U(N)$ theory: 
\begin{equation}
Z_{k,N} = \frac{1}{k!}\frac{\epsilon^{k}}{(2 \pi i \epsilon_{1} \epsilon_{2})^k} \oint 
\prod_{s=1}^k \dfrac{d\sigma_{s}}{P(\sigma_{s}) P(\sigma_{s}+\epsilon)}  \prod_{s<t}^k \dfrac{\sigma_{st}^2 
(\sigma_{st}^2 - \epsilon^{2} )}{(\sigma_{st}^2 - \epsilon_{1}^2 )(\sigma_{st}^2 - \epsilon_{2}^2 )}
\label{nek}
\end{equation} 
Here we defined $P(\sigma_{s})  = \prod_{j=1}^N (\sigma_{s}-a_{j} - \frac{\epsilon}{2})$, while $\Lambda$ is the RGE invariant scale. In fact in \cite{2014JHEP...01..038B} it has been shown that the lowest order term in the $r$ expansion of \eqref{sipp} coincides with \eqref{nek}, and the energy scale is naturally set to $\Lambda = r^{-1}$.


\subsection{Equivariant Gromov-Witten invariants of $\mathcal{M}_{k,N}$} \label{sub3.2}

The explicit evaluation of \eqref{sipp} requires to classify the poles in the integrand; this has been done in \cite{2014JHEP...01..038B}. To summarize, we have to consider the ADHM phase which corresponds to $\xi>0$; this forces us to close the contour integral in the lower half plane. It turns out that the poles can be classified by $N$ Young tableaux $\{\vec{Y}\}_k=\left(Y_1,\ldots,Y_N\right)$ such that $\sum_{j=1}^N |Y_j|=k$, which describe coloured partitions of the instanton number $k$; these are the same ones used in the pole classification of $Z_{k,N}$, with the difference that to every box is associated not just a pole, but an infinite tower of poles, labelled by a positive integer $n$. These towers of poles correspond to D(-1)-branes describing the effective dynamics of the $k$ D1-branes, and represent the vortex/anti-vortex contributions to the spherical partition function; we can easily deal with them by rewriting near each pole \cite{2012arXiv1208.6244J}
\begin{equation}
\sigma_{s} = - \frac{i}{r}\left(n_s + \frac{\vert m_s \vert}{2}\right) + i \lambda_s \label{cov}
\end{equation} 
In this way we resum the contributions coming from the ``third direction'' of the Young tableaux, and the poles for $\lambda_s$ are now given in terms of usual two-dimensional partitions.
With the change of variables \eqref{cov} we can explicitly show how $Z_{k,N}^{S^2}$ can be factorized before integration as in \eqref{ciao}; this is important since we discussed how we can extract the $\mathcal{I}$-function of the GLSM target space  from this factorized form. If we define $z = e^{-2\pi \xi+i\theta}$ and $d_s = n_s +\frac{m_s + \vert m_s \vert}{2}$, $\tilde{d}_s = d_s - m_s$ so that $\sum_{m_s\in \mathbb{Z}}\sum_{n_s\geqslant 0} = \sum_{\tilde{d}_s\geqslant 0}\sum_{d_s\geqslant 0}$ we obtain the following expression:
\begin{equation}
Z_{k,N}^{S^2} = 
\dfrac{1}{k!} \oint \prod_{s=1}^k \dfrac{d (r \lambda_{s})}{2\pi i} (z \bar{z})^{-r \lambda_{s}} Z_{\text{1l}} Z_{\text{v}} Z_{\text{av}} \label{int}
\end{equation}
where
\begin{eqnarray}
Z_{\text{1l}} &=& \left(\dfrac{\Gamma(1-i r \epsilon)\Gamma(i r \epsilon_{1})\Gamma(i r \epsilon_{2})}{\Gamma(i r \epsilon)\Gamma(1-i r \epsilon_{1})\Gamma(1-i r \epsilon_{2})}\right)^k 
\prod_{s=1}^k \prod_{j=1}^N \dfrac{\Gamma \left(r \lambda_{s}+i r a_{j} + i r \frac{\epsilon}{2}\right)\Gamma \left(-r \lambda_{s}-i r a_{j}+i r \frac{\epsilon}{2}\right)}{\Gamma \left(1 - r \lambda_{s}-i r a_{j}-i r \frac{\epsilon}{2}\right)\Gamma \left(1+r \lambda_{s}+i r a_{j}- i r \frac{\epsilon}{2}\right)}\nonumber\\
&&\prod_{s\neq t}^k (r\lambda_{s}-r\lambda_{t})\dfrac{\Gamma(1+ r \lambda_{s}- r\lambda_{t}-i r \epsilon)\Gamma( r \lambda_{s}- r\lambda_{t} +i r \epsilon_{1})\Gamma( r \lambda_{s}- r\lambda_{t}+i r \epsilon_{2})}{\Gamma(-r\lambda_{s} + r \lambda_{t} + i r \epsilon)\Gamma(1-r\lambda_{s} + r \lambda_{t}-i r \epsilon_{1})\Gamma(1-r\lambda_{s} + r \lambda_{t}-i r \epsilon_{2})}\nonumber\\\label{1l}
\end{eqnarray}
\begin{eqnarray}
Z_{\text{v}} &=& \sum_{\tilde{d}_1,\ldots , \tilde{d}_k \,\geq \,0} ((-1)^N z)^{\tilde{d}_1+ \ldots +\tilde{d}_k}  \prod_{s=1}^k \prod_{j=1}^N \dfrac{\left(-r \lambda_{s}-i r a_{j}+ i r \frac{\epsilon}{2}\right)_{\tilde{d}_s}}{\left(1-r \lambda_{s}-i r a_{j}-i r \frac{\epsilon}{2}\right)_{\tilde{d}_s}}
\prod_{s<t}^k \dfrac{\tilde{d}_t - \tilde{d}_s - r \lambda_{t} + r \lambda_{s}}{- r \lambda_{t} + r \lambda_{s}}\nonumber\\
&&\dfrac{(1+ r \lambda_{s}- r\lambda_{t}-i r \epsilon)_{\tilde{d}_t - \tilde{d}_s}}{( r \lambda_{s}- r\lambda_{t}+i r \epsilon)_{\tilde{d}_t - \tilde{d}_s}} 
\dfrac{( r \lambda_{s}- r\lambda_{t}+i r \epsilon_{1})_{\tilde{d}_t - \tilde{d}_s}}{(1+ r \lambda_{s}- r\lambda_{t}-i r \epsilon_{1})_{\tilde{d}_t - \tilde{d}_s}}
\dfrac{(r \lambda_{s}- r\lambda_{t}+i r \epsilon_{2})_{\tilde{d}_t - \tilde{d}_s}}{(1+ r \lambda_{s}- r\lambda_{t}-i r \epsilon_{2})_{\tilde{d}_t - \tilde{d}_s}}\nonumber\\ \label{v}
\end{eqnarray}
\begin{eqnarray}
Z_{\text{av}} &=& \sum_{d_1,\ldots , d_k \,\geq \,0} ((-1)^N \bar{z})^{d_1+ \ldots +d_k}  \prod_{s=1}^k \prod_{j=1}^N \dfrac{\left(-r \lambda_{s}-i r a_{j} + i r \frac{\epsilon}{2}\right)_{d_s}}{\left(1-r \lambda_{s}-i r a_{j} - i r \frac{\epsilon}{2}\right)_{d_s}}
\prod_{s<t}^k \dfrac{d_t - d_s - r \lambda_{t} + r \lambda_{s}}{- r \lambda_{t} + r \lambda_{s}}\nonumber\\
&&\dfrac{(1+ r \lambda_{s}- r\lambda_{t}-i r \epsilon)_{d_t - d_s}}{( r \lambda_{s}- r\lambda_{t}+i r \epsilon)_{d_t - d_s}} 
\dfrac{( r \lambda_{s}- r\lambda_{t}+i r \epsilon_{1})_{d_t - d_s}}{(1+ r \lambda_{s}- r\lambda_{t}-i r \epsilon_{1})_{d_t - d_s}}
\dfrac{(r \lambda_{s}- r\lambda_{t}+i r \epsilon_{2})_{d_t - d_s}}{(1+ r \lambda_{s}- r\lambda_{t}-i r \epsilon_{2})_{d_t - d_s}}\nonumber\\\label{av}
\end{eqnarray}
The Pochhammer symbol $(a)_d$ is defined as 
\begin{equation}
(a)_k = \left\{ 
\begin{array}{cc}
\prod_{i=0}^{k-1} (a+i) & \,\,\text{for}\,\, k>0\\
1 & \,\,\text{for}\,\, k=0\\
\prod_{i=1}^{-k} \dfrac{1}{a-i} & \,\,\text{for}\,\, k<0
\end{array}
\right. \label{poc}
\end{equation}
which implies the identity
\begin{equation}
(a)_{-d} = \dfrac{(-1)^d }{(1-a)_d} \label{poch}
\end{equation}
The $\frac{1}{k!}$ in \eqref{int} is cancelled by the $k!$ possible orderings of the $\lambda$s, so in the rest of this paper we will always choose an ordering and remove the factorial.

As discussed in subsection \ref{sub2.3}, the vortex partition function $Z_v$ appearing in \eqref{v} provides a
conjectural formula for Givental's ${\cal I}$-functions of the ADHM instanton moduli space:
\begin{eqnarray}
{\cal I}_{k,N}&=& \sum_{d_1,\ldots , d_k \,\geq \,0} ((-1)^N z)^{d_1+ \ldots +d_k}  \prod_{s=1}^k \prod_{j=1}^N \dfrac{(-r \lambda_{s}-i r a_{j}+ i r \epsilon)_{d_s}}{(1-r \lambda_{s}-i r a_{j})_{d_s}}
\prod_{s<t}^k \dfrac{d_t - d_s - r \lambda_{t} + r \lambda_{s}}{- r \lambda_{t} + r \lambda_{s}}\nonumber\\
&&\dfrac{(1+ r \lambda_{s}- r\lambda_{t}-i r \epsilon)_{d_t - d_s}}{( r \lambda_{s}- r\lambda_{t}+i r \epsilon)_{d_t - d_s}} 
\dfrac{( r \lambda_{s}- r\lambda_{t}+i r \epsilon_{1})_{d_t - d_s}}{(1+ r \lambda_{s}- r\lambda_{t}-i r \epsilon_{1})_{d_t - d_s}}
\dfrac{(r \lambda_{s}- r\lambda_{t}+i r \epsilon_{2})_{d_t - d_s}}{(1+ r \lambda_{s}- r\lambda_{t}-i r \epsilon_{2})_{d_t - d_s}}\nonumber\\
\label{I-inst}\end{eqnarray}
The $\lambda_s$ are to be interpreted as the Chern roots of the tautological bundle of $\mathcal{M}_{k,N}$.

For Nakajima quiver varieties, the notion of quasi-maps and ${\cal I}$-function were introduced in \cite{2011arXiv1106.3724C}; our ${\cal I}_{k,N}$ \eqref{I-inst} should match the quasi-map ${\cal I}$-function and therefore, according to \cite{2012arXiv1211.1287M}, should compute the ${\cal J}$-function of the instanton moduli space. In Section \ref{chap5} we will apply the same supersymmetric localization approach to other Nakajima quiver varieties in order to produce conjectural formula for Givental's ${\cal I}$-functions of moduli space of instantons on ALE spaces of type A and D; type E can be obtained in a similar way.
 
From \eqref{I-inst} we find that the asymptotic behaviour in $\hbar = r^{-1}$ is 
\begin{equation}
{\cal I}_{k,N}= 1 + \frac{I^{(N)}}{\hbar^N} +\ldots
\end{equation}
Comparing with \eqref{Iexp} we find that $I^{(0)} = 1$ for every $k, N$, while $I^{(1)} = 0$ when $N>1$; this implies that the equivariant mirror map is trivial, namely ${\cal I}_{k,N}={\cal J}_{k,N}$, for $N>1$. In the case $N=1$ we will have to invert the equivariant mirror map. 

As a final comment, let us remark that in the limit $\epsilon \rightarrow 0$ one can show \cite{2014JHEP...01..038B} that all the world-sheet instanton corrections to $Z^{S^2}_{k,N}$ vanish; this is in agreement with the results of \cite{2012arXiv1211.1287M} about equivariant Gromov-Witten invariants of the ADHM moduli space. 


\subsection{Example: ${\cal M}_{1,2}$ versus ${\cal M}_{2,1}$} \label{sub3.3}

In this subsection we will explicitly compute the K\"{a}hler potential for the cases ${\cal M}_{1,2}$ and ${\cal M}_{2,1}$. Since ${\cal M}_{1,2} \simeq {\cal M}_{2,1}$ we expect the two results to be the same, after appropriately identifying the equivariant parameters; we will see that this is indeed the case once the equivariant mirror map has been handled correctly. \\

\noindent \textit{The ${\cal M}_{1,2}$ case}\\

\noindent The instanton moduli space $\mathcal{M}_{1,N}$ is equivalent to $\mathbb{C}^2\times T^*\mathbb{P}^{N-1}$; in the $N=2$ case we have $\mathcal{M}_{1,2} \simeq \mathbb{C}^2\times T^*\mathbb{P}^{1}$. There are only two 2-coloured partitions of $k=1$ labelling the poles of \eqref{int}, given by

\vspace*{0.2 cm}
\begin{center}
\begin{tabular}{ccl}
$({\tiny\yng(1)}, \bullet)$ & $\Longleftrightarrow$ & $\lambda_{1} = -i a_1 - i \frac{\epsilon}{2} $ \\ \vspace*{0.3 cm}
$(\bullet, {\tiny\yng(1)})$ & $\Longleftrightarrow$ & $\lambda_{1} = -i a_2 - i \frac{\epsilon}{2} $
\end{tabular}  
\end{center}
\vspace*{0.2 cm}

\noindent The partition function can be written as
\begin{eqnarray}
Z_{1,2}^{S^2} = (z\bar{z})^{i r (a_1 + \frac{\epsilon}{2})} Z_{\text{1l}}^{(1)} Z_{\text{v}}^{(1)} Z_{\text{av}}^{(1)} + 
(z\bar{z})^{i r (a_2 + \frac{\epsilon}{2})} Z_{\text{1l}}^{(2)} Z_{\text{v}}^{(2)} Z_{\text{av}}^{(2)}\label{12} 
\end{eqnarray}
where
\begin{eqnarray}
Z^{(1)}_{\text{1l}} &=& \frac{\Gamma\left(i r \epsilon_{1}\right)\Gamma\left(i r \epsilon_{2}\right)}{\Gamma\left(1-i r \epsilon_{1}\right)\Gamma\left(1-i r \epsilon_{2}\right)}  
\frac{\Gamma\left(i r a_{21}\right)\Gamma\left(-i r a_{21} + i r \epsilon\right)}{\Gamma\left(1 - i r a_{21}\right)\Gamma\left(1 + i r a_{21} - i r \epsilon \right)}\nonumber\\
Z^{(1)}_{\text{v}} &=& {}_{2}F_{1}\left(\begin{array}{cc}\left\{i r \epsilon , -i r a_{21} + i r \epsilon \right\}\\ \left\{1 - i r a_{21}\right\}\end{array}; z \right)\nonumber\\
Z^{(1)}_{\text{av}} &=& {}_{2}F_{1}\left(\begin{array}{cc}\left\{i r \epsilon , -i r a_{21} + i r \epsilon \right\}\\ \left\{1 - i r a_{21}\right\}\end{array}; \bar{z} \right)
\end{eqnarray}
The other contribution is obtained by exchanging $a_{1} \longleftrightarrow a_{2}$. By identifying $Z^{(1)}_{\text{v}}$ as the Givental ${\cal I}$-function, we expand it in $r = \frac{1}{\hslash}$ in order to find the equivariant mirror map; this gives
\begin{equation}
Z^{(1)}_{\text{v}} = 1 + o(r^2),
\end{equation}
which means there is no equivariant mirror map and ${\cal I} = {\cal J}$. The same applies to $Z^{(2)}_{\text{v}}$. 

We still have to properly normalize the symplectic pairing $Z_{1{\rm l}}$. 
This problem comes from the choice of renormalization scheme used to regularize the infinite products in the 1-loop determinant \eqref{matter1l}. In \cite{BC,DFGL} the $\zeta$-function renormalization scheme has been used.
The ambiguity amounts to replacing the Euler-Mascheroni constant $\gamma$ appearing in the Weierstrass form of the Gamma-function 
\begin{equation}
\frac{1}{\Gamma(x)}=x e^{\gamma x} \prod_{n=1}^\infty\left(1+\frac{x}{n}\right)e^{-\frac{x}{n}}
\label{gamma}
\end{equation}
with a finite holomorphic (because of supersymmetry) function of the parameters, namely $\gamma \to \text{Re}f(z)$. We will fix this normalization by requiring the cancellation of the Euler-Mascheroni constants; moreover we require the normalization to reproduce the correct intersection numbers in classical cohomology, and to start from 1 in the $rM$ expansion in order not to modify the regularized equivariant volume of the target. In our case, in \eqref{12} $Z^{(1)}_{\text{1l}}$ and $Z^{(2)}_{\text{1l}}$ contain an excess of $4 i r (\epsilon_{1}+\epsilon_{2})$ in the argument of the Gamma functions; to eliminate the Euler-Mascheroni constant, we normalize the partition function multiplying it by
\begin{equation}
(z \bar{z})^{-i r \frac{a_1 + a_2}{2}} \left(\dfrac{\Gamma(1- i r \epsilon_{1})\Gamma(1- i r \epsilon_{2})}{\Gamma(1+ i r \epsilon_{1})\Gamma(1+ i r \epsilon_{2})}\right)^2 \label{norm}
\end{equation} 
Expanding the normalized partition function in $r$ up to order $r^{-1}$, we obtain 
\begin{equation}
\begin{split}
Z_{1,2}^{\text{norm}} & = \dfrac{1}{r^2 \epsilon_1 \epsilon_2} \Big[ \dfrac{2}{r^2 (\epsilon^2 - a_{12}^2)} +\dfrac{1}{4}\ln^{2} (z \bar{z}) -i r(\epsilon_1 + \epsilon_2) \Big( -\dfrac{1}{12} \ln^{3} (z \bar{z}) \\
& \hspace{1.6 cm} - \ln (z \bar{z})(\text{Li}_2(z) + \text{Li}_2(\bar{z})) + 2 (\text{Li}_3(z) + \text{Li}_3(\bar{z})) + 4 \left( 1 - \dfrac{\epsilon_1 \epsilon_2}{\epsilon^2 - a_{12}^2} \right) \zeta(3) \Big) \Big] \label{12kahler}
\end{split}
\end{equation}
The first term in \eqref{12kahler} correctly reproduces the Nekrasov partition function of ${\cal M}_{1,2}$ as expected, while the other terms compute the $H^2_T(X)$ part of the genus zero Gromov-Witten potential in agreement with \cite{2006math.....10129B}. We remark that, as a consequence of what inferred at the end of subsection \ref{sub3.2}, the quantum part of the Gromov-Witten potential is linear in the equivariant parameter $\epsilon_1+\epsilon_2$. \\

\noindent \textit{The ${\cal M}_{2,1}$ case} \\

\noindent For ${\cal M}_{2,1}$ there are two poles of \eqref{int}, given by the two partitions of $k=2$ defined by (modulo permutations)

\vspace*{0.2 cm}
\begin{center}
\begin{tabular}{ccl}
${\tiny\yng(2)}$ & $\Longleftrightarrow$ & $\lambda_{1} = -i a - i \frac{\epsilon}{2}$\;,\; $\lambda_{2} = -i a - i \frac{\epsilon}{2} - i \epsilon_1$\\ \vspace*{0.3 cm}
${\tiny\yng(1,1)}$ & $\Longleftrightarrow$ & $\lambda_{1} = -i a - i \frac{\epsilon}{2}$\;,\; $\lambda_{2} = -i a - i \frac{\epsilon}{2} - i \epsilon_2$
\end{tabular}  
\end{center}
\vspace*{0.2 cm}
The permutations of the $\lambda$'s cancel the $\frac{1}{2!}$ in front of \eqref{sipp}. Evaluation of the residues gives
\begin{eqnarray}
Z^{S^2}_{2,1} = (z\bar{z})^{i r (2 a + \epsilon + \epsilon_{1})} Z_{\text{1l}}^{(\text{row})} Z_{\text{v}}^{(\text{row})} Z_{\text{av}}^{(\text{row})} + (z\bar{z})^{i r (2 a + \epsilon + \epsilon_{2})} Z_{\text{1l}}^{(\text{col})} Z_{\text{v}}^{(\text{col})} Z_{\text{av}}^{(\text{col})} \label{21} 
\end{eqnarray}
with
\begin{equation}
\begin{split}
Z_{\text{1l}}^{(\text{row})} \,=&\, \dfrac{\Gamma(i r \epsilon_{1})\Gamma(i r \epsilon_{2})}{\Gamma(1-i r \epsilon_{1})\Gamma(1-i r \epsilon_{2})}  \dfrac{\Gamma(2i r \epsilon_{1})\Gamma(i r \epsilon_{2}-i r \epsilon_{1})}{\Gamma(1-2i r \epsilon_{1})\Gamma(1+i r \epsilon_{1}-i r \epsilon_{2})} \\
Z_{\text{v}}^{(\text{row})}\,=&\, \sum_{\tilde{d} \geqslant 0}(-z)^{\tilde{d}} \sum_{\tilde{d}_1 = 0}^{\tilde{d}/2}\dfrac{(1+i r \epsilon_{1})_{\tilde{d}-2 \tilde{d}_1}}{(i r \epsilon_{1})_{\tilde{d}-2 \tilde{d}_1}}\dfrac{(i r \epsilon)_{\tilde{d}_{1}}}{\tilde{d}_1 !} \dfrac{(i r \epsilon_{1}+i r \epsilon)_{\tilde{d}- \tilde{d}_1}}{(1+i r \epsilon_{1})_{\tilde{d}- \tilde{d}_1}}\\
&\dfrac{(2 i r \epsilon_{1})_{\tilde{d}-2 \tilde{d}_1}}{ (\tilde{d}-2 \tilde{d}_1) !} \dfrac{(1-i r \epsilon_{2})_{ \tilde{d}-2 \tilde{d}_1}}{(i r \epsilon_{1}+i r \epsilon)_{\tilde{d}-2 \tilde{d}_1}} \dfrac{(i r \epsilon)_{ \tilde{d}-2 \tilde{d}_1}}{(1+i r \epsilon_{1}-i r \epsilon_{2})_{ \tilde{d}-2 \tilde{d}_1}}\\
Z_{\text{av}}^{(\text{row})} \,=&\, \sum_{d \geqslant 0} (-\bar{z})^d \sum_{d_1 = 0}^{d/2} \dfrac{(1+i r \epsilon_{1})_{d-2 d_1}}{(i r \epsilon_{1})_{d-2 d_1}}\dfrac{(i r \epsilon)_{d_{1}}}{d_1 !} \dfrac{(i r \epsilon_{1}+i r \epsilon)_{d- d_1}}{(1+i r \epsilon_{1})_{d- d_1}}\\
&\dfrac{(2 i r \epsilon_{1})_{d-2 d_1}}{ (d-2 d_1) !} \dfrac{(1-i r \epsilon_{2})_{ d-2 d_1}}{(i r \epsilon_{1}+i r \epsilon)_{d-2 d_1}} \dfrac{(i r \epsilon)_{ d-2 d_1}}{(1+i r \epsilon_{1}-i r \epsilon_{2})_{ d-2 d_1}} \label{ff}
\end{split}
\end{equation}  
Here we defined $d = d_1 + d_2$ and changed the sums accordingly. The column contribution can be obtained from the row one by exchanging $\epsilon_{1} \longleftrightarrow \epsilon_{2}$. From the expansion
\begin{equation}
Z_{\text{v}}^{(\text{row, col})} = 1 + 2 i r \epsilon \text{Li}_1(-z) + o(r^2)
\end{equation}
we recover the equivariant mirror map, which can be inverted by replacing 
\begin{eqnarray}
Z_{\text{v}}^{(\text{row, col})} \longrightarrow e^{-2 i r \epsilon  \text{Li}_1(-z)} Z_{\text{v}}^{(\text{row, col})} = (1+z)^{2 i r \epsilon}Z_{\text{v}}^{(\text{row, col})} \nonumber\\
Z_{\text{av}}^{(\text{row, col})} \longrightarrow e^{-2 i r \epsilon   \text{Li}_1(-\bar{z})} Z_{\text{av}}^{(\text{row, col})} = (1+\bar{z})^{2 i r \epsilon}Z_{\text{av}}^{(\text{row, col})}
\end{eqnarray}
Now we can prove the equivalence $\mathcal{M}_{1,2} \simeq \mathcal{M}_{2,1}$: after identifying
\begin{equation}
a_1 = 2 a + \dfrac{\epsilon}{2} + \epsilon_{1} \;\;\;\;\;\;,\;\;\;\;\;\; a_2 = 2 a + \dfrac{\epsilon}{2} + \epsilon_{2}
\end{equation} 
so that $a_{12} = \epsilon_{1} - \epsilon_{2}$ and by expanding in $z$, it can be shown that $Z^{(1)}_{\text{v}}(z) = (1+z)^{2 i r \epsilon}Z_{\text{v}}^{(\text{row})}(z)$ and similarly for the antivortex part; since $Z_{\text{1l}}^{(1)} =  Z_{\text{1l}}^{(\text{row})}$ we conclude that $Z^{(1)}(z, \bar{z}) =  (1+z)^{2 i r \epsilon} (1+\bar{z})^{2 i r \epsilon} Z^{(\text{row})}(z, \bar{z})$. The same is valid for $Z^{(2)}$ and $Z^{(\text{col})}$, so in the end we obtain
\begin{equation}
Z^{S^2}_{1,2}(z, \bar{z}) = (1+z)^{2 i r \epsilon} (1+\bar{z})^{2 i r \epsilon} Z^{S^2}_{2,1}(z,\bar{z})
\end{equation}
Taking into account the appropriate normalizations for both the vortex/antivortex partition functions and the 1-loop factors, this implies
\begin{equation}
Z_{1,2}^{\text{norm}}(z, \bar{z}) = Z_{2,1}^{\text{norm}}(z,\bar{z}) \ \ .
\end{equation}
The K\"{a}hler potential will therefore be given by \eqref{12kahler}. 
The same procedure works for generic $\mathcal{M}_{k,N}$; see \cite{2014JHEP...01..038B} for further examples. 


\subsection{Quantum cohomology in oscillator formalism} \label{sub3.4}

In \cite{2004math.....11210O} the quantum multiplication in the Hilbert scheme of points $\mathcal{M}_{k,1}$ has been computed in terms of operators constructed from oscillators satisfying a Heisenberg algebra. In this subsection we will review that construction and show that the Gromov-Witten potential computed for $\mathcal{M}_{2,1}$ in subsection \ref{sub3.3} is in agreement with what obtained in \cite{2004math.....11210O}.

In \cite{2006math.....10129B} and \cite{2004math.....11210O} the equivariant cohomology of the Hilbert scheme of points of $\mathbb{C}^2$ has been given a Fock space description in terms of creation-annihilation operators $\alpha_p$, $p\in \mathbb{Z}$ obeying the Heisenberg algebra
\begin{equation}
[\alpha_p,\alpha_q] = p\delta_{p+q}
\label{o}
\end{equation}
The positive modes annihilate the vacuum 
\begin{equation}
\alpha_p |\emptyset\rangle = 0 \ \ , p>0
\end{equation}
The natural basis of the Fock space is labelled by partitions:
\begin{equation}
|Y\rangle = \frac{1}{|Aut(Y)|\prod_i Y_i}\prod_i \alpha_{Y_i}|\emptyset\rangle
\label{basis}
\end{equation}
Here $|Aut(Y)|$ is the order of the automorphism group of the partition $Y$ and $Y_i$ are the lengths of the columns of the Young tableau $Y$. The total number of boxes of the Young tableau is counted by the eigenvalue of the energy 
\begin{equation}
K=\sum_{p>0}\alpha_{-p}\alpha_p \label{K}
\end{equation} 
We can now consider the subspace ${\rm Ker}(K-k)$ for $k\in\mathbb{Z}_+$ and allow linear combinations with coefficients being rational functions of the equivariant weights; this space is identified with $H^*_T\left({\cal M}_{k,1},\mathbb{Q}\right)$, and in particular
\begin{equation}
|Y\rangle\in H^{2k-2\ell(Y)}_T\left({\cal M}_{k,1},\mathbb{Q}\right),
\end{equation} 
where $\ell(Y)$ denotes the length of the partition $Y$.

According to \cite{2004math.....11210O} the generator of the small quantum cohomology is given by the state $|D\rangle=-|2,1^{k-2}\rangle$ which describes the divisor corresponding to the collision of two point-like instantons.
The operator of quantum multiplication by $|D\rangle$ reads
\begin{equation}
\begin{split}
H_D \, \equiv \, & \left(\epsilon_1+\epsilon_2\right)\sum_{p>0}\frac{p}{2}\frac{(-q)^p+1}{(-q)^p-1}\alpha_{-p}\alpha_p \\
& + \sum_{p,q >0}\left[\epsilon_1\epsilon_2\alpha_{p+q}\alpha_{-p}\alpha_{-q}-\alpha_{-p-q}\alpha_{p}\alpha_{q}\right]
- \frac{\epsilon_1+\epsilon_2}{2}\frac{(-q)+1}{(-q)-1} K
\label{ham}
\end{split}
\end{equation}
The three-point function can be computed as $\langle D|H_D|D\rangle$, where the inner product is normalized to be
\begin{equation}
\langle Y|Y'\rangle = \frac{(-1)^{K-\ell(Y)}}{\left(\epsilon_1\epsilon_2\right)^{\ell(Y)}
|Aut(Y)|\prod_i Y_i}\delta_{YY'}
\label{form}
\end{equation}
This gives
$$
\langle D|H_D|D\rangle=(\epsilon_1+\epsilon_2)\left(\frac{(-q)^2+1}{(-q)^2-1}-\frac{1}{2}\frac{(-q)+1}{(-q)-1}\right)
\langle D|\alpha_{-2}\alpha_2|D\rangle
=
(-1)(\epsilon_1+\epsilon_2)\frac{1+q}{1-q}\langle D|D\rangle ,
$$
where we used $\langle D|\alpha_{-2}\alpha_2|D\rangle=2\langle D|D\rangle$. By (\ref{form}), we finally get
\begin{equation}
\langle D|H_D|D\rangle=\frac{\epsilon_1+\epsilon_2}{\left(\epsilon_1\epsilon_2\right)^{k-1}}\frac{1}{2(k-2)!}
\left(1+2\frac{q}{1-q}\right)
\end{equation}
If we rewrite $1+2\frac{q}{1-q}=\left(q\partial_q\right)^3\left[\frac{\left({\rm ln}q\right)^3}{3!}+2{\rm Li}_3(q)\right]$, 
we obtain the genus zero prepotential 
\begin{equation}
F^0=F^0_{cl}+\frac{\epsilon_1+\epsilon_2}{\left(\epsilon_1\epsilon_2\right)^{k-1}}\frac{1}{2(k-2)!}\left[\frac{\left({\rm ln}q\right)^3}{3!}+2{\rm Li}_3(q)\right]
\end{equation}
which agrees with the prepotential one can extract from \eqref{12kahler}.
In \cite{2014JHEP...01..038B} this comparison has been extended to $\mathcal{M}_{3,1}$ and $\mathcal{M}_{4,1}$. 

The generalization of the Fock space formalism to $\mathcal{M}_{k,N}$ with generic $N$ was given by Baranovsky in \cite{1998math.....11092B} in terms of $N$ copies of Nakajima operators as $\beta_k = \sum_{i=1}^N \alpha_k^{(i)}$. For  example, in the case $N=2$ the operator of quantum multiplication becomes (modulo terms proportional to $K = \sum_{i=1}^2 \sum_{p>0}\alpha_{-p}^{(i)}\alpha_p^{(i)}$) \cite{2012arXiv1211.1287M}
\begin{equation}
\begin{split}
H_D \,=\, & \dfrac{1}{2} \sum_{i=1}^2 \sum_{n,k>0} [\E\EE \alpha_{-n}^{(i)}\alpha_{-k}^{(i)}\alpha_{n+k}^{(i)} - \alpha_{-n-k}^{(i)}\alpha_{n}^{(i)}\alpha_{k}^{(i)}] \\
&\, - \dfrac{\E + \EE}{2} \sum_{k>0} k[ \alpha_{-k}^{(1)}\alpha_{k}^{(1)} + \alpha_{-k}^{(2)}\alpha_{k}^{(2)} + 2 \alpha_{-k}^{(2)}\alpha_{k}^{(1)}] \\
&\, -(\E + \EE)\sum_{k>0} k \dfrac{q^{k}}{1-q^{k}} [ \alpha_{-k}^{(1)}\alpha_{k}^{(1)} + \alpha_{-k}^{(2)}\alpha_{k}^{(2)} + \alpha_{-k}^{(2)}\alpha_{k}^{(1)} + \alpha_{-k}^{(1)}\alpha_{k}^{(2)}] \label{oqm2}
\end{split}
\end{equation}
We will see in Section \ref{chap4} how these operators of quantum multiplication are related to the Hamiltonians of quantum integrable systems of hydrodynamic type.


\subsection{Orbifold cohomology of the ADHM moduli space} \label{sub3.5}

The moduli space ${\cal M}_{k,N}$ of $k$ $SU(N)$ instantons on ${\mathbb C}^2$ is non-compact for two reasons: 
first of all, the manifold ${\mathbb C}^2$ is non-compact; the second source of non-compactness is due to point-like instantons. 
The first problem can be solved by introducing the so-called $\Omega$-background which corresponds to work in the equivariant cohomology with respect to the maximal torus of rotations $U(1)_{\epsilon_1} \times U(1)_{\epsilon_2}$ on ${\mathbb C}^2$. The second one can be approached in different ways. A compactification scheme is the Uhlembeck one:
\beq
{\cal M}^U_{k,N}=\bigsqcup_{l=0}^k {\cal M}_{k-l,N} \times S^l\left({\mathbb C}^2\right)
\eeq
however, due to the presence of the symmetric product factors this space contains orbifold singularities.
A desingularization is provided by the moduli space of torsion free sheaves on ${\mathbb P}^2$ with a framing on the line at infinity. This is described in terms of the ADHM complex linear maps $(B_1,B_2): {\mathbb C}^k\to {\mathbb C}^k$ and 
$(I, J^\dagger): {\mathbb C}^N \to {\mathbb C}^k$ which satisfy the F-term equation
$$[B_1,B_2]+IJ=0$$
and the D-term equation
$$[B_1,B_1^\dagger]+[B_2,B_2^\dagger]+II^\dagger - J^\dagger J = \xi {\mathbb I}$$
where $\xi$ is a parameter that gets identified with the FI parameter of the GLSM and that ensures the stability condition of the sheaf.

Notice that the ADHM equations are symmetric under the reflection $\xi\to - \xi$ and 
$$
(B_i,I,J)\to (B_i^\dagger,-J^\dagger,I^\dagger) 
$$
The Uhlembeck compactification is recovered in the $\xi\to 0$ limit, which allows pointlike instantons. This amounts to set the vortex expansion parameter as 
\beq
(-1)^Nz=e^{i\theta}
\label{map}
\eeq
giving therefore the orbifold ${\cal I}$-function 
\begin{equation}
\begin{split}
{\cal I}_{k,N}^U & =
\sum_{d_1,\ldots , d_k \,\geq \,0} (e^{i\theta})^{d_1+ \ldots +d_k}  \prod_{r=1}^k \prod_{j=1}^N \dfrac{(-r \lambda_{r}-i r a_{j}+ i r \epsilon)_{d_r}}{(1-r \lambda_{r}-i r a_{j})_{d_r}}
\prod_{r<s}^k \dfrac{d_s - d_r - r \lambda_{s} + r \lambda_{r}}{- r \lambda_{s} + r \lambda_{r}} \\
& \dfrac{(1+ r \lambda_{r}- r\lambda_{s}-i r \epsilon)_{d_s - d_r}}{( r \lambda_{r}- r\lambda_{s}+i r \epsilon)_{d_s - d_r}} 
\dfrac{( r \lambda_{r}- r\lambda_{s}+i r \epsilon_{1})_{d_s - d_r}}{(1+ r \lambda_{r}- r\lambda_{s}-i r \epsilon_{1})_{d_s - d_r}}
\dfrac{(r \lambda_{r}- r\lambda_{s}+i r \epsilon_{2})_{d_s - d_r}}{(1+ r \lambda_{r}- r\lambda_{s}-i r \epsilon_{2})_{d_s - d_r}}
\label{IKNU}
\end{split}
\end{equation}
In the abelian case $N=1$ (Hilbert schemes of points) the above ${\cal I}$-function reproduces the results of \cite{2006math.....10129B} for the equivariant quantum cohomology of the symmetric product of $k$ points in ${\mathbb C}^2$. Indeed, by using the map to the Fock space formalism for the equivariant quantum cohomology reviewed in subsection \ref{sub3.4}, it is easy to see that both approaches produce the same small equivariant quantum cohomology. Notice that the map \eqref{map} reproduces in the $N=1$ case the one of \cite{2006math.....10129B}.


\subsection{D5-branes dynamics and Donaldson-Thomas theory} \label{sub3.6}

Let us consider for a moment the brane construction of subsection \ref{sub3.1} with $p=-1$; this is the original setting considered by Nekrasov. The complete partition function for the four-dimensional $\mathcal{N} = 2$ pure $U(N)$ Yang-Mills theory living on the D3-branes will have the form
\begin{equation}
\mathcal{Z}^{(N)}_{\text{4d}} = \mathcal{Z}^{(N)}_{\text{4d,1l}} \mathcal{Z}^{(N)}_{\text{4d,np}} \label{totalpf4d}
\end{equation}
Here $\mathcal{Z}^{(N)}_{\text{4d,np}}$ is the instanton term coming from the D(-1) branes, which contains the non-perturbative corrections to the D3-brane dynamics. The perturbative part of the D3-brane dynamics is contained in $\mathcal{Z}^{(N)}_{\text{4d,1l}}$; this has been computed in \cite{2003hep.th....6238N} and reads
\begin{equation}
\mathcal{Z}^{(N)}_{\text{4d,1l}} = \prod_{l \neq m}^N \Gamma_2(a_{lm}, \epsilon_1, \epsilon_2) \label{pert4d}
\end{equation}
The partition function \eqref{totalpf4d} computes the free energy $\mathcal{E}_{4d}$ of the system according to
\begin{equation}
\mathcal{Z}^{(N)}_{\text{4d}} = \text{exp} \Bigg\{ -\dfrac{1}{\epsilon_1 \epsilon_2} \mathcal{E}_{4d}(\vec{a}, \epsilon_1, \epsilon_2, \Lambda) \Bigg\}
\end{equation}
with $\Lambda$ instanton counting parameter. $\mathcal{E}_{4d}$ is a regular function as $\epsilon_{1,2} \rightarrow 0$, and in this limit becomes the Seiberg-Witten prepotential of the IR four-dimensional theory. \\ 

We can now return to the D5-D1 branes case ($p=1$). In analogy with the familiar four-dimensional case, the total partition function of our six-dimensional $\mathcal{N} = 1$ pure $U(N)$ Yang-Mills theory in $\Omega$-background will be
\begin{equation}
\mathcal{Z}^{(N)}_{\text{6d}} = \mathcal{Z}^{(N)}_{\text{6d,1l}} \mathcal{Z}^{(N)}_{\text{6d,np}} \label{totalpf6d}
\end{equation}
The non-perturbative term
\begin{equation}
\mathcal{Z}^{(N)}_{\text{6d,np}} = \sum_{k \geqslant 0} Q^k Z_{k,N}^{S^2} (\vec{a}, \epsilon_1, \epsilon_2, z, \bar{z}, r)
\end{equation}
is just the partition function of the D1-branes computed in Section \ref{sub3.2} once we resum over $k$, and provides the
non-perturbative corrections to the D5-brane dynamics; it takes into account the contributions of the topological sectors
of the gauge theory labelled by the second and third Chern character of the gauge bundle, with counting parameters $Q$ and $(z, \bar{z})$ respectively.

The perturbative term $\mathcal{Z}^{(N)}_{\text{6d,1l}}$ has been computed in \cite{2014JHEP...01..038B} and reads
\begin{equation}
\mathcal{Z}^{(N)}_{\text{1l}} = \prod_{l \neq m}^N \Gamma_2(a_{lm}, \epsilon_1, \epsilon_2) \dfrac{\Gamma_3\left(a_{lm}, \epsilon_1, \epsilon_2, \frac{1}{i r}\right)}{\Gamma_3\left(a_{lm}, \epsilon_1, \epsilon_2, -\frac{1}{i r}\right)}
= \prod_{l \neq m}^N \Gamma_3\left(a_{lm}, \epsilon_1, \epsilon_2, \frac{i}{r}\right)^{-2}
\end{equation}
This deforms the standard expression for the perturbative part of the Nekrasov partition function \eqref{pert4d} by implementing the finite $r$ corrections, related to the resummation over the Kaluza-Klein modes.

The expression \eqref{totalpf6d} will compute the free energy $\mathcal{E}_{6d}$ of the six-dimensional theory on $\mathbb{C}^2\times \mathbb{P}^1$ via
\begin{equation}
\mathcal{Z}^{(N)}_{\text{6d}} = \text{exp} \Bigg\{ -\dfrac{1}{\epsilon_1 \epsilon_2} \mathcal{E}_{6d}(\vec{a}, \epsilon_1, \epsilon_2, \Lambda; r,z) \Bigg\}
\end{equation}
Again, $\mathcal{E}_{6d}$ is a regular function as $\epsilon_{1,2} \rightarrow 0$, since $\mathcal{Z}^{(N)}_{\text{6d}}$ has the same divergent behaviour as $\mathcal{Z}^{(N)}_{\text{4d}}$ due to the equivariant regularization of the $\mathbb{C}^2$ volume $\frac{1}{\epsilon_1 \epsilon_2}$. Moreover, $\mathcal{E}_{6d}$ reduces to $\mathcal{E}_{4d}$ in the limit $r \rightarrow 0$; higher order corrections in $r$ encode the effect of stringy corrections due to the blown-up sphere resolving the $\mathbb{C}^2/\mathbb{Z}_2$ singularity. The free energy $\mathcal{E}_{6d}$ is expected to be related to higher rank equivariant Donaldson-Thomas theory on $\mathbb{C}^2\times \mathbb{P}^1$; this would lead to the higher rank analogue of the equivalence between Gromov-Witten theory, Donaldson-Thomas theory and quantum cohomology of the Hilbert scheme of points considered in \cite{2005math.....12573O}.

The mathematical framework hosting our results is the theory of ADHM moduli sheaves as developed in \cite{2012JGP....62..763D}. In such a context one gets ${\cal I}_{k,1}={\cal I}_{DT}$ and therefore the
${\cal I}_{k,N}$ function is the most natural candidate for its higher rank generalization.
On the other hand, one can also show that ${\cal I}_{k,1}$ reproduces the 1-legged Pandharipande-Thomas vertex
as in \cite{2007arXiv0709.3823P}.

Let us finally remark that recently a connection between the classical part of the six dimensional partition function \eqref{totalpf6d} and refined topological vertex has been discussed in \cite{Manabe:2014rma}.


\section{ADHM Gauged Linear Sigma Model:\\ Coulomb branch and quantum hydrodynamics} \label{chap4}

We discussed in subsection \ref{sub2.4} how the twisted LG mirror theory in the Coulomb branch is related to quantum integrable systems. In this Section we will consider the mirror of the ADHM GLSM studied in Section \ref{chap3}; the proposal for the associated QIS is a system of hydrodynamic type, the so-called $gl(N)$ periodic Intermediate Long Wave system (ILW${}_N$ or ILW for $N=1$). After a brief review of the basic facts concerning ILW, we will discuss the details of the correspondence between our mirror LG and this hydrodynamic integrable system.


\subsection{The Intermediate Long Wave system} \label{sub4.1}

The non-periodic ILW equation
\begin{equation}
u_t=2 u u_x + \frac{Q}{\delta}u_x + Q {\cal T}[ u_{xx}]
\label{ilw}
\end{equation}
is an integro-differential equation which describes the wave dynamics at the interface of two fluids in a channel of finite depth $\delta$. Here $Q$ is a parameter related to the ratio of the densities of the fluids, while ${\cal T}$ is the integral operator
\begin{equation}
{\cal T}[f](x) = \dfrac{1}{2 \delta} P.V. \int{\rm coth}\left(\frac{\pi(x-y)}{2\delta}\right)f(y) dy
\label{intop}
\end{equation}
with $P.V. \int$  principal value integral. In the limit $\delta \rightarrow 0$ it reduces to the KdV equation
\begin{equation}
u_t = 2u u_x + \frac{Q \delta}{3} u_{xxx}
\label{kdv}
\end{equation} 
while in the infinite-depth limit $\delta \rightarrow \infty$ it becomes the Benjamin-Ono (BO) equation
\begin{equation}
u_t = 2 u u_x + Q H[u_{xx}]
\label{bo1}
\end{equation}
with $H$ Hilbert transform on the real line:
\begin{equation}
H[f](x)=P.V.\int\frac{1}{x-y}f(y) \frac{dy}{\pi}
\label{hilbop}
\end{equation}
The KdV equation \eqref{kdv} is a famous integrable differential equation; \eqref{ilw} can be seen as an integrable deformation of KdV and in fact the form of the integral kernel in \eqref{intop} is fixed by the requirement of integrability \cite{0305-4470-37-32-L02}. \\

What we will be interested in is the periodic version of ILW, in which we identify $x \sim x + 2 \pi$; this is obtained by simply replacing \eqref{intop} with
\begin{equation}
{\cal T}[f](x)=\frac{1}{2\pi} P.V.\int_0^{2\pi}
\frac{\theta_1'}{\theta_1}\left(\frac{y-x}{2},q\right)f(y) dy
\label{perintop}
\end{equation}
where $q=e^{-\delta}$. Equation \eqref{ilw} is Hamiltonian with respect to the Poisson bracket
\begin{equation}
\{u(x),u(y)\}=\delta'(x-y)
\label{1poisson}
\end{equation}
and in particular can be written as
\begin{equation}
u_t(x)=\{I_2,u(x)\}
\label{hamilw}
\end{equation}
with respect to the Hamiltonian $I_2=\int \frac{1}{3} u^3 + \frac{Q}{2} u {\cal T}[u_{x}]$.
The other conserved quantities are given by $I_1=\int\frac{1}{2}u^2$ and $I_{n-1}=\int \frac{1}{n}u^n+\ldots$ for $n>3$, where the missing pieces are determined by the involution condition $\{I_n,I_m\}=0$. These have been computed explicitly in \cite{LR}. 

The ILW${}_N$ system is described in \cite{LR2} in terms of a system of $N$ coupled integrable integro-differential PDEs in $N$ fields; more explicit formulae for the ILW${}_2$ case can be found in \cite{2013JHEP...11..155L}. \\

An important class of solutions of the periodic BO system is represented by \textit{solitons}. Soliton solutions are waves whose profile does not change with time, apart from the instants in which two or more solitons scatter. A $k$-soliton can be written in terms of a rational function whose poles dynamics satisfies the $k$-particles trigonometric Calogero-Sutherland system \cite{Abanov:2008ft}. 

This has been generalized in \cite{2014JHEP...07..141B} by considering $k$-soliton solutions for ILW: in this case the dynamics of the position of the poles turns out to be described by the $k$-particles elliptic Calogero-Sutherland system. Let us review the argument here. The $k$-particle elliptic Calogero-Sutherland model is defined by the Hamiltonian \begin{equation}
H_{eCM}=\frac{1}{2}\sum_{j=1}^k p_j^2+Q^2\sum_{l<j}\wp(x_l-x_j;\omega_1,\omega_2) \label{ecmham}
\end{equation}
Here $\wp$ is the elliptic Weierstrass $\wp$-function and the periods are chosen as $2\omega_1=L$ and $2\omega_2=i\delta$; we usually set $L=2\pi$. From \eqref{ecmham} we can extract the Hamilton equations 
\begin{align}
\label{eq:HEoM}
\dot{x}_j&=p_j \notag \\
\dot{p}_j&=-G^2\partial_j\sum_{l\neq j}\wp(x_j-x_l),
\end{align}
which can be recast as a second order equation of motion
\begin{equation}
\label{eq:LEoM}
\ddot{x}_j=-Q^2\partial_j\sum_{l\neq j}\wp(x_j-x_l).
\end{equation}
It can be shown that equation \eqref{eq:LEoM} is equivalent to the auxiliary system  
\begin{align}
\label{eq:auxsystem}
\dot{x}_j&=iQ\Bigg\{\sum_{l=1}^k\frac{\theta_1^{\prime}\left(\frac{\pi}{L}(x_j-y_l)\right)}{\theta_1\left(\frac{\pi}{L}(x_j-y_l)\right)}-\sum_{l\neq j}\frac{\theta_1^{\prime}\left(\frac{\pi}{L}(x_j-x_l)\right)}{\theta_1\left(\frac{\pi}{L}(x_j-x_l)\right)} \Bigg\} \notag \\ 
\dot{y}_j&=-iQ\Bigg\{\sum_{l=1}^k\frac{\theta_1^{\prime}\left(\frac{\pi}{L}(y_j-x_l)\right)}{\theta_1\left(\frac{\pi}{L}(y_j-x_l)\right)}-\sum_{k\neq j}\frac{\theta_1^{\prime}\left(\frac{\pi}{L}(y_j-y_l)\right)}{\theta_1\left(\frac{\pi}{L}(y_j-y_l)\right)} \Bigg\}.
\end{align}
Notice that in the limit $\delta\to \infty$ ($q\to 0$), the equation of motion \eqref{eq:LEoM} reduces to
\begin{equation}
\label{eq:EoMtrig}
\ddot{x}_j=-Q^2\left(\frac{\pi}{L}\right)^2\partial_j\sum_{l\neq j}\cot^2\left(\frac{\pi}{L}(x_j-x_l)\right),
\end{equation}
while the auxiliary system goes to
\begin{alignat}{3}
\label{eq:auxsystemtrig}
\dot{x}_j&=iQ\frac{\pi}{L}\Bigg\{\sum_{l=1}^k\cot\left(\frac{\pi}{L}(x_j-y_l)
\right)-\sum_{l\neq j}\cot\left(\frac{\pi}{L}(x_j-x_l)
\right)\Bigg\} \notag \\
\dot{y}_j&=-iQ\frac{\pi}{L}\Bigg\{\sum_{l=1}^k\cot\left(\frac{\pi}{L}(y_j-x_l)
\right)-\sum_{l\neq j}\cot\left(\frac{\pi}{L}(y_j-y_l)
\right)\Bigg\} 
\end{alignat}
and we recover the BO soliton solution obtained in \cite{Abanov:2008ft}.
In analogy with \cite{Abanov:2008ft} we now define a pair of functions encoding the particle positions as simple poles
\begin{align}
u_1(z)&=-iQ\sum_{j=1}^k\frac{\theta_1^{\prime}\left(\frac{\pi}{L}(z-x_j)\right)}{\theta_1\left(\frac{\pi}{L}(z-x_j)\right)} \notag \\
u_0(z)&=iQ\sum_{j=1}^k\frac{\theta_1^{\prime}\left(\frac{\pi}{L}(z-y_j)\right)}{\theta_1\left(\frac{\pi}{L}(z-y_j)\right)}
\end{align}
The linear combinations
\begin{equation}
u=u_0+u_1, \;\; \widetilde{u}=u_0-u_1.
\end{equation}
satisfy the differential equation
\begin{equation}
\label{eq:diffeq}
u_t+uu_z+i\frac{Q}{2}\widetilde{u}_{zz}=0,
\end{equation}
as long as $x_j$ and $y_j$ are governed by the dynamical equations \eqref{eq:auxsystem}. When the lattice of periodicity is rectangular, \eqref{eq:diffeq} is nothing but the ILW equation: in fact under the condition $x_i=\bar y_i$ one can show that $\tilde u = -i \mathcal{T}u$ \cite{LR}. 
To recover \eqref{ilw} one can shift $u\to u+1/2\delta$ and rescale parameters.
We stress that \eqref{eq:diffeq} does not explicitly depend on the number of particles $k$ and therefore holds also in the hydrodynamical limit $k,L\to\infty$, with $k/L$ fixed. 
\\

The periodic ILW system can be canonically quantized by first expanding the field $u$ in Fourier modes $\alpha_k$ and then promoting the $\alpha_k$ modes to creation/annihilation operators; from \eqref{1poisson} we get a quantum commutator
\begin{equation}
\left[\alpha_k,\alpha_l\right]=k\delta_{k+l}
\end{equation}
representing a Heisenberg algebra. The quantum Hamiltonians $\hat{I}_n$ can be recovered from the classical ones $I_n$ after an appropriate quantization procedure which also involve normal ordering \cite{2013JHEP...11..155L}; for example we have 
\begin{equation}
\hat{I}_1 = 2\sum_{k>0}\alpha_{-k}\alpha_k-\frac{1}{24} \label{I21}
\end{equation}
\begin{equation}
\begin{split}
\hat{I}_2 = \frac{Q}{2} \sum_{k>0} k \frac{(-q)^k+1}{(-q)^k-1}\alpha_{-k}\alpha_k + 
\sum_{k,l >0} \left[\epsilon_1\epsilon_2\alpha_{k+l}\alpha_{-k}\alpha_{-l} - \alpha_{-k-l}\alpha_{k}\alpha_{l}\right]
- \frac{Q}{2}\frac{(-q)+1}{(-q)-1} \sum_{k>0} \alpha_{-k}\alpha_k \label{I31}
\end{split}
\end{equation}
Here we introduced a complexified depth parameter $2 \pi t = \delta - i \theta$ entering $q = e^{-2\pi t}$. By doing this we get a first hint of why the ILW system might have something to do with the moduli space of instantons: if we identify $Q = \epsilon_1 + \epsilon_2$ we can immediately recognize that the operator of quantum multiplication for the Hilbert scheme of points $\mathcal{M}_{k,1}$ \eqref{ham} coincides with the $\hat{I}_2$ quantum ILW Hamiltonian; the number of points $k$ will be given by the eigenvalue of $\hat{I}_1$, which is related to the energy operator \eqref{K}. The $\alpha_k$ creation and annihilation operators of quantum ILW are identified with the Nakajima operators describing the equivariant cohomology of $\mathcal{M}_{k,1}$: this is why one has to consider {\it periodic} ILW to make a comparison with gauge theory. Notice also that the complexified depth parameter $2\pi t= \delta - i \theta$ gets identified with the complexified K\"ahler parameter $2\pi t= \xi - i \theta$ of the Hilbert scheme of points. In this way the quantum ILW hamiltonian structure reveals to be related to abelian six dimensional gauge theories via BPS/CFT correspondence. The BO limit $t\to \pm\infty$ corresponds to the classical equivariant cohomology of the instanton moduli space, and therefore describes the four-dimensional limit of the abelian gauge theory. \\

Quantization of ILW${}_N$, related to non-abelian gauge theories, will produce the algebra ${ H}\oplus { W_{N}}$ with ${ H}$ Heisenberg algebra of a single chiral $U(1)$ current. The case $N=2$ has been studied in detail in \cite{2013JHEP...11..155L}, while its BO$_2$ limit has been shown in \cite{Alba:2010qc} to appear in the AGT realization of the 4d ${\cal N}=2$ $SU(2)$ gauge theory with $N_f=4$: in fact the expansion of the conformal blocks proposed in \cite{Alday:2009aq} coincides with the particular basis of descendants in CFT which diagonalizes the BO$_2$ Hamiltonians.

The Fourier modes of ILW${}_N$ correspond to the Baranovsky operators acting on the equivariant cohomology of $\mathcal{M}_{k,N}$. For the case $N=2$ \cite{2014JHEP...07..141B} showed that \eqref{oqm2} can be rewritten in terms of the $\hat{I}_2$ quantum Hamiltonian for ILW${}_2$ given in \cite{2013JHEP...11..155L}: 
\begin{equation}
\begin{split}
\hat{I}_2 = \sum_{k\neq 0} L_{-k}a_k + 2 i Q \sum_{k>0} k a_{-k}a_k \dfrac{1+q^{k}}{1-q^{k}} + \dfrac{1}{3} \sum_{n+m+k=0} a_n a_m a_k \label{I32}
\end{split}
\end{equation}
Here $a_k$, $L_k$ are the modes corresponding to a Heisenberg, Virasoro algebra respectively. The idea is to rewrite the Virasoro generators in terms of a second set of Heisenberg generators $c_k$; we can then make the substitution
\begin{equation}
a_k = -\dfrac{i}{\sqrt{\E\EE}}\dfrac{\alpha_k^{(1)} + \alpha_k^{(2)}}{2} \;\;\;,\;\;\; c_k = - \dfrac{i}{\sqrt{\E\EE}}\dfrac{\alpha_k^{(1)} - \alpha_k^{(2)}}{2}
\end{equation}
for positive modes and 
\begin{equation}
a_{-k} = i \sqrt{\E\EE} \ \dfrac{\alpha_{-k}^{(1)} + \alpha_{-k}^{(2)}}{2} \;\;\;,\;\;\; c_{-k} = i \sqrt{\E\EE}\ \dfrac{\alpha_{-k}^{(1)} - \alpha_{-k}^{(2)}}{2}
\end{equation}
for negative modes and at the end we obtain
\begin{equation}
\begin{split}
\hat{I}_2 =& \dfrac{i}{2\sqrt{\E\EE}} \sum_{n,k>0} [\E\EE \alpha_{-n}^{(1)} \alpha_{-k }^{(1)}\alpha_{n+k}^{(1)} - \alpha_{-n-k}^{(1)} \alpha_n^{(1)} \alpha_k^{(1)}
 + \E\EE \alpha_{-n}^{(2)} \alpha_{-k }^{(2)}\alpha_{n+k}^{(2)} - \alpha_{-n-k}^{(2)} \alpha_n^{(2)} \alpha_k^{(2)} ]\\
& + \dfrac{ i Q}{2} \sum_{k>0} k [\alpha_{-k }^{(1)}\alpha_{k }^{(1)} + \alpha_{-k }^{(2)}\alpha_{k }^{(2)} + 2 \alpha_{-k }^{(2)}\alpha_{k }^{(1)}] \\
& + iQ \sum_{k>0} k \dfrac{q^{k}}{1-q^{k}} [\alpha_{-k }^{(1)}\alpha_{k }^{(1)} + \alpha_{-k }^{(2)}\alpha_{k }^{(2)} + \alpha_{-k }^{(1)}\alpha_{k }^{(2)} + \alpha_{-k }^{(2)}\alpha_{k }^{(1)}]
\end{split}
\end{equation}
which reproduces \eqref{oqm2} after an appropriate rescaling of the $\alpha_{k}^{(i)}$.
 

\subsection{The ADHM mirror LG theory} \label{sub4.2}

In the last subsection we had evidence of the fact that the quantum ILW${}_N$ Hamiltonians coincide with the operators of quantum multiplication in the small equivariant quantum cohomology of $\mathcal{M}_{k,N}$. As we know from Section \ref{chap3} this enumerative problem can be studied by computing the partition function of the ADHM GLSM on $S^2$ interpreted in the Higgs branch. On the other hand, in subsection \ref{sub2.4} we discussed a more direct connection between quantum integrable systems and GLSMs by considering the partition function of the mirror LG theory on the Coulomb branch; here we will further elaborate on this point giving the details for the ADHM theory.

As explained in subsection \ref{sub2.4}, by taking a large $r$ limit of \eqref{sipp} we obtain an expression for $Z^{S^2}_{k,N}$ from which we can extract the twisted effective superpotential describing the IR Coulomb branch of the twisted LG model mirror to the ADHM GLSM. In particular we obtain
\begin{equation}
\begin{split}
Z_{k,N}^{S^2} &= \frac{1}{k!} \left( \dfrac{ \epsilon}{r \epsilon_1 \epsilon_2 } \right)^{k} \Bigg{\vert} 
\int \prod_{s=1}^{k} \frac{ d (r \Sigma_{s})}{2\pi} \left( \dfrac{\prod_{s=1}^{k}\prod_{t \neq s = 1}^{k} D( \Sigma_{st})}{\prod_{s=1}^{k} Q( \Sigma_s)} \right)^{\frac{1}{2}} e^{-\mathcal{W}_{\text{eff}}(\Sigma)} \Bigg{\vert}^2 \label{mir}
\end{split}
\end{equation}
The integration measure is expressed in terms of the functions
\begin{equation}
\begin{split}
Q(\Sigma_s) = r^{2N}\prod_{j=1}^{N}(\Sigma_s - a_j- \frac{\epsilon}{2})(- \Sigma_s + a_j -  \frac{\epsilon}{2})
\;\;\;,\;\;\;
D(\Sigma_{st}) = \dfrac{( \Sigma_{st})( \Sigma_{st} +  \epsilon)}{( \Sigma_{st} - \epsilon_1)( \Sigma_{st} - \epsilon_2)}\label{factors}
\end{split}
\end{equation}
while the twisted effective superpotential reads
\begin{equation}
\begin{split}
\mathcal{W}_{\text{eff}}(\Sigma) \;=\; & (2 \pi t - i(k-1)\pi) \sum_{s=1}^k i r \Sigma_s \\
& + \sum_{s=1}^k \sum_{j=1}^{N} \left[ \omega(i r \Sigma_s - i r a_j -  i r \frac{\epsilon}{2}) + \omega(- i r \Sigma_s + i r a_j - i r \frac{\epsilon}{2}) \right] \\
& + \sum_{\substack{s,t = 1}}^k \left[ \omega(i r \Sigma_{st} + i r \epsilon) + \omega(i r \Sigma_{st} - i r \epsilon_1) + \omega(i r \Sigma_{st} - i r \epsilon_2) \right] \label{YY}
\end{split}
\end{equation}
We remind here the definitions $2 \pi t = 2 \pi \xi - i \theta$ and $\omega(x) = x(\ln x - 1)$.
By extremizing \eqref{YY} according to \eqref{randomeq} we obtain the equations describing the supersymmetric quantum vacua of the Coulomb branch:
\begin{equation}
\begin{split}
&\prod_{j=1}^{N} (\Sigma_s - a_j - \frac{\epsilon}{2}) \prod_{\substack{t = 1 \\ t\neq s}}^k \dfrac{( \Sigma_{st} - \epsilon_1)(\Sigma_{st} - \epsilon_2)}{(\Sigma_{st})(\Sigma_{st} - \epsilon)}
\\&= e^{-2\pi t}  \prod_{j=1}^{N} (- \Sigma_s + a_j -  \frac{\epsilon}{2}) \prod_{\substack{t = 1 \\ t\neq s}}^k \dfrac{(- \Sigma_{st} - \epsilon_1)(- \Sigma_{st} - \epsilon_2)}{(- \Sigma_{st})(- \Sigma_{st} - \epsilon)} \label{bethe}
\end{split}
\end{equation}
We already know that \eqref{bethe} can be thought as Bethe Ansatz Equations for some quantum integrable system; the proposal of \cite{2013JHEP...11..155L} is that the relevant system is ILW${}_N$. The motivation lies in an explicit computation of eigenstates and eigenvalues of the first few quantum ILW${}_N$ Hamiltonians $\hat{I}_n$ (for example \eqref{I21}, \eqref{I31} for $N=1$ and \eqref{I32} for $N=2$) in a perturbative expansion around the BO${}_N$ point $q = 0$. The key observation is noticing that the spectrum can be written in terms of symmetric combinations of the $\Sigma_s$ solutions to \eqref{bethe}, as we will see in the next subsection.

To conclude this subsection, let us perform the semiclassical approximation of \eqref{mir} around the saddle points of \eqref{bethe}. First of all we notice that around the BO point $t\to\infty$ the solutions to \eqref{bethe} can be labelled by $N$-partitions $\vec{\lambda} = (\lambda^{(1)}, \ldots, \lambda^{(N)})$ of $k$, i.e. such that $\sum_{l=1}^N \vert \lambda^{(l)} \vert$ is equal to $k$. In this limit the roots of the Bethe equations are given by
\begin{equation}
\Sigma_m^{(l)} = a_l + \dfrac{\epsilon}{2} + (i-1) \epsilon_1 + (j-1) \epsilon_2 \;\;\;,\;\;\; m = 1, \ldots , \vert \lambda^{(l)} \vert \label{solution}
\end{equation} 
with $i,j$ running over all possible rows and columns of the tableau $\lambda^{(l)} $. These are exactly the poles appearing in the contour integral representation for the 4d Nekrasov partition function \cite{2002hep.th....6161N}. We expect that for large $t$ the roots will be given in terms of a series expansion in powers of $e^{-2 \pi t}$, so that we can still associate an $N$-partitions $\vec{\lambda}$ to each eigenstate of the system, which we will call $\vert \vec{\lambda} (t) \rangle$. Then from \eqref{no}, \eqref{fin} we expect the semiclassical approximation to give the norm of the eigenstates; in fact we obtain
\begin{equation}
\dfrac{1}{\langle \vec{\lambda} (t) \vert \vec{\lambda} (t) \rangle} = \Bigg{\vert} \left( \dfrac{ \epsilon}{r \epsilon_1 \epsilon_2 } \right)^{\frac{k}{2}} \left( \dfrac{\prod_{s=1}^{k}\prod_{t \neq s = 1}^{k} D( \Sigma_{st})}{\prod_{s=1}^{k} Q( \Sigma_s)} \right)^{\frac{1}{2}} \left(\text{Det} \dfrac{\partial^2 \mathcal{W}_{\text{eff}}}{r^2 \partial \Sigma_s \partial \Sigma_t} \right)^{-\frac{1}{2}} \Bigg{\vert}^2_{\Sigma = \Sigma_{\text{cr}}^{(\vec{\lambda})}} 
\end{equation}
which is the formula proposed in \cite{2013JHEP...11..155L} for $\langle \vec{\lambda} (t) \vert \vec{\lambda} (t) \rangle^{-1}$ (when $t$ is real).


\subsection{Quantum ILW Hamiltonians from gauge theory} \label{sub4.3}

As remarked above \eqref{above}, from the Bethe/Gauge correspondence we expect the chiral observables of the ADHM GLSM to provide a basis for the quantum Hamiltonians of ILW${_N}$ \cite{2009NuPhS.192...91N,2009PThPS.177..105N,2010maph.conf..265N}:
\begin{equation}\label{antonio}
\text{ILW quantum Hamiltonians} \;\;\; \longleftrightarrow \;\;\; \text{Tr}\, \Sigma^n (t) \Big \vert_{\text{solution BAE}}
\end{equation}
Actually we should consider the chiral ring observables of the 6d $U(N)$ theory, but due to R-symmetry selection rules these  vanish in the perturbative sector and are therefore completely determined by their non-perturbative contributions, given by our GLSM describing D1-branes dynamics in presence of D(-1)s.    

The calculation of the local chiral ring observables of the $U(N)$ gauge theory on $\mathbb{C}^2\times S^2$ is analogous to the one on $\mathbb{C}^2$, apart from an extra dependence on the $S^2$ coordinates of bosonic and fermionic zero-modes in the instanton background. The sum over fixed points is replaced by the sum over the GLSM vacua and we get
\beq
\text{tr} \ e^{\Phi} = \sum_{l=1}^N  \left(e^{a_l}- e^{-\frac{\E+\EE}{2}}(1-e^{\E})(1-e^{\EE})\sum_m e^{\Sigma_m^{(l)}(t)}\right)
\label{ch}
\eeq 
where $\Sigma_m^{(l)} (t)$ are the solutions of \eqref{bethe}. We expect one can give a mathematical proof of \eqref{ch} in the context of ADHM moduli sheaves introduced in \cite{2012JGP....62..763D}. 

A check of the proposal \eqref{antonio} can be obtained by considering the four dimensional limit $t\to\pm\infty$ where explicit formulae are already known. We saw in \eqref{solution} that in this limit the roots of the Bethe equations are given by \cite{2013JHEP...11..155L} 
\begin{equation}\label{BOBR}
\Sigma_m^{(l)} \;=\; a_l + \dfrac{\epsilon}{2} + (i-1) \epsilon_1 + (j-1) \epsilon_2 \;\;\;,\;\;\; i,j \geqslant 1 \;\;\;,\;\;\; m = 1, \ldots, \vert \lambda^{(l)} \vert \ \ .
\end{equation}
therefore \eqref{ch} reduces to the known formula for the chiral ring observables of four-dimensional $U(N)$ SYM \cite{2003hep.th....2191L,Flume:2004rp}
\begin{equation}
\begin{split}
\text{Tr} \Phi^{n+1} = \sum_{l=1}^{N} a_l^{n+1} + \sum_{l=1}^{N} \sum_{j=1}^{k_1^{(l)}} \Big[ & \left( a_l + \E \lambda_j^{(l)} + \EE(j-1) \right)^{n+1} - \left( a_l + \E \lambda_j^{(l)} + \EE j \right)^{n+1} \\
& - \left( a_l + \EE(j-1) \right)^{n+1} + \left( a_l + \EE j \right)^{n+1} \Big] \label{tr}
\end{split}
\end{equation}
Here $k_1^{(l)}$ is the number of boxes in the first row of the partition $\lambda^{(l)}$, while $\lambda_j^{(l)}$ is the number of boxes in the $j$-th column.
Since the $t\to\pm\infty$ limit corresponds to Benjamin-Ono, we expect the above chiral observables to be related to the quantum Hamiltonians of the BO system. Let us consider the case $N = 2$; in this case the Young tableaux correspond to bipartitions ($\lambda$, $\mu$) $= (\lambda_1 \geqslant \lambda_2 \geqslant \ldots, \mu_1 \geqslant \mu_2 \geqslant \ldots)$  such that $\vert \lambda \vert + \vert \mu \vert = k$. 
The eigenvalues of the BO Hamiltonians $\hat{I}_n$ are given by linear combinations of the eigenvalues of two copies of trigonometric Calogero-Sutherland system \cite{2013JHEP...11..155L,Alba:2010qc} as  
\begin{equation}
h^{(n)}_{\lambda, \mu} = h^{(n)}_{\lambda}(a) + h^{(n)}_{\mu}(-a)
\end{equation}
with
\begin{equation}\label{h}
h^{(n)}_{\lambda}(a) = \EE \sum_{j=1}^{k_1^{(\lambda)}} \left[ \left( a + \E \lambda_j + \EE \left(j-\frac{1}{2}\right) \right)^n - \left( a + \EE \left(j-\frac{1}{2}\right) \right)^n \right]
\end{equation}
In particular, $h^{(1)}_{\lambda, \mu} = \E \EE k$. In terms of \eqref{h}, the $N=2$ chiral observables \eqref{tr} read
\begin{equation}
\begin{split}
\dfrac{\text{Tr} \Phi^{n+1}}{n+1} = 
\dfrac{a^{n+1} + (-a)^{n+1}}{n+1} - \sum_{i=1}^n \dfrac{1+(-1)^{n-i}}{2} \dfrac{n!}{i!(n+1-i)!} \left( \dfrac{\epsilon_2}{2} \right)^{n-i} h^{(i)}_{\lambda, \mu}
\end{split}
\end{equation}
The contributions from $i= 0$, $ i = n+1$ are zero, so they were not considered in the sum. 
The first few cases are:
\begin{equation}\label{few}
\begin{split}
\dfrac{\text{Tr} \Phi^{2}}{2} = a^2 - \E \EE k \hspace{0.5 cm} &, \hspace{0.5 cm}
\dfrac{\text{Tr} \Phi^{3}}{3} = -h^{(2)}_{\lambda, \mu}\\
\dfrac{\text{Tr} \Phi^{4}}{4} = \dfrac{a^4}{2} -h^{(3)}_{\lambda, \mu} - \dfrac{\epsilon_2^2}{4} \E \EE k \hspace{0.5 cm} &, \hspace{0.5 cm}
\dfrac{\text{Tr} \Phi^{5}}{5} = -h^{(4)}_{\lambda, \mu} - \dfrac{\epsilon_2^2}{2} h^{(2)}_{\lambda, \mu}
\end{split}
\end{equation}
We can now rewrite the above formulae in terms of the BO Bethe roots \eqref{BOBR} via the combinations $ \text{Tr}\, \Sigma^n$; in fact from \eqref{ch} we have
\begin{equation*}
\begin{split}
& \dfrac{\text{Tr} \Phi^{2}}{2} = a^2 - \E \EE \left( \sum_{m=1}^{\vert \lambda \vert} 1 + \sum_{n=1}^{\vert \mu \vert}  1 \right) \\
& \dfrac{\text{Tr} \Phi^{3}}{3} = -2 \E \EE \left( \sum_{m=1}^{\vert \lambda \vert} \Sigma_m + \sum_{n=1}^{\vert \mu \vert}  \Sigma_n \right) 
\end{split}
\end{equation*}
\begin{equation}
\begin{split}
& \dfrac{\text{Tr} \Phi^{4}}{4} = \dfrac{a^4}{2} - 3 \E \EE \left( \sum_{m=1}^{\vert \lambda \vert} \Sigma^2_m  + \sum_{n=1}^{\vert \mu \vert}  \Sigma^2_n \right) - \E \EE\dfrac{\E^2 + \EE^2}{4} \left( \sum_{m=1}^{\vert \lambda \vert} 1  + \sum_{n=1}^{\vert \mu \vert} 1 \right)\\
& \dfrac{\text{Tr} \Phi^{5}}{5} = - 4 \E \EE \left( \sum_{m=1}^{\vert \lambda \vert} \Sigma^3_m  + \sum_{n=1}^{\vert \mu \vert}  \Sigma^3_n \right) - \E \EE(\E^2 + \EE^2) \left( \sum_{m=1}^{\vert \lambda \vert} \Sigma_m  + \sum_{n=1}^{\vert \mu \vert} \Sigma_n \right) \ \ .
\end{split}
\label{nome}
\end{equation}
from which
\begin{equation}
\begin{split}
& h^{(1)}_{\lambda} \;=\; \E \EE \sum_{m=1}^{\vert \lambda \vert} 1 \\
& h^{(2)}_{\lambda} \;=\; 2 \E \EE \sum_{m=1}^{\vert \lambda \vert} \Sigma_m \\
& h^{(3)}_{\lambda} \;=\; 3 \E \EE \sum_{m=1}^{\vert \lambda \vert} \Sigma_m^2 + \E \EE \dfrac{\E^2}{4} \sum_{n=1}^{\vert \lambda \vert} 1 \\
& h^{(4)}_{\lambda} \;=\; 4 \E \EE \sum_{m=1}^{\vert \lambda \vert} \Sigma_m^3 + \E \EE \E^2 \sum_{n=1}^{\vert \lambda \vert} \Sigma_m \, .
\end{split}
\end{equation}
In the ILW case we expect the same expressions to be true, the only difference being the dependence on $t$ in the Bethe roots $\Sigma_m(t)$ of the full system \eqref{bethe}.


\section{Generalization to ALE quivers} \label{chap5}

We can apply the procedure described in the previous Sections to more general $\mathcal{N} = (2,2)$ gauge theories on $S^2$ in order to extract information about the equivariant quantum cohomology of the relevant target space. A particularly interesting class of theories is given by Nakajima quiver varieties describing the $k$-instanton moduli space for $U(N)$ gauge theories on ALE spaces $\mathbb{C}^2/\Gamma$, $\Gamma$ being a finite subgroup of $SU(2)$ \cite{Nakajima}. If we consider a system of D1$-$D5 branes on $\mathbb{C}^2/\Gamma \times T^* S^2 \times \mathbb{C}^2$ we can think of Nakajima quivers as GLSMs on $S^2$, whose partition function will give us information about the quantum cohomology of the corresponding target ALE space; similar results were discussed in \cite{emanuel}. 

On the other hand, by analogy with the ADHM quiver case discussed in the previous section, 
we expect the mirror LG theory of a general Nakajima quiver variety to be 
related to quantum integrable systems of hydrodynamic type providing a spin generalization of ILW.
Indeed, the 
Bethe Ansatz Equations that we find reads
\begin{equation}
\boxed{
\prod_{j=1}^{N_b} \dfrac{\Sigma_s^{(b)} - a_j^{(b)} - \frac{\epsilon}{2}}{-\Sigma_s^{(b)} + a_j^{(b)} - \frac{\epsilon}{2}} \prod_{c = 0}^{p-1} \prod_{\substack{t = 1 \\ (c,t) \neq (b,s)}}^{k_c}
\dfrac{\Sigma_s^{(b)} - \Sigma_t^{(c)} + \mathbf{C}_{bc}^T}{\Sigma_s^{(b)} - \Sigma_t^{(c)} - \mathbf{C}_{bc}} 
= e^{-2 \pi t_b} }\label{baep0}
\end{equation}
where $\mathbf{C}_{bc}$ is the adjacency matrix of the quiver graph. For $\epsilon_1=\epsilon_2$, 
this reduces to the Cartan matrix of the corresponding affine Dynkin diagram.
In the $A_{p-1}$ case, \eqref{baep0} reduces exactly to the Bethe Ansatz Equations of 
the spin periodic Intermediate Long Wave quantum system proposed in \cite{2014arXiv1411.3313A}, 
which we will call ILW${}_{\vec{N},p}$. 

Observe that eq.\eqref{baep0} extends the one of the XXX spin chains with higher rank spin group
discussed in \cite{2009NuPhS.192...91N}. Indeed, it coincides with that if one considers the
Cartan matrix of the algebra associated to the classical spin group.
In this sense, eq.\eqref{baep0} might be interpreted as the Bethe Ansatz Equation
for a spin chain with an affine spin group.

We will mainly focus on ALE spaces of type $A_{p-1}$, with some comments also on the ALE spaces of type $D_p$. For the $A_{p-1}-$type we will compute the Gromov-Witten invariants in some examples, finding agreement with the mathematical literature when the comparison is possible. We will then discuss the associated QIS, which can be derived from a hydrodynamic limit of spin Calogero-Sutherland; our results coincide with the generalization of the ILW system proposed in \cite{2014arXiv1411.3313A}.


\subsection{The $A_{p-1}$-type ALE space: Gauged Linear Sigma Model on $S^2$} \label{sub5.1}

An ALE space of type $A_{p-1}$ corresponds to a gauge theory on the space $\mathbb{C}^2/\Gamma$ with $\Gamma = \mathbb{Z}_{p}$, $p \geqslant 2$. The moduli space $\mathcal{M}(\vec{k},\vec{N},p)$ of instantons on this space can be obtained via an ADHM-like construction, whose data are encoded in the associated Nakajima quiver, which in this case is the affine quiver $\widehat{A}_{p-1}$ with framing at all nodes. The vector $\vec{k} = (k_0 , \ldots , k_{p-1})$ parametrize the dimensions of the vector spaces at the nodes of the quiver, while the vector $\vec{N} = (N_0 , \ldots , N_{p-1})$ gives the dimensions of the framing vector spaces; the extra node of the affine Dynkin diagram corresponds to $k_0$. The choice of $\vec{N}$ determines $\vec{k}$ once the Chern class of the gauge vector bundle has been fixed \cite{Nakajima}. \\

The Nakajima quiver can be easily transposed to a GLSM on $S^2$. This theory will have gauge group $G = \prod_{b=0}^{p-1} U(k_b)$, flavour group $G_F = \prod_{b=0}^{p-1} U(N_b) \times U(1)^2$ and matter content summarized in the following table:

\begin{center}
\begin{tabular}{c|c|c|c|c|c}
  & $\chi^{(b)}$ & $B^{(b,b+1)}$ & $B^{(b,b-1)}$ & $I^{(b)}$ & $J^{(b)}$ \\ 
\hline gauge $G$ & $Adj^{(b)}$ & $(\mathbf{\overline{k}}^{(b)},\mathbf{k}^{(b+1)})$ & $(\mathbf{\overline{k}}^{(b)},\mathbf{k}^{(b-1)})$ & $\mathbf{k}^{(b)}$ & $\mathbf{\overline{k}}^{(b)}$ \\ 
\hline flavor $G_F$ & $\mathbf{1}_{(-1,-1)}$ & $\mathbf{1}_{(1,0)}$ & $\mathbf{1}_{(0,1)}$ & $\mathbf{\overline{N}}^{(b)}_{(1/2,1/2)}$ & $\mathbf{N}^{(b)}_{(1/2,1/2)}$ \\ 
\hline twisted masses & $\epsilon = \epsilon_1 + \epsilon_2$ & $-\epsilon_1$ & $-\epsilon_2$ & $-a^{(b)}_j - \frac{\epsilon}{2}$ & $a^{(b)}_j - \frac{\epsilon}{2}$ \\ 
\hline $R$-charge & 2 & 0 & 0 & 0 & 0 \\  
\end{tabular}
\end{center} 
If we consider the superpotential
\begin{equation}
W = \sum_{b=0}^{p-1} \text{Tr}_b [\chi^{(b)}(B^{(b,b+1)}B^{(b+1,b)} - B^{(b,b-1)}B^{(b-1,b)} + I^{(b)} J^{(b)})]
\end{equation} 
(assuming the identification $b \sim b + p$), the $F-$ and $D-$term equations describing the classical space of supersymmetric vacua in the Higgs branch coincide with the ADHM-like equations characterizing $\mathcal{M}(\vec{k},\vec{N},p)$.

\begin{figure}[h!]
  \centering
\includegraphics[width=0.5\linewidth]{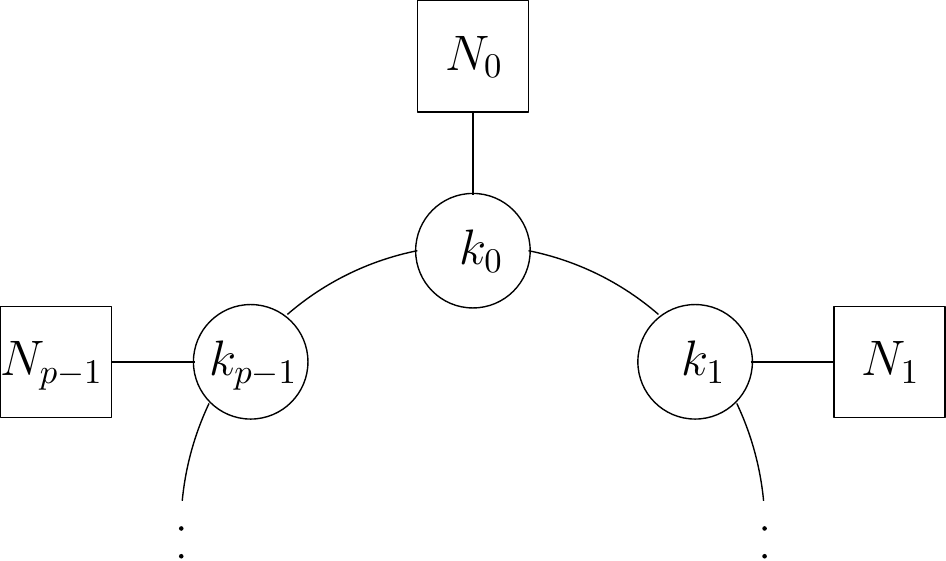} 
  \caption{The affine $\widehat{A}_{p-1}$ quiver.\label{fig:A}}
\end{figure}

We can now write down the partition function on $S^2$ for this GLSM by applying the prescription described in the previous Sections. This is simply given by
\begin{equation}
Z_{\vec{k},\vec{N},p}^{S^2} = \dfrac{1}{k_0! \ldots k_{p-1}!} \sum_{\vec{m}^{(0)},\ldots, \vec{m}^{(p-1)} \in \mathbb{Z}} \int \prod_{b=0}^{p-1} \prod_{s=1}^{k_b} \dfrac{d (r \sigma^{(b)}_s)}{2 \pi} e^{- 4 \pi i \xi_b r \sigma^{(b)}_s - i \theta_{b} m^{(b)}_s} Z_{\text{vec}} Z_{\text{adj}} Z_{\text{bif}} Z_{\text{f} + \text{af}} \label{Z} 
\end{equation}
where the various pieces of the integrand are 
\begin{equation}
\begin{split}
Z_{\text{vec}} &\;=\; \prod_{b=0}^{p-1} \prod_{s<t}^{k_b} (-1)^{m^{(b)}_{s}-m^{(b)}_{t}} \left( \left( r \sigma^{(b)}_{s} - r \sigma^{(b)}_{t} \right)^2 + \left( \frac{m^{(b)}_{s}}{2} - \frac{m^{(b)}_{t}}{2} \right)^2 \right) \\
& \,=\, \prod_{b=0}^{p-1} \prod_{s \neq t}^{k_b} \dfrac{\Gamma \left( 1 - i r \sigma_{s}^{(b)} + i r \sigma_{t}^{(b)} - \frac{m_{s}^{(b)}}{2} + \frac{m_{t}^{(b)}}{2} \right)}{\Gamma \left(  i r \sigma_{s}^{(b)} - i r \sigma_{t}^{(b)} - \frac{m_{s}^{(b)}}{2} + \frac{m_{t}^{(b)}}{2} \right)} \\
Z_{\text{adj}} &\;=\; \prod_{b=0}^{p-1} \prod_{s,t=1}^{k_b} \dfrac{\Gamma \left(1 - i r \sigma_{s}^{(b)} + i r \sigma_{t}^{(b)} - i r \epsilon - \frac{m_{s}^{(b)}}{2} + \frac{m_{t}^{(b)}}{2} \right)}{\Gamma \left( i r \sigma_{s}^{(b)} - i r \sigma_{t}^{(b)} + i r \epsilon - \frac{m_{s}^{(b)}}{2} + \frac{m_{t}^{(b)}}{2} \right)}\\
\end{split}
\end{equation}
\begin{equation}
\begin{split}
Z_{\text{bif}} &\;=\; \prod_{b=0}^{p-1} \prod_{s=1}^{k_b} \prod_{t = 1}^{k_{b-1}} \dfrac{\Gamma \left(- i r \sigma_s^{(b)} + i r \sigma_t^{(b-1)} + i r \epsilon_1 - \frac{m_s^{(b)}}{2} + \frac{m_t^{(b-1)}}{2} \right)}{\Gamma \left(1 + i r \sigma_s^{(b)} - i r \sigma_t^{(b-1)} - i r \epsilon_1 - \frac{m_s^{(b)}}{2} + \frac{m_t^{(b-1)}}{2} \right)} \\
& \hspace{2.8 cm} \dfrac{\Gamma \left(i r \sigma_s^{(b)} - i r \sigma_t^{(b-1)} + i r \epsilon_2 + \frac{m_s^{(b)}}{2} - \frac{m_t^{(b-1)}}{2} \right)}{\Gamma \left(1 - i r \sigma_s^{(b)} + i r \sigma_t^{(b-1)} - i r \epsilon_2 + \frac{m_s^{(b)}}{2} - \frac{m_t^{(b-1)}}{2} \right)}\\
Z_{\text{f+af}} &\;=\; \prod_{b=0}^{p-1} \prod_{s=1}^{k_b} \prod_{j=1}^{N_b} \dfrac{\Gamma \left(- i r \sigma_s^{(b)} + i r a_j^{(b)} + i r \frac{\epsilon}{2} -\frac{m_s^{(b)}}{2} \right)}{\Gamma \left(1 + i r \sigma_s^{(b)} - i r a_j^{(b)} - i r \frac{\epsilon}{2} - \frac{m_s^{(b)}}{2} \right)} \\
& \hspace{2.7 cm} \dfrac{\Gamma \left( i r \sigma_s^{(b)} - i r a_j^{(b)} + i r \frac{\epsilon}{2} + \frac{m_s^{(b)}}{2} \right)}{\Gamma \left(1 - i r \sigma_s^{(b)} + i r a_j^{(b)} - i r \frac{\epsilon}{2} + \frac{m_s^{(b)}}{2} \right)}
\end{split}
\end{equation}
Notice that we included the shift of the $\theta$ angles in $Z_{\text{vec}}$. 
Again, in the limit $r \rightarrow 0$ \eqref{Z} reduces to the contour integral expression for the equivariant volume of $\mathcal{M}(\vec{k},\vec{N},p)$ presented in \cite{2000CMaPh.209...97M}.


\subsection{The $A_{p-1}$-type ALE space: equivariant quantum cohomology} \label{sub5.2}

In order to extract the K\"{a}hler potential and Gromov-Witten invariants of $\mathcal{M}_{\vec{k},\vec{N},p}$ from \eqref{Z}, we need to explicitly evaluate the contour integral. As we did in subsection \ref{sub3.2}, we start by performing the change of variables 
\begin{equation}
\sigma_s^{(b)} = - \dfrac{i}{r} \left( l_s^{(b)} - \frac{m_s^{(b)}}{2} \right) + i \lambda_s^{(b)}
\end{equation}
and define $k_s^{(b)} = l_s^{(b)} - m_s^{(b)}$, $z_b = e^{-2 \pi \xi_b - i \theta_b} = e^{-2 \pi t_b}$ with $t_b = \xi_b + i \theta_b/2\pi$ complexified Fayet-Iliopoulos parameter. Thanks to this, the partition function can be factorized before integration and we get
\begin{equation}
Z_{\vec{k},\vec{N},p}^{S^2} = \dfrac{1}{k_0! \ldots k_{p-1}!} \oint \prod_{b=0}^{p-1} \prod_{s=1}^{k_b} \dfrac{d (r \lambda^{(b)}_s)}{2 \pi i} Z_{\text{1l}} Z_{\text{v}} Z_{\text{av}}  \label{part}
\end{equation}
where
\begin{equation}
\begin{split}
Z_{\text{1l}} \;=&\; \prod_{b=0}^{p-1} \prod_{s=1}^{k_b} \left( \dfrac{\Gamma(1- i r \epsilon)}{\Gamma(i r \epsilon)} (z_b \bar{z}_b)^{-r \lambda_s^{(b)}} \right) 
\prod_{b=0}^{p-1} \prod_{s=1}^{k_b} \prod_{t \neq s}^{k_b} (r \lambda_s^{(b)} - r \lambda_t^{(b)})  \dfrac{\Gamma(1 + r \lambda_s^{(b)} - r \lambda_t^{(b)} - i r \epsilon)}{\Gamma(- r \lambda_s^{(b)} + r \lambda_t^{(b)} + i r \epsilon)}\\
& \prod_{b=0}^{p-1} \prod_{s=1}^{k_b} \prod_{t = 1}^{k_{b-1}} \dfrac{\Gamma(r \lambda_s^{(b)} - r \lambda_t^{(b-1)} + i r \epsilon_1)}{\Gamma(1 - r \lambda_s^{(b)} + r \lambda_t^{(b-1)} - i r \epsilon_1)} \dfrac{\Gamma(- r \lambda_s^{(b)} + r \lambda_t^{(b-1)} + i r \epsilon_2)}{\Gamma(1 + r \lambda_s^{(b)} - r \lambda_t^{(b-1)} - i r \epsilon_2)}\\
& \prod_{b=0}^{p-1} \prod_{s=1}^{k_b} \prod_{j=1}^{N_b} \dfrac{\Gamma(r \lambda_s^{(b)} + i r a_j^{(b)} + i r \frac{\epsilon}{2})}{\Gamma(1 - r \lambda_s^{(b)} - i r a_j^{(b)} - i r \frac{\epsilon}{2})} \dfrac{\Gamma(- r \lambda_s^{(b)} - i r a_j^{(b)} + i r \frac{\epsilon}{2})}{\Gamma(1 + r \lambda_s^{(b)} + i r a_j^{(b)} - i r \frac{\epsilon}{2})}
\end{split}
\end{equation}
\begin{equation}
\begin{split}
Z_{\text{v}} \;=&\; \sum_{\{\vec{l}\}} \prod_{b=0}^{p-1} \prod_{s=1}^{k_b} (-1)^{N_b l_s^{(b)}} \prod_{b=0}^{p-1} z_b^{l_s^{(b)}} 
\prod_{b=0}^{p-1} \prod_{s < t}^{k_b} \dfrac{l_t^{(b)} - l_s^{(b)} - r \lambda_t^{(b)} + r \lambda_s^{(b)}}{-r \lambda_t^{(b)} + r \lambda_s^{(b)}}  \dfrac{(1 + r \lambda_s^{(b)} - r \lambda_t^{(b)} - i r \epsilon)_{l_t^{(b)} - l_s^{(b)}}}{(r \lambda_s^{(b)} - r \lambda_t^{(b)} + i r \epsilon)_{l_t^{(b)} - l_s^{(b)}}}\\
& \prod_{b=0}^{p-1} \prod_{s=1}^{k_b} \prod_{t = 1}^{k_{b-1}} \dfrac{1}{(1 - r \lambda_s^{(b)} + r \lambda_t^{(b-1)} - i r \epsilon_1)_{l_s^{(b)} - l_t^{(b-1)}}} \dfrac{1}{(1 + r \lambda_s^{(b)} - r \lambda_t^{(b-1)} - i r \epsilon_2)_{l_t^{(b-1)} - l_s^{(b)} }}\\
& \prod_{b=0}^{p-1} \prod_{s=1}^{k_b} \prod_{j=1}^{N_b} \dfrac{(- r \lambda_s^{(b)} - i r a_j^{(b)} + i r \frac{\epsilon}{2})_{l_s^{(b)}}}{(1 - r \lambda_s^{(b)} - i r a_j^{(b)} - i r \frac{\epsilon}{2})_{l_s^{(b)}}} 
\end{split}
\end{equation}
\begin{equation}
\begin{split}
Z_{\text{av}} \;=&\; \sum_{\{\vec{k}\}} \prod_{b=0}^{p-1} \prod_{s=1}^{k_b} (-1)^{N_b k_s^{(b)}} \prod_{b=0}^{p-1} \bar{z}_b^{k_s^{(b)}} \prod_{b=0}^{p-1} \prod_{s < t}^{k_b} \dfrac{k_t^{(b)} - k_s^{(b)} - r \lambda_t^{(b)} + r \lambda_s^{(b)}}{-r \lambda_t^{(b)} + r \lambda_s^{(b)}}  \dfrac{(1 + r \lambda_s^{(b)} - r \lambda_t^{(b)} - i r \epsilon)_{k_t^{(b)} - k_s^{(b)}}}{(r \lambda_s^{(b)} - r \lambda_t^{(b)} + i r \epsilon)_{k_t^{(b)} - k_s^{(b)}}}\\
& \prod_{b=0}^{p-1} \prod_{s=1}^{k_b} \prod_{t = 1}^{k_{b-1}} \dfrac{1}{(1 - r \lambda_s^{(b)} + r \lambda_t^{(b-1)} - i r \epsilon_1)_{k_s^{(b)} - k_t^{(b-1)}}} \dfrac{1}{(1 + r \lambda_s^{(b)} - r \lambda_t^{(b-1)} - i r \epsilon_2)_{k_t^{(b-1)} - k_s^{(b)} }}\\
& \prod_{b=0}^{p-1} \prod_{s=1}^{k_b} \prod_{j=1}^{N_b} \dfrac{(- r \lambda_s^{(b)} - i r a_j^{(b)} + i r \frac{\epsilon}{2})_{k_s^{(b)}}}{(1 - r \lambda_s^{(b)} - i r a_j^{(b)} - i r \frac{\epsilon}{2})_{k_s^{(b)}}} \\
\end{split}
\end{equation}
As we saw, the vortex partition function $Z_{\text{v}}$ is interpreted in quantum cohomology as Givental's $\mathcal{I}$ function. Moreover, in order to extract the Gromov-Witten prepotential we have to normalize in an appropriate way $Z_{\text{1l}}$ and invert the equivariant mirror map in $Z_{\text{v}}$. For ALE spaces the equivariant  mirror map is known explicitly \cite{2009JAMS...22.1055M}: it only appears when $N = \sum_{b=0}^{p-1} N_b = 1$, in which case the construction in \cite{Nakajima} forces the vectors $\vec{N}$, $\vec{k}$ to be $\vec{N} = (1, 0, \ldots, 0)$ and $\vec{k} = (k, k, \ldots, k)$, and it consists in multiplying $Z_{\text{v}}$ by $(1+\prod_{b=0}^{p-1}z_b)^{i k r \epsilon}$ (and similarly for $Z_{\text{av}}$). On the other hand the normalization factor for $Z_{\text{1l}}$ is not known, and we will have to find it case by case according to the discussion in subsection \ref{sub3.3}; this is equivalent to require the intersection numbers $\langle 1, 1, \ln z \rangle = 0$, with $\ln z$ combination of K\"{a}hler moduli of the target space.

All we need to do now is to classify the poles in terms of $\lambda_s^{(b)}$ variables. They will coincide with the poles of the $r \rightarrow 0$ limit of \eqref{Z}, as for the ADHM partition function. It turns out that the poles can be labelled by ``colored'' Young tableaux, in which each box has a number associated to it according to its $\mathbb{Z}_p$ representation and the values of $\vec{k}$ and $\vec{N}$; see \cite{2004NuPhB.703..518F} for more details, or the examples below. \\

\noindent \textit{An example: the $N=1$, $k=1$ case} \\

\noindent In the rest of this subsection we will study in detail the case $N=1$ in which, as mentioned above, we have $\vec{N} = (1, 0, \ldots, 0)$ and $\vec{k} = (k, k, \ldots, k)$; we will refer to the $N=1$ instanton moduli space as $\mathcal{M}(k,1,p)$ and denote the corresponding spherical partition function with $Z^{S^2}_{k,1,p}$. Here we will only consider $k=1$; other cases will be explored in Appendix \ref{appA}.

When $k=1$ the instanton moduli space is known in the mathematical literature as $\mathcal{M}(1,1,p) = \mathbb{Z}_{p}$-Hilb$(\mathbb{C}^2)$. The corresponding equivariant quantum Gromov-Witten potential $F_{1,1,p}$ has been computed for $\epsilon_1$, $\epsilon_2$ generic in \cite{2006math.....10129B} ($p=2$) and \cite{2005math.....10335B} ($p=3$); in the special limit $\epsilon_1 = \epsilon_2 = \tilde{\epsilon}$ explicit computations are provided in \cite{2007arXiv0707.1337B} in terms of the (inverse) Cartan matrix and root system of the non-affine algebra $A_{p-1}$ for generic $p$. More in detail, let $C_i^j$ be the $A_{p-1}$ Cartan matrix, $i,j = 1 \ldots p-1$, let $\alpha_i$ be the basis of fundamental weights for the $A_{p-1}$ algebra, and define $R^+$ as the set of $p(p-1)/2$ positive roots. Then we have
\begin{equation}
\begin{split}
F_{1,1,p}& = \dfrac{1}{p \, \tilde{\epsilon}^2} - \dfrac{1}{2} \sum_{i,j=1}^{p-1} \langle \alpha_i , \alpha_j \rangle \ln z_i \ln z_j + \dfrac{\tilde{\epsilon}}{6} \sum_{i,j,k=1}^{p-1} \sum_{\beta \in R^+} \langle \alpha_i , \beta \rangle \langle \alpha_j , \beta \rangle \langle \alpha_k , \beta \rangle \ln z_i \ln z_j \ln z_k \\
& + 2 \tilde{\epsilon} \sum_{\beta \in R^+} \text{Li}_3 \left( \prod_{i=1}^{p-1} z_i^{\langle \alpha_i , \beta \rangle} \right) \label{math} 
\end{split}
\end{equation}
with the product $\langle \alpha_i , \alpha_j \rangle = \alpha_i^{T} C^{-1} \alpha_j$ expressed in terms of the inverse Cartan matrix.

Let us show how these results can be recovered from our spherical partition function.

\begin{itemize}

\item \textbf{Case $p = 2$} \\

Here we are considering the $A_{1}$ singularity. The $A_{1}$ algebra data are just
\begin{equation}
C = 2 \;\;\;,\;\;\; C^{-1} = \frac{1}{2} \;\;\;,\;\;\; \alpha_1 = 1
\end{equation}
therefore $\langle \alpha_1 , \alpha_1 \rangle = \frac{1}{2}$. The only positive root corresponds to $\beta = C \alpha_1 = 2$, which implies $\langle \alpha_1 , \beta \rangle = 1$. From \eqref{math} we then expect 
\begin{equation}
F_{1,1,2} = \dfrac{1}{2 \, \tilde{\epsilon}^2} - \dfrac{1}{4}\ln^2 z_1 + \dfrac{\tilde{\epsilon}}{6}\ln^3 z_1 + 2 \tilde{\epsilon} \,\text{Li}_3(z_1) \label{112}
\end{equation}  
We can compare this expression with what we obtain from the evaluation of the partition function $Z_{1,1,2}^{S^2}$. The poles of \eqref{part} are labelled by colored partitions of $\widehat{k} = \sum_{b=0}^{p-1} k_b = p k$; in our case, for positive Fayet-Iliopoulos parameters we have the two poles

\vspace*{0.2 cm}
\begin{center}
\begin{tabular}{ccl}
\ytableausetup{mathmode, boxsize=0.7em, aligntableaux=center} 
\begin{ytableau} \scriptstyle 0 & \scriptstyle 1 \end{ytableau} & $\Longleftrightarrow$ & $\lambda^{(0)}_1 = -i a_1^{(0)} - i \frac{\epsilon}{2}$\;,\; $\lambda^{(1)}_1 = \lambda^{(0)}_1 - i \epsilon_1$ \\ \vspace*{0.3 cm}
\begin{ytableau} \scriptstyle 1 \\ \scriptstyle 0 \end{ytableau} & $\Longleftrightarrow$ & $\lambda^{(0)}_1 = -i a_1^{(0)} - i \frac{\epsilon}{2}$\;,\; $\lambda^{(1)}_1 = \lambda^{(0)}_1 - i \epsilon_2$ 
\end{tabular}  
\end{center}
\vspace*{0.2 cm}

\noindent Inverting the mirror map simply consists in replacing $Z_{\text{v}} \rightarrow (1+z_0z_1)^{i r \epsilon} Z_{\text{v}}$ and $Z_{\text{av}} \rightarrow (1+\overline{z}_0\overline{z}_1)^{i r \epsilon} Z_{\text{av}}$. For the normalization of the 1-loop part, we find  
\begin{equation}
Z_{\text{1l}} \rightarrow (z_0z_1\overline{z}_0\overline{z}_1)^{-i r a_1^{(0)} - i r \frac{\epsilon}{2}} \frac{\Gamma (1 - i r \epsilon_+)}{\Gamma (1 + i r \epsilon_+)} Z_{\text{1l}}
\end{equation}
to be a consistent choice. All it remains to do now is to evaluate the partition function at the two poles, sum the two contributions, and expand in small $r$. At the end we obtain
\begin{equation}
\begin{split}
Z_{1,1,2}^{S^2, \text{norm}} = & -\dfrac{1}{2 \epsilon_1 \epsilon_2} - \dfrac{1}{4} \ln^2(z_1 \overline{z}_1) + i \epsilon \Bigg( -\dfrac{1}{12}\ln^3(z_1 \overline{z}_1) + 4 \zeta(3) \\
& + 2 (\text{Li}_3(z_1) + \text{Li}_3(\overline{z}_1)) - \ln(z_1 \overline{z}_1)(\text{Li}_2(z_1) + \text{Li}_2(\overline{z}_1)) \Bigg) \label{Z112}
\end{split}
\end{equation}
From this expression we can extract the genus zero Gromov-Witten prepotential (see for example \cite{2012arXiv1208.6244J}). For the sake of comparison we redefine $\epsilon_1 \rightarrow i \epsilon_1$, $\epsilon_2 \rightarrow i \epsilon_2$, so that now 
\begin{equation}
F_{1,1,2} = \dfrac{1}{2 \, \epsilon_1 \epsilon_2} - \dfrac{1}{4}\ln^2 z_1 + \dfrac{\epsilon}{12}\ln^3 z_1 + \epsilon \,\text{Li}_3(z_1)
\end{equation}
This coincides with the expression given in \cite{2006math.....10129B} for generic $\epsilon_1$, $\epsilon_2$ and reduces to \eqref{112} in the special limit $\epsilon_1 = \epsilon_2 = \tilde{\epsilon}$. \\

\item \textbf{Case $p = 3$} \\

For the $A_{2}$ algebra we have
\begin{equation}
C = \left( \begin{array}{cc} 2 & -1 \\ -1 & 2 \end{array} \right)
\;\;\; , \;\;\; C^{-1} = \dfrac{1}{3} \left( \begin{array}{cc} \,2\, & \;1\, \\ \,1\, & \;2\, \end{array} \right)
\;\;\; , \;\;\; \alpha_1 = \left( \begin{array}{c} 1 \\ 0 \end{array} \right) 
\;\;\; , \;\;\; \alpha_2 = \left( \begin{array}{c} 0 \\ 1 \end{array} \right) 
\end{equation}
The three positive roots are $\beta_1 = C \alpha_1$, $\beta_2 = C \alpha_2$ and $\beta_3 = C (\alpha_1 + \alpha_2)$, therefore in this case \eqref{math} gives
\begin{equation}
\begin{split}
F_{1,1,3} &= \dfrac{1}{3 \, \tilde{\epsilon}^2} - \dfrac{1}{3} \left( \ln^2 z_1 + \ln z_1\ln z_2 + \ln^2 z_2 \right) \\
& + \tilde{\epsilon}\,\left( \dfrac{1}{3}\ln^3 z_1 + \dfrac{1}{2}\ln^2 z_1 \ln z_2 + \dfrac{1}{2} \ln z_1 \ln^2 z_2 + \dfrac{1}{3}\ln^3 z_2 \right) \\
& + 2 \tilde{\epsilon} \,\left( \text{Li}_3(z_1) + \text{Li}_3(z_2) + \text{Li}_3(z_1 z_2)\right) \label{113}
\end{split}
\end{equation}

To compare with the gauge theory result, we have to compute $Z^{S^2}_{1,1,3}$. The relevant poles are at

\vspace*{0.2 cm}
\begin{center}
\begin{tabular}{ccl}
\ytableausetup{mathmode, boxsize=0.7em, aligntableaux=center} 
\begin{ytableau} \scriptstyle 0 & \scriptstyle 1 & \scriptstyle 2 \end{ytableau} & $\Longleftrightarrow$ & 
$\lambda^{(0)}_1 = -i a_1^{(0)} - i \frac{\epsilon}{2}$ \;,\;
$\lambda^{(1)}_1 = \lambda^{(0)}_1 - i \epsilon_1$  \;,\;
$\lambda^{(2)}_1 = \lambda^{(0)}_1 - 2 i \epsilon_1$ 
\\ 
\begin{ytableau} \scriptstyle 2 \\ \scriptstyle 0 & \scriptstyle 1 \end{ytableau} & $\Longleftrightarrow$ & 
$\lambda^{(0)}_1 = -i a_1^{(0)} - i \frac{\epsilon}{2}$ \;,\;
$\lambda^{(1)}_1 = \lambda^{(0)}_1 - i \epsilon_1$  \;,\;
$\lambda^{(2)}_1 = \lambda^{(0)}_1 - i \epsilon_2$ 
\\ 
\begin{ytableau} \scriptstyle 1 \\ \scriptstyle 2 \\ \scriptstyle 0 \end{ytableau} & $\Longleftrightarrow$ & 
$\lambda^{(0)}_1 = -i a_1^{(0)} - i \frac{\epsilon}{2}$ \;,\;
$\lambda^{(2)}_1 = \lambda^{(0)}_1 - i \epsilon_2$  \;,\;
$\lambda^{(1)}_1 = \lambda^{(0)}_1 - 2 i \epsilon_2$ 
\end{tabular}  
\end{center}
\vspace*{0.2 cm}

Inverting the mirror map by $Z_{\text{v}} \rightarrow (1+z_0z_1z_2)^{i r \epsilon} Z_{\text{v}}$ and $Z_{\text{av}} \rightarrow (1+\overline{z}_0\overline{z}_1\overline{z}_2)^{i r \epsilon} Z_{\text{av}}$, and normalizing the 1-loop part as 
\begin{equation}
Z_{\text{1l}} \rightarrow (z_0z_1z_2\overline{z}_0\overline{z}_1\overline{z}_2)^{-i r a_1^{(0)} - i r \frac{\epsilon}{2}} \frac{\Gamma (1 - i r \epsilon)}{\Gamma (1 + i r \epsilon)} Z_{\text{1l}}
\end{equation}
we obtain
\begin{equation}
\begin{split}
Z_{1,1,3}^{S^2, \text{norm}} &= -\dfrac{1}{3 \epsilon_1 \epsilon_2} - \dfrac{1}{3} \left( \ln^2(z_1 \overline{z}_1) + \ln(z_1 \overline{z}_1)\ln(z_2 \overline{z}_2) + \ln^2(z_2 \overline{z}_2) \right) \\ 
& + i \Bigg( -\dfrac{\epsilon_1 + 2 \epsilon_2}{9}\ln^3(z_1 \overline{z}_1) - \dfrac{\epsilon_1 + 2 \epsilon_2}{6}\ln^2(z_1 \overline{z}_1)\ln(z_2 \overline{z}_2) \\
&\;\;\;\;\;\;\;\; - \dfrac{2\epsilon_1 + \epsilon_2}{6} \ln(z_1 \overline{z}_1) \ln^2(z_2 \overline{z}_2) - \dfrac{2\epsilon_1 + \epsilon_2}{9}\ln^3(z_2 \overline{z}_2) \Bigg) \\
&+ i \epsilon \Bigg(6 \zeta(3) + 2 (\text{Li}_3(z_1) + \text{Li}_3(z_2) + \text{Li}_3(z_1 z_2) + \text{Li}_3(\overline{z}_1) + \text{Li}_3(\overline{z}_2) + \text{Li}_3(\overline{z}_1 \overline{z}_2)) \\
&\;\;\;\;\;\;\;\;\;\;\; - \ln(z_1 \overline{z}_1)(\text{Li}_2(z_1) + \text{Li}_2(\overline{z}_1)) - \ln(z_2 \overline{z}_2)(\text{Li}_2(z_2) + \text{Li}_2(\overline{z}_2)) \\
&\;\;\;\;\;\;\;\;\;\;\; - \ln(z_1 z_2 \overline{z}_1 \overline{z}_2)(\text{Li}_2(z_1 z_2) + \text{Li}_2(\overline{z}_1 \overline{z}_2)) \Bigg) 
\end{split}
\end{equation}
The corresponding genus zero Gromov-Witten prepotential (after the redefinition $\epsilon_1 \rightarrow i \epsilon_1$, $\epsilon_2 \rightarrow i \epsilon_2$) reads 
\begin{equation}
\begin{split}
F_{1,1,3} &= \dfrac{1}{3 \, \epsilon_1 \epsilon_2} - \dfrac{1}{3} \left( \ln^2 z_1 + \ln z_1 \ln z_2 + \ln^2 z_2 \right) \\
& + \left( \dfrac{\epsilon_1 + 2 \epsilon_2}{9}\ln^3 z_1 + \dfrac{\epsilon_1 + 2 \epsilon_2}{6}\ln^2 z_1 \ln z_2 + \dfrac{2\epsilon_1 + \epsilon_2}{6} \ln z_1 \ln^2 z_2 + \dfrac{2\epsilon_1 + \epsilon_2}{9}\ln^3 z_2 \right) \\ 
& + \epsilon \,(\text{Li}_3(z_1) + \text{Li}_3(z_2) + \text{Li}_3(z_1 z_2) )
\end{split}
\end{equation}
and coincides with the expression given in \cite{2005math.....10335B} for generic $\epsilon_1$, $\epsilon_2$, or with \eqref{113} when $\epsilon_1 = \epsilon_2 = \tilde{\epsilon}$.\\

\item \textbf{Case $p = 4$} \\

In the $A_3$ case, the relevant algebra data are
\begin{equation}
C = \left( \begin{array}{ccc} 2 & -1 & 0 \\ -1 & 2 & -1 \\ 0 & -1 & 2 \end{array} \right)
\; , \; 
C^{-1} = \left( \begin{array}{ccc} \,\frac{3}{4}\, & \;\frac{1}{2}\, & \; \frac{1}{4} \, \\ \,\frac{1}{2}\, & \;1\, & \; \frac{1}{2} \, \\ \,\frac{1}{4}\, & \;\frac{1}{2}\, & \; \frac{3}{4} \,  \end{array} \right) 
\; , \;
\alpha_1 = \left( \begin{array}{c} 1 \\ 0 \\ 0 \end{array} \right) 
\; , \; 
\alpha_2 = \left( \begin{array}{c} 0 \\ 1 \\ 0 \end{array} \right) 
\; , \; 
\alpha_3 = \left( \begin{array}{c} 0 \\ 0 \\ 1 \end{array} \right) 
\end{equation}
We have the six positive roots $\beta_1 = C \alpha_1$, $\beta_2 = C \alpha_2$, $\beta_3 = C \alpha_3$, $\beta_4 = C (\alpha_1 + \alpha_2)$, $\beta_5 = C (\alpha_2 + \alpha_3)$, $\beta_6 = C (\alpha_1 + \alpha_2 + \alpha_3)$, which inserted in \eqref{math} lead to 
\begin{equation}
\begin{split}
F_{1,1,4} &= \dfrac{1}{4 \, \tilde{\epsilon}^2} - \dfrac{1}{8} \left( 3 \ln^2 z_1 + 4 \ln^2 z_2 + 3 \ln^2 z_3 + 4 \ln z_1\ln z_2 + 2 \ln z_1\ln z_3 + 4  \ln z_2\ln z_3 \right) \\
& + \tilde{\epsilon}\Big( \dfrac{1}{2}\ln^3 z_1 + \ln^2 z_1 \ln z_2 + \dfrac{1}{2} \ln^2 z_1 \ln z_3 + \dfrac{2}{3}\ln^3 z_2
 + \ln z_1 \ln^2 z_2 + \ln z_1 \ln z_2 \ln z_3 \\
& \;\;\;\; + \ln^2 z_2 \ln z_3 + \dfrac{1}{2} \ln^3 z_3 + \dfrac{1}{2} \ln z_1 \ln^2 z_3 + \ln z_2 \ln^2 z_3 \Big) \\
& + 2 \tilde{\epsilon} \,(\text{Li}_3(z_1) + \text{Li}_3(z_2) + \text{Li}_3(z_3)  + \text{Li}_3(z_1 z_2) + \text{Li}_3(z_2 z_3) + \text{Li}_3(z_1 z_2 z_3)) \label{temp}
\end{split}
\end{equation} 
On the other hand, to compute the partition function $Z_{1,1,4}$ we have to evaluate residues at the four poles

\vspace*{0.2 cm}
\begin{center}
\begin{tabular}{ccl}
\ytableausetup{mathmode, boxsize=0.7em, aligntableaux=center} 
\begin{ytableau} \scriptstyle 0 & \scriptstyle 1 & \scriptstyle 2 & \scriptstyle 3 \end{ytableau} & $\Longleftrightarrow$ & 
\Bigg\{
\begin{tabular}{ll}
$\lambda^{(0)}_1 = -i a_1^{(0)} - i \frac{\epsilon}{2}$ \;, &
$\lambda^{(1)}_1 = \lambda^{(0)}_1 - i \epsilon_1$  \\ 
$\lambda^{(2)}_1 = \lambda^{(0)}_1 - 2 i \epsilon_1$ \;, &
$\lambda^{(3)}_1 = \lambda^{(0)}_1 - 3 i \epsilon_1$ \\ 
\end{tabular} 
\\
\begin{ytableau} \scriptstyle 3 \\ \scriptstyle 0 & \scriptstyle 1 & \scriptstyle 2 \end{ytableau} & $\Longleftrightarrow$ & 
\Bigg\{
\begin{tabular}{ll}
$\lambda^{(0)}_1 = -i a_1^{(0)} - i \frac{\epsilon}{2}$ \;, &
$\lambda^{(1)}_1 = \lambda^{(0)}_1 - i \epsilon_1$ \\
$\lambda^{(2)}_1 = \lambda^{(0)}_1 - 2 i \epsilon_1$ \;, &
$\lambda^{(3)}_1 = \lambda^{(0)}_1 - i \epsilon_2$ \\
\end{tabular} 
\\ 
\begin{ytableau} \scriptstyle 2 \\ \scriptstyle 3 \\ \scriptstyle 0 & \scriptstyle 1 \end{ytableau} & $\Longleftrightarrow$ & 
\Bigg\{
\begin{tabular}{ll}
$\lambda^{(0)}_1 = -i a_1^{(0)} - i \frac{\epsilon}{2}$ \;, &
$\lambda^{(1)}_1 = \lambda^{(0)}_1 - i \epsilon_1$  \\
$\lambda^{(3)}_1 = \lambda^{(0)}_1 - i \epsilon_2$ \;, &
$\lambda^{(2)}_1 = \lambda^{(0)}_1 - 2 i \epsilon_2$ \\
\end{tabular} 
\\ 
\begin{ytableau} \scriptstyle 1 \\ \scriptstyle 2 \\ \scriptstyle 3 \\ \scriptstyle 0 \end{ytableau} & $\Longleftrightarrow$ &
\Bigg\{
\begin{tabular}{ll} 
$\lambda^{(0)}_1 = -i a_1^{(0)} - i \frac{\epsilon}{2}$ \;, &
$\lambda^{(3)}_1 = \lambda^{(0)}_1 - i \epsilon_2$  \\
$\lambda^{(2)}_1 = \lambda^{(0)}_1 - 2 i \epsilon_2$ \;, &
$\lambda^{(1)}_1 = \lambda^{(0)}_1 - 3 i \epsilon_2$ \\ 
\end{tabular} 
\end{tabular}  
\end{center}
\vspace*{0.2 cm}

As usual by now, the mirror map is inverted by $Z_{\text{v}} \rightarrow (1+z_0z_1z_2 z_3)^{i r \epsilon} Z_{\text{v}}$ and $Z_{\text{av}} \rightarrow (1+\overline{z}_0\overline{z}_1\overline{z}_2 \overline{z}_3)^{i r \epsilon} Z_{\text{av}}$, while we normalize the 1-loop part with 
\begin{equation}
Z_{\text{1l}} \rightarrow (z_0z_1z_2z_3\overline{z}_0\overline{z}_1\overline{z}_2\overline{z}_3)^{-i r a_1^{(0)} - i r \frac{\epsilon}{2}} \frac{\Gamma (1 - i r \epsilon)}{\Gamma (1 + i r \epsilon)} Z_{\text{1l}}
\end{equation}
At the end we get
\begin{equation}
\begin{split}
Z_{1,1,4}^{S^2,\text{norm}} &= -\dfrac{1}{4 \epsilon_1 \epsilon_2} -  \dfrac{1}{8} \Bigg( 3 \ln^2(z_1 \overline{z}_1) + 4 \ln^2(z_2 \overline{z}_2) + 3 \ln^2(z_3 \overline{z}_3) \\
& + 4 \ln(z_1 \overline{z}_1)\ln(z_2 \overline{z}_2) + 2 \ln(z_1 \overline{z}_1)\ln(z_3 \overline{z}_3) + 4 \ln(z_2 \overline{z}_2)\ln(z_3 \overline{z}_3) \Bigg) \\ 
& + i \Bigg( -\dfrac{\epsilon_1 + 3 \epsilon_2}{8}\ln^3(z_1 \overline{z}_1) - \dfrac{\epsilon_1 + 3 \epsilon_2}{4}\ln^2(z_1 \overline{z}_1)\ln(z_2 \overline{z}_2)  -\dfrac{\epsilon_1 + 3 \epsilon_2}{8}\ln^2(z_1 \overline{z}_1)\ln(z_3 \overline{z}_3) \\
& \;\;\;\;\;\; - \dfrac{\epsilon_1 + \epsilon_2}{3}\ln^3(z_2 \overline{z}_2) - \dfrac{\epsilon_1 + \epsilon_2}{2} \ln(z_1 \overline{z}_1) \ln^2(z_2 \overline{z}_2)  - \dfrac{\epsilon_1 + \epsilon_2}{2}\ln(z_1 \overline{z}_1)\ln(z_2 \overline{z}_2)\ln(z_3 \overline{z}_3) \\
& \;\;\;\;\;\; - \dfrac{\epsilon_1 + \epsilon_2}{2}\ln^2(z_2 \overline{z}_2)\ln(z_3 \overline{z}_3) - \dfrac{3\epsilon_1 + \epsilon_2}{8}\ln^3(z_3 \overline{z}_3) - \dfrac{3\epsilon_1 + \epsilon_2}{8}\ln(z_1 \overline{z}_1)\ln^2(z_3 \overline{z}_3) \\
& \;\;\;\;\;\; - \dfrac{3\epsilon_1 + \epsilon_2}{4}\ln(z_2 \overline{z}_2)\ln^2(z_3 \overline{z}_3)  \Bigg) \\
&+ i \epsilon \Bigg(8 \zeta(3) + 2 (\text{Li}_3(z_1) + \text{Li}_3(z_2) + \text{Li}_3(z_3)  + \text{Li}_3(z_1 z_2) + \text{Li}_3(z_2 z_3) + \text{Li}_3(z_1 z_2 z_3)) \\
&\;\;\;\;\;\;\;\;\;\; + 2( \text{Li}_3(\overline{z}_1) + \text{Li}_3(\overline{z}_2) + \text{Li}_3(\overline{z}_3) + \text{Li}_3(\overline{z}_1 \overline{z}_2)) + \text{Li}_3(\overline{z}_2 \overline{z}_3)) + \text{Li}_3(\overline{z}_1 \overline{z}_2 \overline{z}_3)) \\
&\;\;\;\;\;\;\;\;\;\; - \ln(z_1 \overline{z}_1)(\text{Li}_2(z_1) + \text{Li}_2(\overline{z}_1)) - \ln(z_2 \overline{z}_2)(\text{Li}_2(z_2) + \text{Li}_2(\overline{z}_2)) \\
&\;\;\;\;\;\;\;\;\;\;  - \ln(z_3 \overline{z}_3)(\text{Li}_2(z_3) + \text{Li}_2(\overline{z}_3)) - \ln(z_1 z_2 \overline{z}_1 \overline{z}_2)(\text{Li}_2(z_1 z_2) + \text{Li}_2(\overline{z}_1 \overline{z}_2)) 
\\
&\;\;\;\;\;\;\;\;\;\; - \ln(z_2 z_3 \overline{z}_2 \overline{z}_3)(\text{Li}_2(z_2 z_3) + \text{Li}_2(\overline{z}_2 \overline{z}_3)) - \ln(z_1 z_2 z_3 \overline{z}_1 \overline{z}_2 \overline{z}_3)(\text{Li}_2(z_1 z_2 z_3) + \text{Li}_2(\overline{z}_1 \overline{z}_2 \overline{z}_3)) \Bigg) 
\end{split}
\end{equation}
which corresponds to a prepotential
\begin{equation}
\begin{split}
F_{1,1,4} &= \dfrac{1}{4 \, \epsilon_1\epsilon_2} - \dfrac{1}{8} \left( 3 \ln^2 z_1 + 4 \ln^2 z_2 + 3 \ln^2 z_3 + 4 \ln z_1\ln z_2 + 2 \ln z_1\ln z_3 + 4  \ln z_2\ln z_3 \right) \\
& + \Bigg( \dfrac{\epsilon_1 + 3 \epsilon_2}{8}\ln^3 z_1 + \dfrac{\epsilon_1 + 3 \epsilon_2}{4}\ln^2 z_1 \ln z_2 + \dfrac{\epsilon_1 + 3 \epsilon_2}{8}\ln^2 z_1 \ln z_3 + \dfrac{\epsilon_1 + \epsilon_2}{3}\ln^3 z_2 \\
&\;\;\;\;\;\; + \dfrac{\epsilon_1 + \epsilon_2}{2} \ln z_1 \ln^2 z_2 + \dfrac{\epsilon_1 + \epsilon_2}{2}\ln z_1 \ln z_2 \ln z_3 + \dfrac{\epsilon_1 + \epsilon_2}{2}\ln^2 z_2 \ln z_3 \\
&\;\;\;\;\;\;\; + \dfrac{3\epsilon_1 + \epsilon_2}{8}\ln^3 z_3 + \dfrac{3\epsilon_1 + \epsilon_2}{8}\ln z_1 \ln^2 z_3 + \dfrac{3\epsilon_1 + \epsilon_2}{4}\ln z_2 \ln^2 z_3 \Bigg) \\
& + \epsilon \,(\text{Li}_3(z_1) + \text{Li}_3(z_2) + \text{Li}_3(z_3)  + \text{Li}_3(z_1 z_2) + \text{Li}_3(z_2 z_3) + \text{Li}_3(z_1 z_2 z_3))
\end{split}
\end{equation} 
This prepotential reduces to \eqref{temp} for $\epsilon_1 = \epsilon_2 = \tilde{\epsilon}$.

\end{itemize}


\subsection{The $A_{p-1}$-type ALE space: quantum hydrodynamics} \label{sub5.3}

Let us now consider the mirror LG theory in the Coulomb branch, again following the procedure described in subsection \ref{sub2.4}. We start by defining $\Sigma_s^{(b)} = \sigma_s^{(b)} - i \frac{m_s^{(b)}}{2 r}$; we can then take the large radius limit $r \rightarrow \infty$ of \eqref{Z} to arrive at
\begin{equation}
Z_{\vec{k},\vec{N},p}^{S^2} = \prod_{b=0}^{p-1}  \dfrac{(r \epsilon)^{k_b}}{k_b!} \Bigg\vert \int \prod_{b=0}^{p-1} \prod_{s=1}^{k_b} \dfrac{d (r \Sigma^{(b)}_s)}{2 \pi}\left( \prod_{b=0}^{p-1} \prod_{s=1}^{k_b} \dfrac{\prod_{t\neq s}^{k_b} D(\Sigma_{s}^{(b)} - \Sigma_{t}^{(b)})}{Q(\Sigma_{s}^{(b)})\prod_{t=1}^{k_{b-1}} F(\Sigma_{s}^{(b)} - \Sigma_t^{(b-1)})} \right)^{\frac{1}{2}} e^{-\mathcal{W}_{\text{eff}}(\Sigma)} \Bigg\vert^2 \label{lastZ}
\end{equation}
where the functions entering the integration measure are
\begin{equation}
\begin{split} 
D(\Sigma_{s}^{(b)} - \Sigma_{t}^{(b)}) &\;=\; r^2(\Sigma_{s}^{(b)} - \Sigma_{t}^{(b)})(\Sigma_{s}^{(b)} - \Sigma_{t}^{(b)} + \epsilon) \\
F(\Sigma_{s}^{(b)} - \Sigma_{t}^{(b-1)}) &\;=\; r^2(\Sigma_{s}^{(b)} - \Sigma_{t}^{(b-1)} - \epsilon_1)(\Sigma_{s}^{(b)} - \Sigma_{t}^{(b-1)} + \epsilon_2)\\
Q(\Sigma_{s}^{(b)}) &\;=\; \prod_{j=1}^{N_b} r^2 \left(\Sigma_s^{(b)} - a_j^{(b)} - \dfrac{\epsilon}{2} \right) \left(\Sigma_s^{(b)} - a_j^{(b)} + \dfrac{\epsilon}{2}\right)
\end{split}
\end{equation}
while the twisted effective superpotential reads
\begin{equation}
\begin{split} 
\mathcal{W}_{\text{eff}}(\Sigma) &\;=\; 2 \pi \sum_{b=0}^{p-1} \sum_{s=1}^{k_b} i r t_b \Sigma_s^{(b)} + \sum_{b=0}^{p-1} \sum_{s=1}^{k_b} \sum_{j=1}^{N_b} \left[ \omega(i r \Sigma_s^{(b)} - i r a_j^{(b)} - i r \dfrac{\epsilon}{2}) + \omega(- i r \Sigma_s^{(b)} + i r a_j^{(b)} - i r \dfrac{\epsilon}{2}) \right] \\
&\;+\; \sum_{b=0}^{p-1} \sum_{s,t \neq s}^{k_b} \left[ \omega(i r \Sigma_s^{(b)} - i r \Sigma_t^{(b)}) + \omega(i r \Sigma_s^{(b)} - i r \Sigma_t^{(b)} + i r \epsilon) \right] \\
&\;+\; \sum_{b=0}^{p-1} \sum_{s=1}^{k_b} \sum_{t=1}^{k_{b-1}} \left[ \omega( i r \Sigma_s^{(b)} - i r \Sigma_t^{(b-1)} - i r \epsilon_1) + \omega( - i r \Sigma_s^{(b)} + i r \Sigma_t^{(b-1)} - i r \epsilon_2) \right]
\end{split}
\end{equation}
From the Bethe/Gauge correspondence, the equations determining the supersymmetric vacua in the Coulomb branch (saddle points of $\mathcal{W}_{\text{eff}}$)
\begin{equation}
\text{exp}\left( \dfrac{\partial \mathcal{W}_{\text{eff}}}{\partial \Sigma_s^{(b)}} \right) = 1
\end{equation}
correspond to Bethe Ansatz Equations for a quantum integrable system. For our theory, the equations are
\begin{equation}
\prod_{j=1}^{N_b} \dfrac{\Sigma_s^{(b)} - a_j^{(b)} - \frac{\epsilon}{2}}{-\Sigma_s^{(b)} + a_j^{(b)} - \frac{\epsilon}{2}} \prod_{\substack{t=1 \\ t \neq s}}^{k_b} \dfrac{\Sigma_s^{(b)} - \Sigma_t^{(b)} + \epsilon}{\Sigma_s^{(b)} - \Sigma_t^{(b)} - \epsilon}
\prod_{t=1}^{k_{b-1}} \dfrac{\Sigma_s^{(b)} - \Sigma_t^{(b-1)} - \epsilon_1}{\Sigma_s^{(b)} - \Sigma_t^{(b-1)} + \epsilon_2} 
\prod_{t=1}^{k_{b+1}} \dfrac{\Sigma_s^{(b)} - \Sigma_t^{(b+1)} - \epsilon_2}{\Sigma_s^{(b)} - \Sigma_t^{(b+1)} + \epsilon_1}
= e^{-2 \pi t_b} \label{baeporiginal}
\end{equation}
These are exactly the Bethe Ansatz Equations for the generalization of the periodic Intermediate Long Wave quantum system ILW${}_{\vec{N},p}$ proposed in \cite{2014arXiv1411.3313A}. 
These equations can be rewritten in a form which generalizes to any quiver 
\begin{equation}
\prod_{j=1}^{N_b} \dfrac{\Sigma_s^{(b)} - a_j^{(b)} - \frac{\epsilon}{2}}{-\Sigma_s^{(b)} + a_j^{(b)} - \frac{\epsilon}{2}} \prod_{c = 0}^{p-1} \prod_{\substack{t = 1 \\ (c,t) \neq (b,s)}}^{k_c}
\dfrac{\Sigma_s^{(b)} - \Sigma_t^{(c)} + \mathbf{C}_{bc}^T}{\Sigma_s^{(b)} - \Sigma_t^{(c)} - \mathbf{C}_{bc}} 
= e^{-2 \pi t_b} \label{baep}
\end{equation}
where
\begin{equation}
\mathbf{C}_{bc}=
\begin{bmatrix}
    \epsilon & -\epsilon_1 & 0 & \dots & 0 & -\epsilon_2 \\
    -\epsilon_2 & \epsilon & -\epsilon_1 & \dots & 0 & 0 \\
    0 & -\epsilon_2 & \epsilon & \ddots & \vdots & \vdots \\
    \vdots & \vdots & \ddots & \ddots & -\epsilon_1 & 0 \\
    0 & 0 & \vdots & -\epsilon_2 & \epsilon & -\epsilon_1 \\
    -\epsilon_1 & 0 & \dots & 0 & -\epsilon_2 & \epsilon
\end{bmatrix} \label{matrix}
\end{equation}
is the adjacency matrix of the quiver graph. 

The solutions to \eqref{baeporiginal} are in one to one correspondence with the supersymmetric vacua in the Coulomb branch and with the eigenstates of the infinite set of integrals of motion for the generalized ILW${}_{\vec{N},p}$ system. In general, these equations are extremely hard to solve. However, significant simplification appears in the BO${}_{\vec{N},p}$ limit $(q_0,\ldots,q_{p-1})=(0,\ldots,0)$. In this case solutions can be actually expressed explicitly in terms of $N$-tuples of colored Young diagrams ($N = \sum_{b=0}^{p-1} N_b$) whose boxes are associated to one out of the $p$ colors, the total number of boxes being $k = \sum_{b=0}^{p-1} k_b$. They coincide with the fixed points on the moduli space $\mathcal{M}_{\vec{k}, \vec{N}, p}$: see the examples in the previous subsection and Appendix \ref{appA}.

We already discussed how BO${}_{N}$ is related to $N$ copies of trigonometric Calogero-Sutherland; in the present case, the analogous proposal put forward in \cite{2014arXiv1411.3313A} is that the BO${}_{\vec{N},p}$ system can be viewed as 
a coupled system of $N$ copies of spin $p$ trigonometric Calogero-Sutherland model (which we will call $\textrm{sCS}(N,p)$), where $k_b$ is the number of particles of spin $b = 0, \ldots, p-1$.
In particular the integral of motion $\hat{I}_2$ for BO${}_{1,p}$ coincides with the Hamiltonian of trigonometric $\textrm{sCS}(1,p)$, and its eigenvalues can be written using the roots of the Bethe equations \eqref{baeporiginal} as
\begin{equation}
\label{eq:IMspectrum}
\hat{I}_2 \propto \frac{1}{p}\sum_{s=1}^{k_0}\Sigma_s^{(0)}. 
\end{equation}
In other words the sum runs just over Bethe roots corresponding to the affine node of the quiver. 
It is then natural to expect ILW${}_{\vec{N},p}$ to be related to $N$ copies of elliptic spin $p$ Calogero-Sutherland.
 
In the following we will show the correspondence between BO${}_{\vec{N},p}$ and trigonometric $\textrm{sCS}(N,p)$ for the special case $p=2$ in full detail. Arbitrary $p$ case presents more difficulties, which will be explained in the course of upcoming discussion. The plan is to compare \eqref{eq:IMspectrum} with $N=1$  to the eigenvalues of the $\textrm{sCS}(1,p)$ Hamiltonian. The spectrum of $\textrm{sCS}(1,p)$ was computed in \cite{Uglov:1997ia}. For convenience we quote the results that will be needed in the following. The normalized Hamiltonian is
\begin{equation}
H_{\beta,p}=W^{-\beta}\thickbar{H}_{\beta,p}W^{\beta}, 
\end{equation}
where $W=\prod_{i<j}^k\sin\frac{\pi}{L}(y_i-y_j)$ and 
\begin{equation}
\thickbar{H}_{\beta,p}=-\frac{1}{2}\sum_{i=1}^k\frac{\partial^2}{\partial y_i^2}+\frac{\pi^2}{2L^2}\sum_{i\neq j}^k\frac{\beta\left(\beta+\mathbf{P}_{ij}\right)}{\sin^2\frac{\pi}{L}\left(y_i-y_j\right)}.
\end{equation}
In the above formulae $y_i$ are the coordinates of $k$ particles placed on a circle of length $L$, $s_i$ are their spins, $\mathbf{P}_{ij}$ permute $s_i$ and $s_j$ and form the $SU(p)$ spin representation of the permutation group $S_k$, while $\beta$ is the coupling constant. The spectrum of $H_{\beta,p}$ reads
\begin{equation}
E_{\beta,p}=\sum_{i=1}^k\bar{K}_i^2+\beta\sum_{i=1}^k\left(2i-k-1\right)\bar{K}_i+\frac{\beta^2k\left(k^2-1\right)}{12},
\end{equation}
where the following notation is used.
Given a strictly decreasing integer sequence $\mathbf{K}=(K_1,\ldots,K_k)$, $K_i>K_{i+1}$, it uniquely decomposes as $\mathbf{K}=\ubar{\mathbf{K}}-p\bar{\mathbf{K}}$, where $\ubar{\mathbf{K}}\in\{1,\ldots,p\}^k$ and $\bar{\mathbf{K}}\in\mathbb{Z}^k$. In other terms
\begin{align}
\ubar{K}_i&=1+\left(K_i-1\right)_{\textrm{mod}\; p} \\
\label{eq:Kbar}
\bar{K}_i&=-\left\lfloor\frac{K_i-1}{p}\right\rfloor.
\end{align}
The spectrum arising from the BAE \eqref{baep} should be compared with the excited energy levels of sCS. It is then crucial to 
find a vacuum state $\mathbf{K}^0$, whose energy should be subtracted. The vacuum state was explicitly given for $p=2$ in \cite{Uglov:1997ia} and it reads
\begin{equation}
\label{eq:vac}
\mathbf{K}^0=\left(M,M-1,\ldots,M-k+1\right),\quad M=\frac{k}{2}+1. 
\end{equation}
By the integrality requirement, this makes sense only for $k$ even. 
Moreover the solution is unique only for $k=4l+2$ while for $k=4l$ it can 
be chosen consistently in this form. 
For $k$ odd the vacuum state is never unique, 
nevertheless by practicing with examples we collected evidence that there is always a choice supporting the results derived below.
Once we have the vacuum, we define $\mathbf{K}=\mathbf{\sigma}+\mathbf{K}^0$. From the definitions given above it follows that $\mathbf{\sigma}$ is a non-increasing sequence. By restricting $\mathbf{\sigma}$ to $\mathbb{Z}^k_{\geq0}$ we obtain a partition $\lambda$. In the rest we are going to focus only on states which are labelled by partitions. The coloring of the partition ($0$-coloring when the box in the first row and first column is colored by $0$ and $1$-coloring when it is colored by $1$) is dependent on $k$. For $k=4l+1$ and $k=4l+2$ we have to apply $0$-coloring while $k=4l$ and $k=4l+3$ requires $1$-coloring. In the following we focus on $k=4l+2$, where we have a unique vacuum and a $0$-coloring. However, the conclusions remain valid for $k$ general, one just needs to do appropriate changes in the derivation. We will study the normalized energy eigenvalue for states corresponding to partitions
\begin{equation}
\label{eq:normenergy} 
\mathcal{E}_{\beta,p}(\lambda)=E_{\beta,p}(\mathbf{K})-E_{\beta,p}(\mathbf{K}^0)=\sum_{i=1}^k\left(\bar{K}_i-\bar{K}_i^0\right)\left(\bar{K}_i+\bar{K}_i^0\right)+\beta\sum_{i=1}^k\left(2i-k-1\right)\left(\bar{K}_i-\bar{K}_i^0\right)
\end{equation}
and show that it can be matched with the spectrum of $\hat{I}_2$. Possible partitions for $N=1$, $p=2$ are
\begin{itemize}
\item $k/p=1$:
$$\ytableausetup{mathmode, boxsize=0.7em, aligntableaux=center} 
\begin{ytableau} \scriptstyle 0 & \scriptstyle 1 \end{ytableau} \;,\;
\begin{ytableau} \scriptstyle 1 \\ \scriptstyle 0 \end{ytableau}$$ 
\item $k/p=2$:
$$\ytableausetup{mathmode, boxsize=0.7em, aligntableaux=center} 
\begin{ytableau} \scriptstyle 0 & \scriptstyle 1 & \scriptstyle 0 & \scriptstyle 1 \end{ytableau} \;,\;
\begin{ytableau} \scriptstyle 1 \\ \scriptstyle 0 & \scriptstyle 1 & \scriptstyle 0 \end{ytableau} \;,\;
\begin{ytableau} \scriptstyle 1 & \scriptstyle 0 \\ \scriptstyle 0 & \scriptstyle 1 \end{ytableau} \;,\;
\begin{ytableau} \scriptstyle 0 \\ \scriptstyle 1 \\ \scriptstyle 0 & \scriptstyle 1 \end{ytableau} \;,\;
\begin{ytableau} \scriptstyle 1 \\ \scriptstyle 0 \\ \scriptstyle 1 \\ \scriptstyle 0 \end{ytableau} $$
\item $k/p=3$:
$$\ytableausetup{mathmode, boxsize=0.7em, aligntableaux=center} 
\begin{ytableau} \scriptstyle 0 & \scriptstyle 1 & \scriptstyle 0 & \scriptstyle 1 & \scriptstyle 0 & \scriptstyle 1 \end{ytableau} \;,\;
\begin{ytableau} \scriptstyle 1 \\ \scriptstyle 0 & \scriptstyle 1 & \scriptstyle 0 & \scriptstyle 1 & \scriptstyle 0 \end{ytableau} \;,\;
\begin{ytableau} \scriptstyle 0 \\ \scriptstyle 1 \\ \scriptstyle 0 & \scriptstyle 1 & \scriptstyle 0 & \scriptstyle 1  \end{ytableau} \;,\;
\begin{ytableau} \scriptstyle 1 & \scriptstyle 0 \\ \scriptstyle 0 & \scriptstyle 1 & \scriptstyle 0 & \scriptstyle 1  \end{ytableau} \;,\;
\begin{ytableau} \scriptstyle 1 & \scriptstyle 0 & \scriptstyle 1  \\ \scriptstyle 0 & \scriptstyle 1 & \scriptstyle 0 \end{ytableau} \;,\;
\begin{ytableau} \scriptstyle 0 & \scriptstyle 1 \\ \scriptstyle 1 & \scriptstyle 0 \\ \scriptstyle 0 & \scriptstyle 1  \end{ytableau} \;,\;
\begin{ytableau} \scriptstyle 1 \\ \scriptstyle 0 \\ \scriptstyle 1 & \scriptstyle 0 \\ \scriptstyle 0 & \scriptstyle 1  \end{ytableau} \;,\;
\begin{ytableau} \scriptstyle 1 \\ \scriptstyle 0 \\ \scriptstyle 1 \\ \scriptstyle 0 & \scriptstyle 1 & \scriptstyle 0  \end{ytableau} \;,\;
\begin{ytableau} \scriptstyle 0 \\ \scriptstyle 1 \\ \scriptstyle 0 \\ \scriptstyle 1 \\ \scriptstyle 0 & \scriptstyle 1  \end{ytableau} \;,\;
\begin{ytableau} \scriptstyle 1 \\ \scriptstyle 0 \\ \scriptstyle 1 \\ \scriptstyle 0 \\ \scriptstyle 1 \\ \scriptstyle 0  \end{ytableau} \;,\;$$
\end{itemize}
At this point we need to introduce some notation about colored Young diagrams. The number of boxes colored by $0$ in the $i$-th row is denoted as $C_i^{(0)}(\lambda)$. Drawing a colored diagram and looking at it for sufficient time,
we can write a formula
\begin{equation}
C_i^{(0)}(\lambda)=1+\left\lfloor\frac{\lambda_i-1-\left(i-1\right)_{\textrm{mod}\;p}}{p}\right\rfloor. 
\end{equation}
On the other hand, using \eqref{eq:Kbar}, we have an expression for $\bar{K}_i-\bar{K}_i^{(0)}$
\begin{equation}
\bar{K}_i-\bar{K}_i^{(0)}=-\left\lfloor\frac{\lambda_i+K_i^{(0)}-1}{p}\right\rfloor+\left\lfloor\frac{K_i^{(0)}-1}{p}\right\rfloor
\end{equation}
and plugging in \eqref{eq:vac} while setting $p=2$ at the same time yields a simple relation
\begin{equation}
\label{eq:keyrelation}
\bar{K}_i-\bar{K}_i^{(0)}=-C_i^{(0)}(\lambda).
\end{equation}
Still, we need to build three more quantities out of $C_i^{(0)}(\lambda)$
\begin{align}
\label{eq:numzeroboxes}
\lvert C^{(0)}(\lambda)\rvert&=\sum_{i=1}^{\# \textrm{rows}(\lambda)}C_i^{(0)}(\lambda) \\
n^{(0)}(\lambda)&=\sum_{i=1}^{\# \textrm{rows}(\lambda)}\left(i-1\right)C_i^{(0)}(\lambda) \\
\label{eq:n0transp}
n^{(0)}(\lambda^{t})&=\sum_{i=1}^{\# \textrm{rows}(\lambda^{t})}\left(i-1\right)C_i^{(0)}(\lambda^{t}),
\end{align}
where $\lambda^t$ is the transposed Young diagram. It will be useful to have a formula for $n^{(0)}(\lambda^{t})$ just in terms of data related to the original partition $\lambda$
\begin{align}
\label{eq:n0transpnew}
 n^{(0)}(\lambda^{t})&=\sum_{i=1}^{\# \textrm{rows}(\lambda)}\sum_{j=1}^{C_i^{(0)}(\lambda)}\left[\left(i-1\right)_{\textrm{mod}\;p}+\left(j-1\right)p\right] \notag \\
&=\sum_{i=1}^{\# \textrm{rows}(\lambda)} C_i^{(0)}(\lambda)\left[\left(i-1\right)_{\textrm{mod}\;p}+\frac{p}{2}\left(C_i^{(0)}(\lambda)-1\right)\right]. 
\end{align}
Equipped with these information we can rewrite the normalized energy eigenvalue \eqref{eq:normenergy} just using the data of  colored Young diagrams. The essential ingredient is equation \eqref{eq:keyrelation} which implies $p=2$. After some algebra, combining \eqref{eq:keyrelation}--\eqref{eq:n0transpnew}, we finally arrive at\footnote{This formula appears in \cite{Uglov:1997ia}, but there are typos present.}
\begin{equation}
\label{eq:energyfinal}
\mathcal{E}_{\beta.p=2}(\lambda)=n^{(0)}(\lambda^{t})-\left(2\beta+1\right)n^{(0)}(\lambda)+\left[\frac{k}{2}\left(2\beta+1\right)-\beta\right]\lvert C^{(0)}(\lambda)\rvert.
\end{equation}
To accomplish the comparison we just have to write the spectrum of $\hat{I}_2$ \eqref{eq:IMspectrum} in terms of \eqref{eq:numzeroboxes}--\eqref{eq:n0transp}. Remind that all the above discussion assumes $N=1$, so only the affine node in the quiver contains a single fundamental/antifundamental pair. We mark this node by a star. Then we have (we freely change between the gauge theory notation and CFT notation: $Q\leftrightarrow\epsilon$, $b\leftrightarrow\epsilon_1$, $b^{-1}\leftrightarrow\epsilon_2$) 
\begin{align*}
&\textrm{contribution from }\frac{\epsilon}{2} + a^{(0)}:\quad \left(\frac{\epsilon}{2} + a^{(0)}\right) \lvert C^{(0)}(\lambda)\rvert \\
&\textrm{contribution from }\epsilon_2:\quad 0\cdot C_1^{(0)}(\lambda)+1\cdot C_2^{(0)}(\lambda)+\cdots+\left(\#\textrm{rows}(\lambda)-1\right)\cdot C_{\#\textrm{rows}(\lambda)}^{(0)}(\lambda) \\
&\textrm{contribution from }\epsilon_1:\quad 0\cdot C_1^{(0)}(\lambda^{t})+1\cdot C_2^{(0)}(\lambda^{t})+\cdots+\left(\#\textrm{rows}(\lambda^t)-1\right)\cdot C_{\#\textrm{rows}(\lambda^t)}^{(0)}(\lambda^{t}) \\
\end{align*}
Consequently, it is straightforward to conclude
\begin{equation}
\label{eq:I2final}
\hat{I}_2 \propto \frac{1}{p}\left[\left(\frac{\epsilon}{2} + a^{(0)}\right) \lvert C^{(0)}(\lambda)\rvert+\epsilon_2 n^{(0)}(\lambda)+\epsilon_1 n^{(0)}(\lambda^{t})\right].
\end{equation}
Note that this equation holds for general $p$. 
Indeed from a preliminary analysis we found that this relation generalizes to arbitrary $p$ by constructing the corresponding vacuum state. However it is not clear to us whether all excited states of sCS(1,p) can be described in terms of coloured Young tableaux.
Here we will set $a^{(0)} = 0$, since a global $U(1)$ factor in the flavour group $G_F$ of our GLSM is actually part of the gauge group. The matching between \eqref{eq:energyfinal} and \eqref{eq:I2final} has to be done modulo overall constants and possible linear combinations with lower rank Hamiltonians (in this case $\hat{I}_1$, whose eigenvalue is given by $\lvert C^{(0)}(\lambda)\rvert$); the identification
\begin{align}
\frac{\epsilon_2}{\epsilon_1}&=-\left(2\beta+1\right) 
\end{align}
does the job, ignoring terms proportional to $\hat{I}_1$ and multiplying by appropriate constants.

We expect the $\hat{I}_2$ spectrum of BO${}_{\vec{N},p}$ to be given by $N$ copies of \eqref{eq:I2final}, with the constraint that the sum of the $\vec{a}^{(b)}$ parameters has to be zero.

For the ILW case the spectrum will be still given by \eqref{eq:IMspectrum}, but the $\Sigma^{(0)}$ variables will depend on the Fayet-Iliopoulos parameters, as for the spinless case. \\

To conclude, let us just write down the formulae for the norm of the ILW${}_{\vec{k}, \vec{N}, p}$ eigenstates which can be obtained from \eqref{lastZ}. We saw that in the BO limit, eigenstates are labelled by colored partitions; we expect this to be true also in the ILW case. By performing a semiclassical analysis of the partition function around a vacuum $\vec{\lambda}_{\text{col}}(\vec{t})$ we obtain
\begin{equation}
Z_{\vec{k},\vec{N},p}^{S^2,\,\vec{\lambda}_{\text{col}}} = \Bigg\vert e^{-\mathcal{W}_{\text{eff,cr}}} \prod_{b=0}^{p-1} (r \epsilon)^{\frac{k_b}{2}} \left( \prod_{b=0}^{p-1} \prod_{s=1}^{k_b} \dfrac{\prod_{t\neq s}^{k_b} D(\Sigma_{s}^{(b)} - \Sigma_{t}^{(b)})}{Q_b(\Sigma_{s}^{(b)})\prod_{t=1}^{k_{b-1}} F(\Sigma_{s}^{(b)} - \Sigma_t^{(b-1)})} \right)^{\frac{1}{2}} \left(\text{Det}\dfrac{\partial^2 \mathcal{W}_{\text{eff}}}{r^2 \partial \Sigma_s^{(a)}\partial \Sigma_t^{(b)}} \right)^{-\frac{1}{2}} \Bigg\vert^2_{\Sigma = \Sigma_{\text{cr}}^{\vec{\lambda}_{\text{col}}}}
\end{equation}
where we chose an ordering for the saddle points in order to eliminate the factorials; here the $\Sigma$'s are the solutions corresponding to the vacuum $\vec{\lambda}_{\text{col}}(\vec{t})$. The expression for the norm of the state $\vert \vec{\lambda}_{\text{col}}(\vec{t}) \rangle$ is then
\begin{equation}
\dfrac{1}{\langle \vec{\lambda}_{\text{col}}(\vec{t}) \vert \vec{\lambda}_{\text{col}}(\vec{t}) \rangle} = \Bigg\vert \prod_{b=0}^{p-1} (r \epsilon)^{\frac{k_b}{2}} \left( \prod_{b=0}^{p-1} \prod_{s=1}^{k_b} \dfrac{\prod_{t\neq s}^{k_b} D(\Sigma_{s}^{(b)} - \Sigma_{t}^{(b)})}{Q_b(\Sigma_{s}^{(b)})\prod_{t=1}^{k_{b-1}} F(\Sigma_{s}^{(b)} - \Sigma_t^{(b-1)})} \right)^{\frac{1}{2}} \left(\text{Det}\dfrac{\partial^2 \mathcal{W}_{\text{eff}}}{r^2 \partial \Sigma_s^{(a)}\partial \Sigma_t^{(b)}} \right)^{-\frac{1}{2}} \Bigg\vert^2_{\Sigma = \Sigma_{\text{cr}}^{\vec{\lambda}_{\text{col}}}}
\end{equation}  


\subsection{The $D_p$-type ALE space: comments} \label{sub5.4}

ALE spaces of type $D_p$ ($p \geqslant 4$) correspond to gauge theories living on the space $\mathbb{C}^2/\Gamma$ with $\Gamma = BD_{4(p-2)}$ binary dihedral group. This discrete group has the presentation
\begin{equation}
\langle g, \tau \; \vert \, g^{2(p-2)} = \tau^4 = 1, \; g^{p-2} = \tau^2, \; \tau g \tau^{-1} = g^{-1} \rangle \label{definition}
\end{equation}
and order $4(p-2)$. A possible realization is given by
\begin{equation}
g = \left( \begin{array}{cc}
\,\alpha\, & \;0\, \\ 
\,0\, & \;\alpha^{-1}\,
\end{array} \right) 
\;\;\;,\;\;\; 
\tau = \left( \begin{array}{cc} \,0\, & \;1\, \\ \,-1\, & \;0\, \end{array} \right)
\end{equation} 
with $\alpha$ a primitive $2(p-2)$-th root of unity. As for the $A_{p-1}$ case, the ADHM-like construction of the instanton moduli space is associated to an affine quiver, which in this case is $\widehat{D}_{p}$. The quiver data are contained in the vectors $\vec{k} = (k^{(O)}, k^{(A)}, k^{(1)}, \ldots, k^{({p-3})}, k^{(B)}, k^{(C)})$ and $\vec{N} = (N^{(O)}, N^{(A)}, N^{(1)}, \ldots, N^{({p-3})}, N^{(B)}, N^{(C)})$, with $k^{(O)}$ affine node. In the following we will only consider the case $N^{(O)} + N^{(A)} + N^{(1)} + \ldots + N^{({p-3})} + N^{(B)} + N^{(C)} = 1$; by \cite{Nakajima} this choice fixes $\vec{N} = (1, 0, \ldots, 0)$ and $\vec{k} = (k, k, 2k, \ldots, 2k, k, k)$. \\

The associated GLSM on $S^2$ for this choice of vectors is a theory with gauge group $G = U(k)^4 \times U(2k)^{p-3}$, flavour group $G_F = U(1)_a \times U(1)^2$ and matter content
\begin{center}
\begin{tabular}{c|c|c|c|c|c}
 & $\chi_b$ & $B_{b,b+1}$ & $B_{b,b-1}$ & $I$ & $J$ \\
\hline gauge $G$ & $Adj^{(b)}$ & $(\mathbf{\overline{k}}^{(b)},\mathbf{k}^{(b+1)})$ & $(\mathbf{\overline{k}}^{(b)},\mathbf{k}^{(b-1)})$ & $\mathbf{k}^{(O)}$ & $\mathbf{\overline{k}}^{(O)}$ \\ 
\hline flavor $G_F$ & $\mathbf{1}_{(-1,-1)}$ & $\mathbf{1}_{(1,0)}$ & $\mathbf{1}_{(0,1)}$ & $\mathbf{\overline{N}}^{(O)}_{(1/2,1/2)}$ & $\mathbf{N}^{(O)}_{(1/2,1/2)}$ \\ 
\hline twisted mass & $\epsilon = \epsilon_1 + \epsilon_2$ & $-\epsilon_1$ & $-\epsilon_2$ & $-a - \frac{\epsilon}{2}$ & $a - \frac{\epsilon}{2}$ \\  
\hline R-charge & 2 & 0 & 0 & 0 & 0 \\ 
\end{tabular} 
\end{center}
Here $b$ is an index running over $O, A, 1, \ldots, p-3, B, C$ and $N^{(O)} = 1$.
The superpotential of the theory is given by
\begin{equation}
\begin{split}
W = & \text{Tr}_O[ \chi_{O}( B_{O,1} B_{1,O} + I J )] + \text{Tr}_A[ \chi_{A}( B_{A,1} B_{1,A} )] \\ 
& + \text{Tr}_1[ \chi_{1}( B_{1,2} B_{2,1} - B_{1,O} B_{O,1} - B_{1,A}B_{A,1} )] \\
& + \sum_{b=2}^{p-4} \text{Tr}_b[ \chi_{b}( B_{b,b+1} B_{b+1,b} - B_{b,b-1}B_{b-1,b} )] \\
& + \text{Tr}_{p-3}[ \chi_{p-3}( - B_{p-3,p-4}B_{p-4,p-3} + B_{p-3,B} B_{B,p-3} + B_{p-3,C} B_{C,p-3} )] \\
& + \text{Tr}_{B}[ \chi_{B}( - B_{B,p-3} B_{p-3,B} )] + \text{Tr}_{C}[ \chi_{C}( - B_{C,p-3} B_{p-3,C} )]
\end{split}
\end{equation}
for $p \geqslant 5$, while in the special case $p = 4$ it reduces to
\begin{equation}
\begin{split}
W = & \text{Tr}_O[ \chi_{O}( B_{O,1} B_{1,O} + I J )] + \text{Tr}_A[ \chi_{A}( B_{A,1} B_{1,A} )] 
+ \text{Tr}_{B}[ \chi_{B}( - B_{B,1} B_{1,B} )] \\
& + \text{Tr}_{C}[ \chi_{C}( - B_{C,1} B_{1,C} )] + \text{Tr}_1[ \chi_{1}( B_{1,B} B_{B,1} + B_{1,C} B_{C,1} - B_{1,O} B_{O,1} - B_{1,A}B_{A,1} )]
\end{split}
\end{equation}
This last case is symmetric under exchange of $A,B,C$, as expected from the associated quiver. With these choices for the superpotential, the moduli space of classical supersymmetric vacua of our GLSM in the Higgs branch coincides with the moduli space of $k$ instantons for a $U(1)$ theory on $\mathbb{C}^2/ BD_{4(p-2)}$. 
 
\begin{figure}[h!]
  \centering
\includegraphics[width=0.7\linewidth]{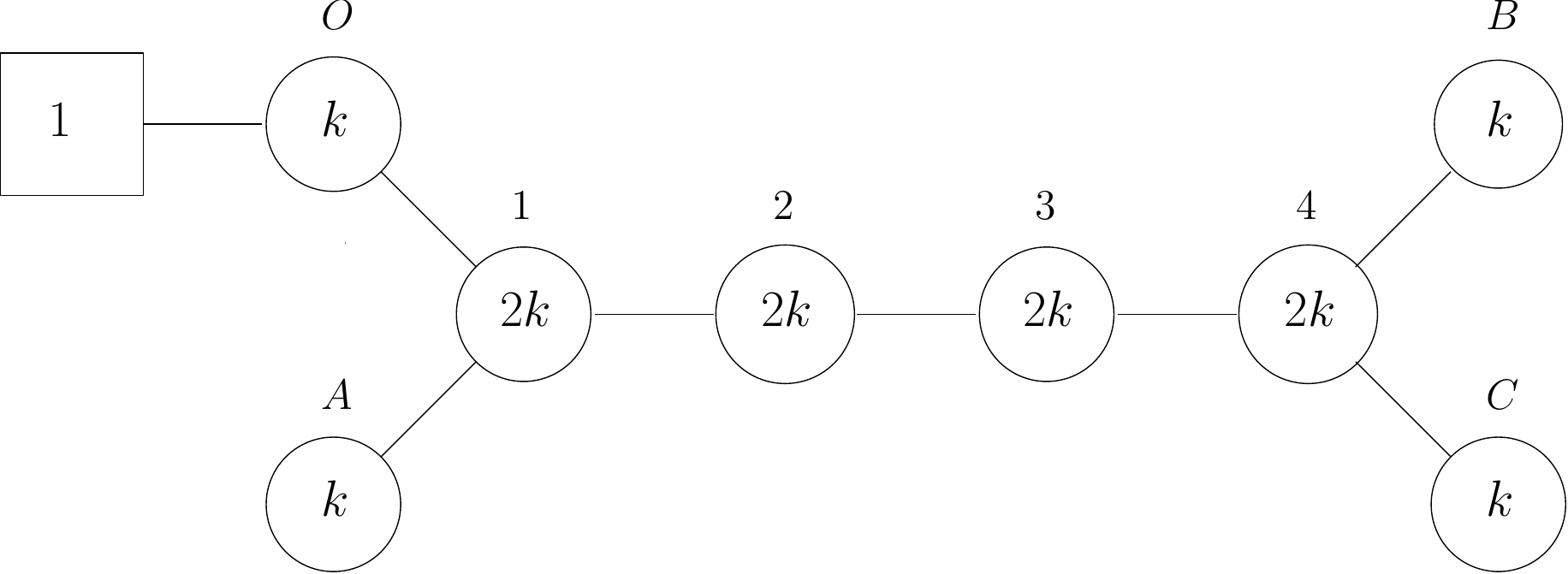}
  \caption{The affine $\widehat{D}_p$ Dynkin diagram, in the case $p = 7$.}
\end{figure}

We can now compute the partition function on $S^2$ for this quiver theory; this will give us information about the quantum cohomology of these ALE spaces. Defining $z = e^{-2 \pi \xi - i \theta}$ as usual, the partition function reads 
\begin{equation}
Z_{k,1,p}^{S^2} = \dfrac{1}{(k!)^4(2k!)^{p-3}} \sum_{\vec{m} \in \mathbb{Z}} \int \prod_{J= O,A,B,C} \prod_{s=1}^{k} \dfrac{d (r \sigma^{(J)}_s)}{2 \pi} \prod_{I=1}^{p-3} \prod_{s=1}^{2k} \dfrac{d (r \sigma^{(I)}_s)}{2 \pi} Z_{\text{cl}} Z_{\text{g,ad}} Z_{\text{f,af}} Z_{\text{bf}} \label{pf}
\end{equation}
where the various pieces in the integrand are given by
\begin{equation}
\begin{split}
Z_{\text{cl}} = & \prod_{I=1}^{p-3} \prod_{s=1}^{2k} z_I^{i r \sigma_s^{(I)} + \frac{m_s^{(I)}}{2}} \overline{z}_I^{i r \sigma_s^{(I)} - \frac{m_s^{(I)}}{2}}
\prod_{J= O,A,B,C} \prod_{s=1}^{k} z_J^{i r \sigma_s^{(J)} + \frac{m_s^{(J)}}{2}} \overline{z}_J^{i r \sigma_s^{(J)} - \frac{m_s^{(J)}}{2}} \\
\end{split}
\end{equation}
\begin{equation}
\begin{split}
Z_{\text{g,ad}} = & \prod_{I=1}^{p-3} \prod_{s<t=1}^{2k} \left( r^2 (\sigma^{(I)}_{s,t})^2 + \dfrac{(m^{(I)}_{s,t})^2}{4} \right) \prod_{J=O,A,B,C} \prod_{s<t=1}^{k} \left( r^2 (\sigma^{(J)}_{s,t})^2 + \dfrac{(m^{(J)}_{s,t})^2}{4} \right) \\
& \prod_{I=1}^{p-3} \prod_{s,t=1}^{2k} \dfrac{\Gamma(1 - i r \sigma_{s,t}^{(I)} - \frac{m_{s,t}^{(I)}}{2} - i r \epsilon)}{\Gamma( i r \sigma_{s,t}^{(I)} - \frac{m_{s,t}^{(I)}}{2} + i r \epsilon)}
\prod_{J=O,A,B,C}  \prod_{s,t=1}^{k} \dfrac{\Gamma(1 - i r \sigma_{s,t}^{(J)} - \frac{m_{s,t}^{(J)}}{2} - i r \epsilon)}{\Gamma( i r \sigma_{s,t}^{(J)} - \frac{m_{s,t}^{(J)}}{2} + i r \epsilon)} \\
\end{split}
\end{equation}
\begin{equation}
\begin{split}
Z_{\text{f,af}} = & \prod_{s=1}^{k} 
\dfrac{\Gamma(- i r \sigma_s^{(O)} - \frac{m_s^{(O)}}{2} + i r a + i r \frac{\epsilon}{2})}{\Gamma(1 + i r \sigma_s^{(O)} - \frac{m_s^{(O)}}{2} - i r a_j - i r \frac{\epsilon}{2})} 
\dfrac{\Gamma(i r \sigma_s^{(O)} + \frac{m_s^{(O)}}{2} - i r a + i r \frac{\epsilon}{2})}{\Gamma(1 - i r \sigma_s^{(O)} + \frac{m_s^{(O)}}{2} + i r a_j - i r \frac{\epsilon}{2})}\\
\end{split}
\end{equation}
\begin{equation}
\begin{split}
Z_{\text{bf}} =& \prod_{I=1}^{p-4} \prod_{s,t=1}^{2k} 
\dfrac{\Gamma(- i r \sigma_{s,t}^{(I+1,I)} - \frac{m_{s,t}^{(I+1,I)}}{2} + i r \epsilon_1)}{\Gamma(1 + i r \sigma_{s,t}^{(I+1,I)} - \frac{m_{s,t}^{(I+1,I)}}{2} - i r \epsilon_1)} 
\dfrac{\Gamma( i r \sigma_{s,t}^{(I+1,I)} + \frac{m_{s,t}^{(I+1,I)}}{2} + i r \epsilon_2)}{\Gamma(1 - i r \sigma_{s,t}^{(I+1,I)} + \frac{m_{s,t}^{(I+1,I)}}{2} - i r \epsilon_2)} \\
& \prod_{J=O,A} \prod_{s=1}^{2k} \prod_{t=1}^{k}
\dfrac{\Gamma(- i r \sigma_{s,t}^{(1,J)} - \frac{m_{s,t}^{(1,J)}}{2} + i r \epsilon_1)}{\Gamma(1 + i r \sigma_{s,t}^{(1,J)} - \frac{m_{s,t}^{(1,J)}}{2} - i r \epsilon_1)} 
\dfrac{\Gamma( i r \sigma_{s,t}^{(1,J)} + \frac{m_{s,t}^{(1,J)}}{2} + i r \epsilon_2)}{\Gamma(1 - i r \sigma_{s,t}^{(1,J)} + \frac{m_{s,t}^{(1,J)}}{2} - i r \epsilon_2)}\\
& \prod_{J=B,C}\prod_{s=1}^{k} \prod_{t=1}^{2k}
\dfrac{\Gamma(- i r \sigma_{s,t}^{(J,p-3)} - \frac{m_{s,t}^{(J,p-3)}}{2} + i r \epsilon_1)}{\Gamma(1 + i r \sigma_{s,t}^{(J,p-3)} - \frac{m_{s,t}^{(J,p-3)}}{2} - i r \epsilon_1)} 
\dfrac{\Gamma( i r \sigma_{s,t}^{(J,p-3)} + \frac{m_{s,t}^{(J,p-3)}}{2} + i r \epsilon_2)}{\Gamma(1 - i r \sigma_{s,t}^{(J,p-3)} + \frac{m_{s,t}^{(J,p-3)}}{2} - i r \epsilon_2)}
\end{split}
\end{equation}
Here we used the compact notation $\sigma_{s,t}^{(I,J)} = \sigma_{s}^{(I)} - \sigma_{t}^{(J)}$ and $\sigma_{s,t}^{(I)} = \sigma_{s}^{(I)} - \sigma_{t}^{(I)}$.

Again, as explained in the previous Sections, the small radius limit $r \rightarrow 0$ produces a contour integral representation for the instanton part of Nekrasov partition function at fixed $k$. In this case, we obtain
\begin{equation}
\begin{split}
Z^{\text{inst}}_{k,1,p} = &  \dfrac{\epsilon^{2k(p-1)}}{(ir)^{2k}} \oint \prod_{J= O,A,B,C} \prod_{s=1}^{k} \dfrac{d  \sigma^{(J)}_s}{2 \pi i} \prod_{I=1}^{p-3} \prod_{s=1}^{2k} \dfrac{d \sigma^{(I)}_s}{2 \pi i} \\
& \prod_{s=1}^{k} \dfrac{1}{(\sigma_s^{(O)} - a - \frac{\epsilon}{2})(-\sigma_s^{(O)} + a - \frac{\epsilon}{2})}  \prod_{I=1}^{p-3} \prod_{\substack{s,t=1 \\ s \neq t}}^{2k} (\sigma_{s,t}^{(I)})(\sigma_{s,t}^{(I)} - \epsilon) \\
& \prod_{J=O,A,B,C} \prod_{\substack{s,t=1 \\ s \neq t}}^{k} (\sigma_{s,t}^{(J)})(\sigma_{s,t}^{(J)} - \epsilon)  \prod_{I=1}^{p-4} \prod_{s,t=1}^{2k} \dfrac{1}{(\sigma_{s,t}^{(I+1,I)} - \epsilon_1)(-\sigma_{s,t}^{(I+1,I)} - \epsilon_2)} \\
& \prod_{J=O,A} \prod_{s=1}^{2k} \prod_{t=1}^{k} \dfrac{1}{(\sigma_{s,t}^{(1,J)} - \epsilon_1)(-\sigma_{s,t}^{(1,J)} - \epsilon_2)} 
\prod_{J=B,C}\prod_{s=1}^{k} \prod_{t=1}^{2k} \dfrac{1}{(\sigma_{s,t}^{(J,p-3)} - \epsilon_1)(-\sigma_{s,t}^{(J,p-3)} - \epsilon_2)}
\end{split}
\end{equation} 
which coincides with the expression of \cite{2000CMaPh.209...97M}. The factorials have been omitted, since they are cancelled by the possible orderings of the integration variables.\\

\noindent \textit{Equivariant quantum cohomology} \\

\noindent For $r$ finite, the partition function computes the equivariant quantum cohomology of the moduli space of $U(1)$ instantons on the $D_p$ ALE space, i.e. of the $BD_{4(p-2)}$-Hilbert scheme of $k$ points. In particular, after factorizing \eqref{pf} as

\begin{equation}
Z_{k,1,p}^{S^2} = \dfrac{1}{(k!)^4(2k!)^{p-3}} \oint \prod_{J=O,A,B,C} \prod_{s=1}^{k} \dfrac{d (r \lambda^{(J)}_s)}{2 \pi i} \prod_{I=1}^{p-3} \prod_{s=1}^{2k} \dfrac{d (r \lambda^{(I)}_s)}{2 \pi i} Z_{\text{1l}} Z_{\text{v}} Z_{\text{av}}   
\end{equation}

\begin{equation}
\begin{split}
Z_{\text{1l}} \,=&\, \left( \dfrac{\Gamma(1- i r \epsilon)}{\Gamma(i r \epsilon)} \right)^{2k(p-1)} 
\prod_{I=1}^{p-3} \prod_{s=1}^{2k} (z_I \bar{z}_I)^{-r \lambda_s^{(I)}} 
\prod_{J=O,A,B,C} \prod_{s=1}^{k} (z_J \bar{z}_J)^{-r \lambda_s^{(J)}} \\
& \prod_{I=1}^{p-3} \prod_{s=1}^{2k} \prod_{t \neq s}^{2k} (r \lambda_{s,t}^{(I)})  \dfrac{\Gamma(1 + r \lambda_{s,t}^{(I)} - i r \epsilon)}{\Gamma(- r \lambda_{s,t}^{(I)} + i r \epsilon)}
\prod_{J=O,A,B,C} \prod_{s=1}^{k} \prod_{t \neq s}^{k} (r \lambda_{s,t}^{(J)})  \dfrac{\Gamma(1 + r \lambda_{s,t}^{(J)} - i r \epsilon)}{\Gamma(- r \lambda_{s,t}^{(J)} + i r \epsilon)}\\
& \prod_{I=1}^{p-4} \prod_{s,t=1}^{2k} \dfrac{\Gamma(r \lambda_{s,t}^{(I+1,I)} + i r \epsilon_1)}{\Gamma(1 - r \lambda_{s,t}^{(I+1,I)} - i r \epsilon_1)} \dfrac{\Gamma(- r \lambda_{s,t}^{(I+1,I)} + i r \epsilon_2)}{\Gamma(1 + r \lambda_{s,t}^{(I+1,I)} - i r \epsilon_2)}\\
& \prod_{J=O,A} \prod_{s=1}^{2k} \prod_{t=1}^{k} \dfrac{\Gamma(r \lambda_{s,t}^{(1,J)} + i r \epsilon_1)}{\Gamma(1 - r \lambda_{s,t}^{(1,J)} - i r \epsilon_1)} \dfrac{\Gamma(- r \lambda_{s,t}^{(1,J)} + i r \epsilon_2)}{\Gamma(1 + r \lambda_{s,t}^{(1,J)} - i r \epsilon_2)}\\
& \prod_{J=B,C} \prod_{s=1}^{k} \prod_{t = 1}^{2k} \dfrac{\Gamma(r \lambda_{s,t}^{(J,p-3)} + i r \epsilon_1)}{\Gamma(1 - r \lambda_{s,t}^{(J,p-3)} - i r \epsilon_1)} \dfrac{\Gamma(- r \lambda_{s,t}^{(J,p-3)} + i r \epsilon_2)}{\Gamma(1 + r \lambda_{s,t}^{(J,p-3)} - i r \epsilon_2)}\\
& \prod_{s=1}^{k} \dfrac{\Gamma(r \lambda_s^{(O)} + i r a + i r \frac{\epsilon}{2})}{\Gamma(1 - r \lambda_s^{(O)} - i r a - i r \frac{\epsilon}{2})} \dfrac{\Gamma(- r \lambda_s^{(O)} - i r a + i r \frac{\epsilon}{2})}{\Gamma(1 + r \lambda_s^{(O)} + i r a - i r \frac{\epsilon}{2})}\\
\end{split}
\end{equation}

\begin{equation}
\begin{split}
Z_{\text{v}} \,=&\, \sum_{\{\vec{l}\} \in \mathbb{N}} \prod_{s=1}^{k} (-1)^{N l_s^{(O)}} \prod_{I=1}^{p-3}\prod_{s=1}^{2k} z_I^{l_s^{(I)}} \prod_{J=O,A,B,C} \prod_{s=1}^{k} z_J^{l_s^{(J)}}\\
&\prod_{I=1}^{p-3} \prod_{s < t}^{2k} \dfrac{l_{t,s}^{(I)} - r \lambda_{t,s}^{(I)} }{-r \lambda_{t,s}^{(I)}} 
\dfrac{(1 + r \lambda_{s,t}^{(I)} - i r \epsilon)_{l_{t,s}^{(I)}}}{(r \lambda_{s,t}^{(I)} + i r \epsilon)_{l_{t,s}^{(I)} }}
\prod_{J=O,A,B,C} \prod_{s < t}^{k} \dfrac{l_{t,s}^{(J)} - r \lambda_{t,s}^{(J)} }{-r \lambda_{t,s}^{(J)}} 
\dfrac{(1 + r \lambda_{s,t}^{(J)} - i r \epsilon)_{l_{t,s}^{(J)}}}{(r \lambda_{s,t}^{(J)} + i r \epsilon)_{l_{t,s}^{(J)} }}\\
& \prod_{I=1}^{p-4} \prod_{s=1}^{2k} \prod_{t = 1}^{2k} \dfrac{1}{(1 - r \lambda_{s,t}^{(I+1,I)} - i r \epsilon_1)_{l_{s,t}^{(I+1,I)}}} \dfrac{1}{(1 + r \lambda_{s,t}^{(I+1,I)} - i r \epsilon_2)_{l_{t,s}^{(I,I+1)} }}\\
& \prod_{J=O,A} \prod_{s=1}^{2k} \prod_{t = 1}^{k} \dfrac{1}{(1 - r \lambda_{s,t}^{(1,J)} - i r \epsilon_1)_{l_{s,t}^{(1,J)}}} \dfrac{1}{(1 + r \lambda_{s,t}^{(1,J)} - i r \epsilon_2)_{l_{t,s}^{(J,1)} }}\\
& \prod_{J=B,C} \prod_{s=1}^{k} \prod_{t = 1}^{2k} \dfrac{1}{(1 - r \lambda_{s,t}^{(J,p-3)} - i r \epsilon_1)_{l_{s,t}^{(J,p-3)}}} \dfrac{1}{(1 + r \lambda_{s,t}^{(J,p-3)} - i r \epsilon_2)_{l_{t,s}^{(p-3,J)} }}\\
& \prod_{s=1}^{k} \dfrac{(- r \lambda_s^{(O)} - i r a + i r \frac{\epsilon}{2})_{l_s^{(O)}}}{(1 - r \lambda_s^{(O)} - i r a - i r \frac{\epsilon}{2})_{l_s^{(O)}}} \\
\end{split}
\end{equation}

\begin{equation}
\begin{split}
Z_{\text{av}} \,=&\, \sum_{\{\vec{k}\} \in \mathbb{N}} \prod_{s=1}^{k} (-1)^{N k_s^{(O)}} \prod_{I=1}^{p-3}\prod_{s=1}^{2k} \bar{z}_I^{k_s^{(I)}} \prod_{J=O,A,B,C} \prod_{s=1}^{k} \bar{z}_J^{k_s^{(J)}}\\
&\prod_{I=1}^{p-3} \prod_{s < t}^{2k} \dfrac{k_{t,s}^{(I)} - r \lambda_{t,s}^{(I)} }{-r \lambda_{t,s}^{(I)}} 
\dfrac{(1 + r \lambda_{s,t}^{(I)} - i r \epsilon)_{k_{t,s}^{(I)}}}{(r \lambda_{s,t}^{(I)} + i r \epsilon)_{k_{t,s}^{(I)} }}
\prod_{J=O,A,B,C} \prod_{s < t}^{k} \dfrac{k_{t,s}^{(J)} - r \lambda_{t,s}^{(J)} }{-r \lambda_{t,s}^{(J)}} 
\dfrac{(1 + r \lambda_{s,t}^{(J)} - i r \epsilon)_{k_{t,s}^{(J)}}}{(r \lambda_{s,t}^{(J)} + i r \epsilon)_{k_{t,s}^{(J)} }}\\
& \prod_{I=1}^{p-4} \prod_{s=1}^{2k} \prod_{t = 1}^{2k} \dfrac{1}{(1 - r \lambda_{s,t}^{(I+1,I)} - i r \epsilon_1)_{k_{s,t}^{(I+1,I)}}} \dfrac{1}{(1 + r \lambda_{s,t}^{(I+1,I)} - i r \epsilon_2)_{k_{t,s}^{(I,I+1)} }}\\
& \prod_{J=O,A} \prod_{s=1}^{2k} \prod_{t = 1}^{k} \dfrac{1}{(1 - r \lambda_{s,t}^{(1,J)} - i r \epsilon_1)_{k_{s,t}^{(1,J)}}} \dfrac{1}{(1 + r \lambda_{s,t}^{(1,J)} - i r \epsilon_2)_{k_{t,s}^{(J,1)} }}\\
& \prod_{J=B,C} \prod_{s=1}^{k} \prod_{t = 1}^{2k} \dfrac{1}{(1 - r \lambda_{s,t}^{(J,p-3)} - i r \epsilon_1)_{k_{s,t}^{(J,p-3)}}} \dfrac{1}{(1 + r \lambda_{s,t}^{(J,p-3)} - i r \epsilon_2)_{k_{t,s}^{(p-3,J)} }}\\
& \prod_{s=1}^{k} \dfrac{(- r \lambda_s^{(O)} - i r a + i r \frac{\epsilon}{2})_{k_s^{(O)}}}{(1 - r \lambda_s^{(O)} - i r a - i r \frac{\epsilon}{2})_{k_s^{(O)}}} \\
\end{split}
\end{equation}
we can identify $Z_{\text{v}}$ with Givental's $\mathcal{I}$-function for our target space. 

Explicit evaluation of the Gromov-Witten prepotential requires the analysis of the pole structure of our partition function; we leave this complicated combinatorial problem for future work. For the case $k=1$, we expect the result to only depend on the $D_p$ algebra data \cite{2007arXiv0707.1337B}, similarly to what we discussed in subsection \ref{sub5.2}. Nevertheless, an analysis of the simplest cases gives $(1 + z_O z_A \prod_{I=1}^{p-3} z_I^2 z_{B}z_{C})^{i r k \epsilon}$ as the equivariant mirror map, again in agreement with \cite{2009JAMS...22.1055M}. We therefore expect also the equivariant mirror map for the $E$-type ALE spaces to depend only on the dual Dynkin label of the affine Dynkin diagram for the corresponding algebra. 

As far as the orbifold phase is concerned, by reversing the sign of all Fayet-Iliopoulos parameters one obtains the same phase due to the symmetry of ADHM constraints; the orbifold phase is then reached by analytic continuation on the product
of circles $|z_b|=1$. This provides conjectural formulae for the equivariant $\mathcal{I}$ and $\mathcal{J}$ functions of the Hilbert scheme of points of $D_p$ singularities that will have to be checked against rigorous mathematical results. Similar conjectures are valid for the $A_{p-1}$ singularities discussed in previous subsections. 

As a final comment, let us remark that in the case of ALE spaces of type $D$ and $E$ only a diagonal combination of $U(1)_{\epsilon_1} \times U(1)_{\epsilon_2}$ is preserved: for the $D$ case, this is due to the action of the generator $\tau$ in \eqref{definition}. 
This corresponds to set $\epsilon_1 = \epsilon_2$ in the Gromov-Witten prepotential if one wants to compute the correct equivariant quantum cohomology of the ALE space. \\

\noindent \textit{Quantum hydrodynamics} \\

\noindent As familiar by now, the mirror LG model in the Coulomb branch can be recovered by taking the large radius limit $r \rightarrow \infty$ of \eqref{pf}. We obtain
\begin{equation}
Z_{k,1,p}^{S^2} = \dfrac{(r \epsilon)^{2k(p-1)}}{(k!)^4(2k!)^{p-3}} \Bigg\vert \int \prod_{J=O,A,B,C} \prod_{s=1}^{k} \dfrac{d (r \Sigma^{(J)}_s)}{2 \pi} \prod_{I=0}^{p-3} \prod_{s=1}^{2k} \dfrac{d (r \Sigma^{(I)}_s)}{2 \pi} 
Z_{\text{meas}} (\Sigma) e^{-\mathcal{W}_{\text{eff}}(\Sigma)} \Bigg\vert^2 
\end{equation}
Here the integration measure is given by
\begin{equation}
\begin{split}
Z_{\text{meas}} (\Sigma) &= \left( \dfrac{\prod_{I=1}^{p-3} \prod_{s,t\neq s}^{2k} D(\Sigma_{s}^{(I)} - \Sigma_{t}^{(I)}) \prod_{J=O,A,B,C} \prod_{s,t\neq s}^{k} D(\Sigma_{s}^{(J)} - \Sigma_{t}^{(J)})}{\prod_{s=1}^{k} Q(\Sigma_{s}^{(O)}) \prod_{I=1}^{p-4} \prod_{s=1}^{2k} \prod_{t=1}^{2k} F(\Sigma_{s}^{(I+1)} - \Sigma_t^{(I)})} \right)^{\frac{1}{2}} \\
& \left( \dfrac{1}{ \prod_{J=O,A} \prod_{s=1}^{2k} \prod_{t=1}^{k} F(\Sigma_{s}^{(1)} - \Sigma_t^{(J)}) \prod_{J=B,C} \prod_{s=1}^{k} \prod_{t=1}^{2k} F(\Sigma_{s}^{(J)} - \Sigma_t^{(p-3)})} \right)^{\frac{1}{2}}
\end{split}
\end{equation}
with
\begin{equation}
\begin{split} 
D(\Sigma_{s}^{(I)} - \Sigma_{t}^{(I)}) &\;=\; r^2(\Sigma_{s}^{(I)} - \Sigma_{t}^{(I)})(\Sigma_{s}^{(I)} - \Sigma_{t}^{(I)} + \epsilon) \\
F(\Sigma_{s}^{(I+1)} - \Sigma_{t}^{(I)}) &\;=\; r^2(\Sigma_{s}^{(I+1)} - \Sigma_{t}^{(I)} - \epsilon_1)(\Sigma_{s}^{(I+1)} - \Sigma_{t}^{(I)} + \epsilon_2)\\
Q(\Sigma_{s}^{(O)}) &\;=\; r^2 \left(\Sigma_s^{(O)} - a^{(O)} - \dfrac{\epsilon}{2} \right) \left(\Sigma_s^{(O)} - a^{(O)} + \dfrac{\epsilon}{2}\right)
\end{split}
\end{equation}
The twisted effective superpotential has the form
\begin{equation}
\begin{split} 
\mathcal{W}_{\text{eff}}(\Sigma) &\;=\; 
2 \pi \sum_{I=1}^{p-3} \sum_{s=1}^{2k} i r t_I \Sigma_s^{(I)} 
+ 2 \pi \sum_{J=O,A,B,C} \sum_{s=1}^{k} i r t_J \Sigma_s^{(J)} \\		
&\;+\; \sum_{s=1}^{k} \left[ \omega(i r \Sigma_s^{(O)} - i r a^{(O)} - i r \dfrac{\epsilon}{2}) + \omega(- i r \Sigma_s^{(O)} + i r a^{(O)} - i r \dfrac{\epsilon}{2}) \right] \\
&\;+\; \sum_{I=1}^{p-3} \sum_{s,t \neq s}^{2k} \left[ \omega(i r \Sigma_s^{(I)} - i r \Sigma_t^{(I)}) + \omega(i r \Sigma_s^{(I)} - i r \Sigma_t^{(I)} + i r \epsilon) \right] \\
&\;+\; \sum_{J=O,A,B,C} \sum_{s,t \neq s}^{k} \left[ \omega(i r \Sigma_s^{(J)} - i r \Sigma_t^{(J)}) + \omega(i r \Sigma_s^{(J)} - i r \Sigma_t^{(J)} + i r \epsilon) \right] \\
&\;+\; \sum_{I=1}^{p-4} \sum_{s=1}^{2k} \sum_{t=1}^{2k} \left[ \omega( i r \Sigma_s^{(I+1)} - i r \Sigma_t^{(I)} - i r \epsilon_1) + \omega( - i r \Sigma_s^{(I+1)} + i r \Sigma_t^{(I)} - i r \epsilon_2) \right] \\
&\;+\; \sum_{J=O,A} \sum_{s=1}^{2k} \sum_{t=1}^{k} \left[ \omega( i r \Sigma_s^{(1)} - i r \Sigma_t^{(J)} - i r \epsilon_1) + \omega( - i r \Sigma_s^{(1)} + i r \Sigma_t^{(J)} - i r \epsilon_2) \right] \\
&\;+\; \sum_{J=B,C} \sum_{s=1}^{k} \sum_{t=1}^{2k} \left[ \omega( i r \Sigma_s^{(J)} - i r \Sigma_t^{(p-3)} - i r \epsilon_1) + \omega( - i r \Sigma_s^{(J)} + i r \Sigma_t^{(p-3)} - i r \epsilon_2) \right] \label{lastTES}
\end{split}
\end{equation}
From \eqref{lastTES} we recover a set of Bethe Ansatz Equations, which can be written as
\begin{equation}
\prod_{j=1}^{N_b} \dfrac{\Sigma_s^{(b)} - a_j^{(b)} - \frac{\epsilon}{2}}{-\Sigma_s^{(b)} + a_j^{(b)} - \frac{\epsilon}{2}} \prod_{c} \prod_{\substack{t = 1 \\ (c,t) \neq (b,s)}}^{k_c}
\dfrac{\Sigma_s^{(b)} - \Sigma_t^{(c)} + \mathbf{C}_{bc}^T}{\Sigma_s^{(b)} - \Sigma_t^{(c)} - \mathbf{C}_{bc}} 
= e^{-2 \pi t_b} \label{baeD}
\end{equation}
Here $c = O, A, 1, \ldots, p-3, B, C$, while $\vec{N} = (1,0,\ldots, 0)$ and $\vec{k} = (k,k,, 2k,\ldots, 2k,k,k)$ as discusses earlier. The matrix 
\begin{equation}
\mathbf{C}_{bc}=
\begin{bmatrix}
\epsilon & 0 & -\epsilon_1 & 0 & 0 & \cdots & 0 \\ 
0 & \epsilon & -\epsilon_1 & 0 & \ddots & \ddots & 0  \\ 
-\epsilon_2 & -\epsilon_2 & \epsilon & -\epsilon_1 & \ddots & \ddots & 0  \\ 
0 & 0 & -\epsilon_2 & \epsilon & -\epsilon_1 & \ddots & 0 \\ 
0 & 0 & \ddots & -\epsilon_2 & \epsilon & -\epsilon_1 & -\epsilon_1 \\
\vdots & \vdots & \ddots & 0 & -\epsilon_2 & \epsilon & 0 \\
0 & 0 & \cdots & 0 & -\epsilon_2 & 0 & \epsilon
\end{bmatrix} \label{matrix2}
\end{equation}
is again the adjacency matrix of the quiver graph, and reduces to the Cartan matrix of the affine $\widehat{D}_{p}$ algebra for $\epsilon_1 = \epsilon_2$. We expect \eqref{baeD} to be related to a quantum hydrodynamical integrable system, a sort of $D_p$-type generalization of ILW. Solutions to \eqref{baeD} will correspond to eigenvalues of the QIS; expressions for the norm of the eigenstates can be obtained by performing a semiclassical approximation of the partition function around the corresponding vacua, as we already discussed in the previous Sections.


\section{Acknowledgments}  

We thank D.E. Diaconescu, N. Nekrasov and V. Roubtsov for discussions.
This research was partly supported by the INFN Research Projects GAST and ST\&FI and by PRIN 
``Geometria delle variet\`a algebriche''.


\appendix

\section{Appendix A} \label{appA}

In this Appendix we will give some more explicit computations of the K\"{a}hler potential for the instanton moduli space $\mathcal{M}_{\vec{k}, \vec{N}, p}$. We will skip all the intermediate computations and provide only the basic ingredients: the relevant poles of the partition function, the equivariant mirror map, the normalization of the 1-loop factor, and the final result. \\

\noindent \textit{The $N=1$, $k=2$ case} \\

\noindent When $N=1$ but $k\geqslant 2$ there no longer is a general expression for the Gromov-Witten prepotential in terms of the Cartan matrix and positive roots of the algebra $A_{p-1}$. We will make good use of our partition function and compute the prepotential in the simplest cases; certainly this procedure can be pursued further, the only difficulty being an integral which becomes more and more complicated. We notice that for $k\geqslant 2$ also $\ln z_0$ enters in the prepotential, thus making impossible a description of the quantum cohomology purely in terms of $A_{p-1}$ algebra data. The results of this case should be compared with \cite{2009JAMS...22.1055M}.
 
\begin{itemize}

\item \textbf{Case $p = 2$} \\

Poles:

\vspace*{0.2 cm}
\begin{center}
\begin{tabular}{ccl}
\ytableausetup{mathmode, boxsize=0.7em, aligntableaux=center} 
\begin{ytableau} \scriptstyle 0 & \scriptstyle 1 & \scriptstyle 0 & \scriptstyle 1 \end{ytableau} & $\Longleftrightarrow$ & 
\Bigg\{
\begin{tabular}{ll}
$\lambda^{(0)}_1 = -i a_1^{(0)} - i \frac{\epsilon}{2}$ \;, &
$\lambda^{(1)}_1 = \lambda^{(0)}_1 - i \epsilon_1$  \\ 
$\lambda^{(0)}_2 = \lambda^{(0)}_1 - 2 i \epsilon_1$ \;, &
$\lambda^{(1)}_2 = \lambda^{(0)}_1 - 3 i \epsilon_1$ \\ 
\end{tabular} 
\\
\begin{ytableau} \scriptstyle 1 \\ \scriptstyle 0 & \scriptstyle 1 & \scriptstyle 0 \end{ytableau} & $\Longleftrightarrow$ & 
\Bigg\{
\begin{tabular}{ll}
$\lambda^{(0)}_1 = -i a_1^{(0)} - i \frac{\epsilon}{2}$ \;, &
$\lambda^{(1)}_1 = \lambda^{(0)}_1 - i \epsilon_1$ \\
$\lambda^{(0)}_2 = \lambda^{(0)}_1 - 2 i \epsilon_1$ \;, &
$\lambda^{(1)}_2 = \lambda^{(0)}_1 - i \epsilon_2$ \\
\end{tabular} 
\\ 
\begin{ytableau} \scriptstyle 1 & \scriptstyle 0 \\ \scriptstyle 0 & \scriptstyle 1 \end{ytableau} & $\Longleftrightarrow$ & 
\Bigg\{
\begin{tabular}{ll}
$\lambda^{(0)}_1 = -i a_1^{(0)} - i \frac{\epsilon}{2}$ \;, &
$\lambda^{(1)}_1 = \lambda^{(0)}_1 - i \epsilon_1$ \\
$\lambda^{(1)}_2 = \lambda^{(0)}_1 - i \epsilon_2$ \;, &
$\lambda^{(0)}_2 = \lambda^{(0)}_1 - i \epsilon_1 - i \epsilon_2$ \\
\end{tabular} 
\\ 
\begin{ytableau} \scriptstyle 0 \\ \scriptstyle 1 \\ \scriptstyle 0 & \scriptstyle 1 \end{ytableau} & $\Longleftrightarrow$ & 
\Bigg\{
\begin{tabular}{ll}
$\lambda^{(0)}_1 = -i a_1^{(0)} - i \frac{\epsilon}{2}$ \;, &
$\lambda^{(1)}_1 = \lambda^{(0)}_1 - i \epsilon_1$  \\
$\lambda^{(1)}_2 = \lambda^{(0)}_1 - i \epsilon_2$ \;, &
$\lambda^{(0)}_2 = \lambda^{(0)}_1 - 2 i \epsilon_2$ \\
\end{tabular} 
\\ 
\begin{ytableau} \scriptstyle 1 \\ \scriptstyle 0 \\ \scriptstyle 1 \\ \scriptstyle 0 \end{ytableau} & $\Longleftrightarrow$ &
\Bigg\{
\begin{tabular}{ll} 
$\lambda^{(0)}_1 = -i a_1^{(0)} - i \frac{\epsilon}{2}$ \;, &
$\lambda^{(1)}_1 = \lambda^{(0)}_1 - i \epsilon_2$  \\
$\lambda^{(0)}_2 = \lambda^{(0)}_1 - 2 i \epsilon_2$ \;, &
$\lambda^{(1)}_2 = \lambda^{(0)}_1 - 3 i \epsilon_2$ \\ 
\end{tabular} 
\end{tabular}  
\end{center}
\vspace*{0.2 cm}

Equivariant mirror map: 
\begin{equation}
Z_{\text{v}} \rightarrow (1+z_0z_1)^{2 i r \epsilon} Z_{\text{v}} \;\;\;,\;\;\; 
Z_{\text{av}} \rightarrow (1+\overline{z}_0\overline{z}_1)^{2 i r \epsilon} Z_{\text{av}}
\end{equation}
Normalization of the 1-loop factor: 
\begin{equation}
Z_{\text{1l}} \rightarrow (z_0z_1\overline{z}_0\overline{z}_1)^{-2 i r a_1^{(0)} - i r \epsilon} \left( \dfrac{\Gamma (1 - i r \epsilon)}{\Gamma (1 + i r \epsilon)} \right)^2 Z_{\text{1l}}
\end{equation}
Partition function:
\begin{equation}
\begin{split}
Z_{2,1,2}^{S^2,\text{norm}} &= \dfrac{1}{8 \epsilon_1^2 \epsilon_2^2} + \dfrac{1}{2 \epsilon_1 \epsilon_2} \Bigg(\dfrac{1}{4}\ln^2(z_0 \overline{z}_0) + \dfrac{1}{2} \ln(z_0 \overline{z}_0)\ln(z_1 \overline{z}_1) + \dfrac{1}{2} \ln^2(z_1 \overline{z}_1) \Bigg) \\ 
&- i \dfrac{\epsilon}{2 \epsilon_1 \epsilon_2} \Bigg( -\dfrac{1}{12}\ln^3(z_0 \overline{z}_0) - \dfrac{1}{4}\ln^2(z_0 \overline{z}_0)\ln(z_1 \overline{z}_1) \\
& - \dfrac{1}{4}\ln(z_0 \overline{z}_0)\ln^2(z_1 \overline{z}_1) - \dfrac{1}{6}\ln^3(z_1 \overline{z}_1) + 7 \zeta(3) \Bigg) \\
&- i \dfrac{\epsilon}{2 \epsilon_1 \epsilon_2} \Bigg( 2 (\text{Li}_3(z_1) + \text{Li}_3(z_0 z_1) + \text{Li}_3(\overline{z}_1) + \text{Li}_3(\overline{z}_0\overline{z}_1))\\
& - \ln(z_1 \overline{z}_1)(\text{Li}_2(z_1) + \text{Li}_2(\overline{z}_1)) - \ln(z_0 z_1 \overline{z}_0 \overline{z}_1)(\text{Li}_2(z_0 z_1) + \text{Li}_2(\overline{z}_0 \overline{z}_1)) \Bigg) \label{Z212}
\end{split}
\end{equation}
Prepotential (after $\epsilon_1 \rightarrow i \epsilon_1$, $\epsilon_2 \rightarrow i \epsilon_2$):
\begin{equation}
\begin{split}
F_{2,1,2} &= \dfrac{1}{8 \epsilon_1^2 \epsilon_2^2} - \dfrac{1}{2 \epsilon_1 \epsilon_2} \left(\dfrac{1}{4} \ln^2 z_0 + \dfrac{1}{2} \ln z_0 \ln z_1 + \dfrac{1}{2} \ln^2 z_1 \right) \\
& + \dfrac{\epsilon}{2 \epsilon_1 \epsilon_2} \left( \dfrac{1}{12}\ln^3 z_0 + \dfrac{1}{4}\ln^2 z_0 \ln z_1 + \dfrac{1}{4}\ln z_0 \ln^2 z_1  + \dfrac{1}{6}\ln^3 z_1 + \text{Li}_3(z_1) + \text{Li}_3(z_0 z_1) \right) 
\end{split}
\end{equation}

\item \textbf{Case $p = 3$} \\

Poles:

\vspace*{0.2 cm}
\begin{center}
\begin{tabular}{ccl}
\ytableausetup{mathmode, boxsize=0.7em, aligntableaux=center} 
\begin{ytableau} \scriptstyle 0 & \scriptstyle 1 & \scriptstyle 2 & \scriptstyle 0 & \scriptstyle 1 & \scriptstyle 2  \end{ytableau} & $\Longleftrightarrow$ & 
\Bigg\{
\begin{tabular}{l}
$\lambda^{(0)}_1 = -i a_1^{(0)} - i \frac{\epsilon}{2}$ \;, 
$\lambda^{(1)}_1 = \lambda^{(0)}_1 - i \epsilon_1$  \;, 
$\lambda^{(2)}_1 = \lambda^{(0)}_1 - 2 i \epsilon_1$ \\
$\lambda^{(0)}_2 = \lambda^{(0)}_1 - 3 i \epsilon_1$ \;, 
$\lambda^{(1)}_2 = \lambda^{(0)}_1 - 4 i \epsilon_1$ \;, 
$\lambda^{(2)}_2 = \lambda^{(0)}_1 - 5 i \epsilon_1$ \\
\end{tabular} 
\\
\begin{ytableau} \scriptstyle 2 \\ \scriptstyle 0 & \scriptstyle 1 & \scriptstyle 2 & \scriptstyle 0 & \scriptstyle 1 \end{ytableau} & $\Longleftrightarrow$ & 
\Bigg\{
\begin{tabular}{l}
$\lambda^{(0)}_1 = -i a_1^{(0)} - i \frac{\epsilon}{2}$ \;, 
$\lambda^{(1)}_1 = \lambda^{(0)}_1 - i \epsilon_1$  \;, 
$\lambda^{(2)}_1 = \lambda^{(0)}_1 - 2 i \epsilon_1$ \\
$\lambda^{(0)}_2 = \lambda^{(0)}_1 - 3 i \epsilon_1$ \;, 
$\lambda^{(1)}_2 = \lambda^{(0)}_1 - 4 i \epsilon_1$ \;, 
$\lambda^{(2)}_2 = \lambda^{(0)}_1 - i \epsilon_2$ \\
\end{tabular} 
\\ 
\begin{ytableau} \scriptstyle 1 \\ \scriptstyle 2 \\ \scriptstyle 0 & \scriptstyle 1 & \scriptstyle 2 & \scriptstyle 0 \end{ytableau} & $\Longleftrightarrow$ & 
\Bigg\{
\begin{tabular}{l}
$\lambda^{(0)}_1 = -i a_1^{(0)} - i \frac{\epsilon}{2}$ \;, 
$\lambda^{(1)}_1 = \lambda^{(0)}_1 - i \epsilon_1$  \;, 
$\lambda^{(2)}_1 = \lambda^{(0)}_1 - 2 i \epsilon_1$ \\
$\lambda^{(0)}_2 = \lambda^{(0)}_1 - 3 i \epsilon_1$ \;, 
$\lambda^{(2)}_2 = \lambda^{(0)}_1 - i \epsilon_2$ \;, 
$\lambda^{(1)}_2 = \lambda^{(0)}_1 - 2 i \epsilon_2$ \\
\end{tabular} 
\\ 
\begin{ytableau} \scriptstyle 2 & \scriptstyle 0 & \scriptstyle 1 \\ \scriptstyle 0 & \scriptstyle 1 & \scriptstyle 2 \end{ytableau} & $\Longleftrightarrow$ & 
\Bigg\{
\begin{tabular}{l}
$\lambda^{(0)}_1 = -i a_1^{(0)} - i \frac{\epsilon}{2}$ \;, 
$\lambda^{(1)}_1 = \lambda^{(0)}_1 - i \epsilon_1$  \;, 
$\lambda^{(2)}_1 = \lambda^{(0)}_1 - 2 i \epsilon_1$ \\
$\lambda^{(2)}_2 = \lambda^{(0)}_1 - i \epsilon_2$ \;, 
$\lambda^{(0)}_2 = \lambda^{(0)}_1 - i \epsilon_1 - i \epsilon_2$ \;, 
$\lambda^{(1)}_2 = \lambda^{(0)}_1 - 2 i \epsilon_1 - i \epsilon_2$ \\
\end{tabular} 
\\
\begin{ytableau} \scriptstyle 1 \\ \scriptstyle 2 & \scriptstyle 0 \\ \scriptstyle 0 & \scriptstyle 1 & \scriptstyle 2 \end{ytableau} & $\Longleftrightarrow$ &
\Bigg\{
\begin{tabular}{l}
$\lambda^{(0)}_1 = -i a_1^{(0)} - i \frac{\epsilon}{2}$ \;, 
$\lambda^{(1)}_1 = \lambda^{(0)}_1 - i \epsilon_1$  \;, 
$\lambda^{(2)}_1 = \lambda^{(0)}_1 - 2 i \epsilon_1$ \\
$\lambda^{(2)}_2 = \lambda^{(0)}_1 - i \epsilon_2$ \;, 
$\lambda^{(0)}_2 = \lambda^{(0)}_1 - i \epsilon_1 - i \epsilon_2$ \;, 
$\lambda^{(1)}_2 = \lambda^{(0)}_1 - 2 i \epsilon_2$ \\
\end{tabular} 
\\ 
\begin{ytableau} \scriptstyle 1 & \scriptstyle 2 \\ \scriptstyle 2 & \scriptstyle 0 \\ \scriptstyle 0 & \scriptstyle 1 \end{ytableau} & $\Longleftrightarrow$ &
\Bigg\{
\begin{tabular}{l}
$\lambda^{(0)}_1 = -i a_1^{(0)} - i \frac{\epsilon}{2}$ \;, 
$\lambda^{(1)}_1 = \lambda^{(0)}_1 - i \epsilon_1$  \;, 
$\lambda^{(2)}_1 = \lambda^{(0)}_1 - i \epsilon_2$ \\
$\lambda^{(0)}_2 = \lambda^{(0)}_1 - i \epsilon_1 - i \epsilon_2$ \;, 
$\lambda^{(1)}_2 = \lambda^{(0)}_1 - 2 i \epsilon_2$ \;, 
$\lambda^{(2)}_2 = \lambda^{(0)}_1 - i \epsilon_1 - 2 i \epsilon_2$ \\
\end{tabular} 
\end{tabular}
\end{center}
\begin{center}
\begin{tabular}{ccl}
\begin{ytableau} \scriptstyle 0 \\ \scriptstyle 1 \\ \scriptstyle 2 \\ \scriptstyle 0 & \scriptstyle 1 & \scriptstyle 2 \end{ytableau} & $\Longleftrightarrow$ &
\Bigg\{
\begin{tabular}{l}
$\lambda^{(0)}_1 = -i a_1^{(0)} - i \frac{\epsilon}{2}$ \;, 
$\lambda^{(1)}_1 = \lambda^{(0)}_1 - i \epsilon_1$  \;, 
$\lambda^{(2)}_1 = \lambda^{(0)}_1 - 2 i \epsilon_1$ \\
$\lambda^{(2)}_2 = \lambda^{(0)}_1 - i \epsilon_2$ \;, 
$\lambda^{(1)}_2 = \lambda^{(0)}_1 - 2 i \epsilon_2$ \;, 
$\lambda^{(0)}_2 = \lambda^{(0)}_1 - 3 i \epsilon_2$ \\
\end{tabular} 
\\ 
\begin{ytableau} \scriptstyle 2 \\ \scriptstyle 0 \\ \scriptstyle 1 \\ \scriptstyle 2 \\ \scriptstyle 0 & \scriptstyle 1 \end{ytableau} & $\Longleftrightarrow$ &
\Bigg\{
\begin{tabular}{l}
$\lambda^{(0)}_1 = -i a_1^{(0)} - i \frac{\epsilon}{2}$ \;, 
$\lambda^{(1)}_1 = \lambda^{(0)}_1 - i \epsilon_1$  \;, 
$\lambda^{(2)}_1 = \lambda^{(0)}_1 - i \epsilon_2$ \\
$\lambda^{(1)}_2 = \lambda^{(0)}_1 - 2 i \epsilon_2$ \;, 
$\lambda^{(0)}_2 = \lambda^{(0)}_1 - 3 i \epsilon_2$ \;, 
$\lambda^{(2)}_2 = \lambda^{(0)}_1 - 4 i \epsilon_2$ \\
\end{tabular} 
\\ 
\begin{ytableau} \scriptstyle 1 \\ \scriptstyle 2 \\ \scriptstyle 0 \\ \scriptstyle 1 \\ \scriptstyle 2 \\ \scriptstyle 0 \end{ytableau} & $\Longleftrightarrow$ &
\Bigg\{
\begin{tabular}{l}
$\lambda^{(0)}_1 = -i a_1^{(0)} - i \frac{\epsilon}{2}$ \;, 
$\lambda^{(2)}_1 = \lambda^{(0)}_1 - i \epsilon_2$  \;, 
$\lambda^{(1)}_1 = \lambda^{(0)}_1 - 2 i \epsilon_2$ \\
$\lambda^{(0)}_2 = \lambda^{(0)}_1 - 3 i \epsilon_2$ \;, 
$\lambda^{(2)}_2 = \lambda^{(0)}_1 - 4 i \epsilon_2$ \;, 
$\lambda^{(1)}_2 = \lambda^{(0)}_1 - 5 i \epsilon_2$ \\
\end{tabular}
\end{tabular}  
\end{center}
\vspace*{0.2 cm}

Equivariant mirror map: 
\begin{equation}
Z_{\text{v}} \rightarrow (1+z_0z_1z_2)^{2 i r \epsilon} Z_{\text{v}} \;\;\;,\;\;\; 
Z_{\text{av}} \rightarrow (1+\overline{z}_0\overline{z}_1\overline{z}_2)^{2 i r \epsilon} Z_{\text{av}}
\end{equation}
Normalization of the 1-loop factor: 
\begin{equation}
Z_{\text{1l}} \rightarrow (z_0z_1z_2\overline{z}_0\overline{z}_1\overline{z}_2)^{-2 i r a_1^{(0)} - i r \epsilon} \left( \dfrac{\Gamma (1 - i r \epsilon)}{\Gamma (1 + i r \epsilon)} \right)^2 Z_{\text{1l}}
\end{equation}
Partition function:
\begin{equation}
\begin{split}
Z_{2,1,3}^{S^2,\text{norm}} &= \dfrac{1}{18 \epsilon_1^2 \epsilon_2^2} + \dfrac{1}{3 \epsilon_1 \epsilon_2} \Bigg(\dfrac{1}{4}\ln^2(z_0 \overline{z}_0) + \dfrac{1}{2} \ln(z_0 \overline{z}_0)\ln(z_1 \overline{z}_1) + \dfrac{1}{2} \ln(z_0 \overline{z}_0)\ln(z_2 \overline{z}_2) \\
& + \dfrac{5}{6} \ln(z_1 \overline{z}_1)\ln(z_2 \overline{z}_2) + \dfrac{7}{12} \ln^2(z_1 \overline{z}_1) + \dfrac{7}{12} \ln^2(z_2 \overline{z}_2) \Bigg) \\ 
& - i \dfrac{1}{3 \epsilon_1 \epsilon_2} \Bigg( -\dfrac{7\epsilon_1 + 11\epsilon_2}{36}\ln^3(z_1 \overline{z}_1) - \dfrac{11\epsilon_1 + 7\epsilon_2}{36}\ln^3(z_2 \overline{z}_2) \\
& - \dfrac{5\epsilon_1 + 7\epsilon_2}{12}\ln^2(z_1 \overline{z}_1)\ln(z_2 \overline{z}_2) - \dfrac{7\epsilon_1 + 5\epsilon_2}{12}\ln(z_1 \overline{z}_1)\ln^2(z_2 \overline{z}_2) \Bigg) \\
& - i \dfrac{\epsilon}{3 \epsilon_1 \epsilon_2} \Bigg( 9 \zeta(3) - \dfrac{1}{12}\ln^3(z_0 \overline{z}_0) -\dfrac{1}{4}\ln^2(z_0 \overline{z}_0)\ln(z_1 \overline{z}_1) -\dfrac{1}{4}\ln(z_0 \overline{z}_0)\ln^2(z_1 \overline{z}_1) \\
& - \dfrac{1}{4}\ln^2(z_0 \overline{z}_0)\ln(z_2 \overline{z}_2) -\dfrac{1}{4}\ln(z_0 \overline{z}_0)\ln^2(z_2 \overline{z}_2) -\dfrac{1}{2}\ln(z_0 \overline{z}_0)\ln(z_1 \overline{z}_1)\ln(z_2 \overline{z}_2) \\
& + 2 (\text{Li}_3(z_1) + \text{Li}_3(z_2) + \text{Li}_3(z_1 z_2) + \text{Li}_3(z_0 z_1 z_2)) \\
& + 2 (\text{Li}_3(\overline{z}_1) + \text{Li}_3(\overline{z}_2) + \text{Li}_3(\overline{z}_1\overline{z}_2) + \text{Li}_3(\overline{z}_0 \overline{z}_1\overline{z}_2))\\
& - \ln(z_1 \overline{z}_1)(\text{Li}_2(z_1) + \text{Li}_2(\overline{z}_1)) - \ln(z_2 \overline{z}_2)(\text{Li}_2(z_2) + \text{Li}_2(\overline{z}_2)) \\
& - \ln(z_1 z_2 \overline{z}_1 \overline{z}_2)(\text{Li}_2(z_1 z_2) + \text{Li}_2(\overline{z}_1 \overline{z}_2))\\
& - \ln(z_0 z_1 z_2 \overline{z}_0 \overline{z}_1 \overline{z}_2)(\text{Li}_2(z_0 z_1 z_2) + \text{Li}_2(\overline{z}_0\overline{z}_1 \overline{z}_2)) \Bigg)
\end{split}
\end{equation}
Prepotential (after $\epsilon_1 \rightarrow i \epsilon_1$, $\epsilon_2 \rightarrow i \epsilon_2$):
\begin{equation}
\begin{split}
F_{2,1,3} &= \dfrac{1}{18 \epsilon_1^2 \epsilon_2^2} - \dfrac{1}{3 \epsilon_1 \epsilon_2} \Bigg(\dfrac{1}{4}\ln^2 z_0 + \dfrac{1}{2} \ln z_0  \ln z_1  + \dfrac{1}{2} \ln z_0 \ln z_2 \\
& + \dfrac{5}{6} \ln z_1 \ln z_2 + \dfrac{7}{12} \ln^2 z_1  + \dfrac{7}{12} \ln^2 z_2  \Bigg) \\
& + \dfrac{1}{3 \epsilon_1 \epsilon_2} \Bigg( \dfrac{7\epsilon_1 + 11\epsilon_2}{36}\ln^3 z_1 + \dfrac{11\epsilon_1 + 7\epsilon_2}{36}\ln^3 z_2 \\
& + \dfrac{5\epsilon_1 + 7\epsilon_2}{12}\ln^2 z_1 \ln z_2  + \dfrac{7\epsilon_1 + 5\epsilon_2}{12}\ln z_1 \ln^2 z_2  \Bigg) \\
& + \dfrac{\epsilon}{3 \epsilon_1 \epsilon_2} \Bigg( \dfrac{1}{12}\ln^3 z_0 + \dfrac{1}{4}\ln^2 z_0 \ln z_1 + \dfrac{1}{4}\ln z_0 \ln^2 z_1 \\
& - \dfrac{1}{4}\ln^2 z_0 \ln z_2 -\dfrac{1}{4}\ln z_0 \ln^2 z_2 - \dfrac{1}{2}\ln z_0 \ln z_1 \ln z_2 \Bigg) \\
& + \dfrac{\epsilon}{3 \epsilon_1 \epsilon_2} (\text{Li}_3(z_1) + \text{Li}_3(z_2) + \text{Li}_3(z_1 z_2) + \text{Li}_3(z_0 z_1 z_2))
\end{split}
\end{equation}
\end{itemize} 
 
\noindent \textit{The $N=2$ sector, $p = 2$} \\

\noindent Let us now focus on $p=2$. Consider the case in which $\vec{N} = (N_0, N_1)$ is required to satisfy $N_0 + N_1 = 2$; the construction in \cite{Nakajima} then forces us to the two possibilities $\vec{N} = (0,2)$, $\vec{k} = (k-1,k)$ or $\vec{N} = (2,0)$, $\vec{k} = (k,k)$, corresponding respectively to fractional or integral instanton number $\frac{k_0 + k_1}{2}$. We can compute the Gromov-Witten prepotential for small values of $k$ as we did for in the previous examples, the main difference being the absence of equivariant mirror map; let us present here the final results. \\

\begin{itemize}

\item \textbf{Case $\vec{N} = (0,2)$, $\vec{k} = (0,1)$} \\

Poles:

\vspace*{0.2 cm}
\begin{center}
\begin{tabular}{ccl}
\ytableausetup{mathmode, boxsize=0.7em, aligntableaux=center} 
$\left( (\bullet, \bullet), (\begin{ytableau} \scriptstyle 1 \end{ytableau}, \bullet) \right)$ & $\Longleftrightarrow$ & 
$\lambda^{(1)}_1 = -i a_1^{(1)} - i \frac{\epsilon}{2}$
\\
$\left( (\bullet, \bullet), (\bullet, \begin{ytableau} \scriptstyle 1 \end{ytableau}) \right)$ & $\Longleftrightarrow$ & 
$\lambda^{(1)}_1 = -i a_2^{(1)} - i \frac{\epsilon}{2}$
\\ 
\end{tabular}  
\end{center}
\vspace*{0.2 cm}

Normalization of the 1-loop factor: 
\begin{equation}
Z_{\text{1l}} \rightarrow (z_1\overline{z}_1)^{- i r \frac{a_1^{(1)} + a_2^{(1)}}{2}} \dfrac{\Gamma (1 - i r \epsilon)}{\Gamma (1 + i r \epsilon)} Z_{\text{1l}}
\end{equation}
Partition function:
\begin{equation}
\begin{split}
Z_{(0,1),(0,2),2}^{S^2,\text{norm}} &= \frac{2}{(a^{(1)}_1-a^{(1)}_2)^2 - \epsilon^2} - \dfrac{1}{4} \ln^2 (z_1 \overline{z}_1) \\
& + i \epsilon \left( 4\zeta(3) - \dfrac{1}{12} \ln^3(z_1 \overline{z}_1) + 2 (\text{Li}_3(z_1) + \text{Li}_3(\overline{z}_1)) - \ln (z_1 \overline{z}_1) (\text{Li}_2(z_1) + \text{Li}_2(\overline{z}_1)) \right)
\end{split}
\end{equation}
Prepotential (after $\epsilon_1 \rightarrow i \epsilon_1$, $\epsilon_2 \rightarrow i \epsilon_2$, $a^{(1)}_1 \rightarrow i a^{(1)}_1$, $a^{(1)}_2 \rightarrow i a^{(1)}_2$):
\begin{equation}
\begin{split}
F_{(0,1),(0,2),2} &= \frac{2}{\epsilon^2 - (a^{(1)}_1-a^{(1)}_2)^2} - \dfrac{1}{4} \ln^2 z_1 + \dfrac{\epsilon}{12} \ln^3 z_1 + \epsilon \text{Li}_3(z_1)
\end{split}
\end{equation}

\item \textbf{Case $\vec{N} = (2,0)$, $\vec{k} = (1,1)$} \\

Poles:

\vspace*{0.2 cm}
\begin{center}
\begin{tabular}{ccl}
\ytableausetup{mathmode, boxsize=0.7em, aligntableaux=center} 
$\left( (\begin{ytableau} \scriptstyle 0 & \scriptstyle 1 \end{ytableau}, \bullet) , (\bullet, \bullet) \right)$ & $\Longleftrightarrow$ & 
$\lambda^{(0)}_1 = -i a_1^{(0)} - i \frac{\epsilon}{2}$ \;, 
$\lambda^{(1)}_1 = \lambda^{(0)}_1 - i \epsilon_1$ 
\\
$\left( \left(\begin{ytableau} \scriptstyle 1 \\ \scriptstyle 0 \end{ytableau}, \bullet \right) , (\bullet, \bullet) \right)$ & $\Longleftrightarrow$ & 
$\lambda^{(0)}_1 = -i a_1^{(0)} - i \frac{\epsilon}{2}$ \;, 
$\lambda^{(1)}_1 = \lambda^{(0)}_1 - i \epsilon_2$ 
\\ 
$\left( (\bullet, \begin{ytableau} \scriptstyle 0 & \scriptstyle 1 \end{ytableau}) , (\bullet, \bullet) \right)$ & $\Longleftrightarrow$ & 
$\lambda^{(0)}_1 = -i a_2^{(0)} - i \frac{\epsilon}{2}$ \;, 
$\lambda^{(1)}_1 = \lambda^{(0)}_1 - i \epsilon_1$ 
\\
$\left( \left(\bullet, \begin{ytableau} \scriptstyle 1 \\ \scriptstyle 0 \end{ytableau} \right) , (\bullet, \bullet) \right)$ & $\Longleftrightarrow$ & 
$\lambda^{(0)}_1 = -i a_2^{(0)} - i \frac{\epsilon}{2}$ \;, 
$\lambda^{(1)}_1 = \lambda^{(0)}_1 - i \epsilon_2$ 
\\ 
\end{tabular}  
\end{center}
\vspace*{0.2 cm}

Normalization of the 1-loop factor: 
\begin{equation}
Z_{\text{1l}} \rightarrow (z_1 z_2 \overline{z}_1 \overline{z}_2)^{- i r \frac{a_1^{(0)} + a_2^{(0)}}{2}} \left( \dfrac{\Gamma (1 - i r \epsilon)}{\Gamma (1 + i r \epsilon)} \right)^2 Z_{\text{1l}}
\end{equation}
Partition function:
\begin{equation}
\begin{split}
Z_{(1,1),(2,0),2}^{S^2,\text{norm}} &= \frac{1}{2 \epsilon_1 \epsilon_2}\frac{2}{\left(\epsilon^2 - (a^{(0)}_1-a^{(0)}_2)^2\right)} \\
& + \dfrac{1}{2 \epsilon_1 \epsilon_2} \left( \dfrac{1}{4} \ln^2 (z_0 \overline{z}_0) + \dfrac{1}{2} \ln (z_0 \overline{z}_0) \ln (z_1 \overline{z}_1) + \dfrac{1}{4} \ln^2 (z_1 \overline{z}_1) \right)  \\
& + \frac{1}{2\left(\epsilon^2 - (a^{(0)}_1-a^{(0)}_2)^2\right)} \ln^2 (z_1 \overline{z}_1) \\
& - i \dfrac{\epsilon}{2 \epsilon_1 \epsilon_2} \Bigg(- \dfrac{1}{12} \ln^3 (z_0 \overline{z}_0) - \dfrac{1}{4} \ln^2(z_0 \overline{z}_0)  \ln (z_1 \overline{z}_1) \\
& - \dfrac{1}{4} \ln (z_0 \overline{z}_0) \ln^2 (z_1 \overline{z}_1) - \dfrac{1}{12} \ln^3 (z_1 \overline{z}_1) + 4 \zeta(3) \\
& + 2 (\text{Li}_3 (z_0 z_1) + \text{Li}_3 (\overline{z}_0\overline{z}_1) ) - \ln (z_0 z_1\overline{z}_0\overline{z}_1) (\text{Li}_2 (z_0 z_1) + \text{Li}_2 (\overline{z}_0\overline{z}_1) ) \Bigg) \\
& - i \frac{2 \epsilon}{\left(\epsilon^2 - (a^{(0)}_1-a^{(0)}_2)^2\right)} \Bigg( - \dfrac{1}{12} \ln^3 (z_1 \overline{z}_1) + 4 \zeta(3) + 2 (\text{Li}_3 (z_1) + \text{Li}_3 (\overline{z}_1) ) \\
& - \ln (z_1 \overline{z}_1) (\text{Li}_2 (z_1) + \text{Li}_2 (\overline{z}_1) ) \Bigg)
\end{split}
\end{equation}
Prepotential (after $\epsilon_1 \rightarrow i \epsilon_1$, $\epsilon_2 \rightarrow i \epsilon_2$, $a^{(0)}_1 \rightarrow i a^{(0)}_1$, $a^{(0)}_2 \rightarrow i a^{(0)}_2$):
\begin{equation}
\begin{split}
F_{(1,1),(2,0),2} &= \frac{1}{2 \epsilon_1 \epsilon_2}\frac{2}{\left(\epsilon^2 - (a^{(0)}_1-a^{(0)}_2)^2\right)} 
- \dfrac{1}{2 \epsilon_1 \epsilon_2} \left( \dfrac{1}{4} \ln^2 z_0 + \dfrac{1}{2} \ln z_0 \ln z_1 + \dfrac{1}{4} \ln^2 z_1 \right) \\
& - \frac{1}{2\left(\epsilon^2 - (a^{(0)}_1-a^{(0)}_2)^2\right)} \ln^2 z_1 \\
& + \dfrac{\epsilon}{2 \epsilon_1 \epsilon_2} \Bigg( \dfrac{1}{12} \ln^3 z_0 + \dfrac{1}{4} \ln^2 z_0 \ln z_1 + \dfrac{1}{4} \ln z_0 \ln^2 z_1 + \dfrac{1}{12} \ln^3 z_1 + \text{Li}_3 (z_0 z_1) \Bigg)\\
& + \frac{2 \epsilon}{\left(\epsilon^2 - (a^{(0)}_1-a^{(0)}_2)^2\right)} \Bigg( \dfrac{1}{12} \ln^3 z_1 + \text{Li}_3 (z_1) \Bigg)
\end{split}
\end{equation} 

\end{itemize} 


\bibliography{Bibliography}
\bibliographystyle{JHEP-2}


\end{document}